\documentclass[10pt,fullpage]{article}

\usepackage{fullpage}
\usepackage{booktabs} % For formal tables
\usepackage{array}
\usepackage{listings}
\usepackage{algorithm, algpseudocode}
\usepackage{float}
\usepackage{subcaption}
\usepackage{color,soul}
\usepackage{multirow, makecell}
\usepackage{acronym} % for acronyms
\usepackage{hhline}
\usepackage{enumitem}
\usepackage{comment}
\usepackage{amsmath}
\usepackage{url}
\usepackage{hyperref}
\usepackage{times}
\usepackage{multicol}

\usepackage{lipsum} % Dummytext
\usepackage{xargs} % Use more than one optional parameter in a new commands
\usepackage{xcolor}
\usepackage[colorinlistoftodos,prependcaption,textsize=tiny]{todonotes}

\newcommandx{\unsure}[2][1=]{\todo[linecolor=red,backgroundcolor=red!25,bordercolor=red,#1]{#2}}
\newcommandx{\change}[2][1=]{\todo[linecolor=blue,backgroundcolor=blue!25,bordercolor=blue,#1]{#2}}

\newcommandx{\info}[2][1=]{\todo[linecolor=OliveGreen,backgroundcolor=width=13cm,height=4cmOliveGreen!25,bordercolor=OliveGreen,#1]{#2}}
\newcommandx{\improvement}[2][1=]{\todo[linecolor=Plum,backgroundcolor=Plum!25,bordercolor=Plum,#1]{#2}}
\newcommandx{\thiswillnotshow}[2][1=]{\todo[disable,#1]{#2}}

\algnewcommand{\LineComment}[1]{\State \(\triangleright\) #1}

\makeatletter
\newenvironment{sql}%
 {\vskip 5pt\begin{list}{}{%
  \setlength{\topsep}{0pt}\setlength{\partopsep}{0pt}\setlength{\parskip}{0pt}%
  \setlength{\parsep}{0pt}\setlength{\labelwidth}{0pt}%
  \setlength{\rightmargin}{0pt}\setlength{\leftmargin}{0pt}%
  \setlength{\labelsep}{0pt}%
  \obeylines\@vobeyspaces\normalfont\ttfamily%
  \item[]}}
 {\end{list}\vskip5pt\noindent}
\makeatother
\acrodef{JOB}{Join Order Benchmark}
\acrodef{RDBMS}{Relational Database Management Systems}
\acrodef{QEP}{Query Execution Plan}
\acrodef{ESC}{Exact Selectivity Estimation}
\acrodef{SQL}{Structured Query Language}
\acrodef{eDFS}{Exhaustive Depth First Search}
\acrodef{GPU}{Graphics Processing Units}
\acrodef{CPU}{Central Processing Unit}
\acrodef{AQP}{Approximate Query Processing}
\acrodef{AdQP}{Adaptive Query Processing}
\acrodef{NUMA}{Non-Uniform Memory Access}
\acrodef{NVMe}{Non-Volatile Memory Express}
\acrodef{SIMD}{Single-Instruction, multiple-data)}

\begin{document}

\date{February 2021}

\title{Online Sketch-based Query Optimization}

\author{
Yesdaulet Izenov, Asoke Datta, Florin Rusu, Jun Hyung Shin\\
\{yizenov,adatta2,frusu,jshin33\}@ucmerced.edu\\
University of California Merced
}

\maketitle

%%%%%%%%%%%%%%%%%%%%%%%%%%%%%%%%%%%%%%%%%%%%%%%%%%%%%%%%%%%%%%%%%
%\input{abstract}
\begin{abstract}
Cost-based query optimization remains a critical task in relational databases even after decades of research and industrial development. Query optimizers rely on a large range of statistical synopses -- including attribute-level histograms and table-level samples -- for accurate cardinality estimation. As the complexity of selection predicates and the number of join predicates increase, two problems arise. First, statistics cannot be incrementally composed to effectively estimate the cost of the sub-plans generated in plan enumeration. Second, small errors are propagated exponentially through join operators, which can lead to severely sub-optimal plans.

In this paper, we introduce COMPASS, a novel query optimization paradigm for in-memory databases based on a single type of statistics---Fast-AGMS sketches. In COMPASS, query optimization and execution are intertwined. Selection predicates and sketch updates are pushed-down and evaluated online during query optimization. This allows Fast-AGMS sketches to be computed only over the relevant tuples---which enhances cardinality estimation accuracy. Plan enumeration is performed over the query join graph by incrementally composing attribute-level sketches---not by building a separate sketch for every sub-plan.

We prototype COMPASS in MapD -- an open-source parallel database -- and perform extensive experiments over the complete JOB benchmark. The results prove that COMPASS generates better execution plans -- both in terms of cardinality and runtime -- compared to four other database systems. Overall, COMPASS achieves a speedup ranging from 1.35X to 11.28X in cumulative query execution time over the considered competitors.
\end{abstract}

%%%%%%%%%%%%%%%%%%%%%%%%%%%%%%%%%%%%%%%%%%%%%%%%%%%%%%%%%%%%%%%%%
%\input{introduction}
\section{INTRODUCTION}\label{sec:intro}

\begin{figure*}[htbp]
  \centering
  \includegraphics[width=\textwidth]{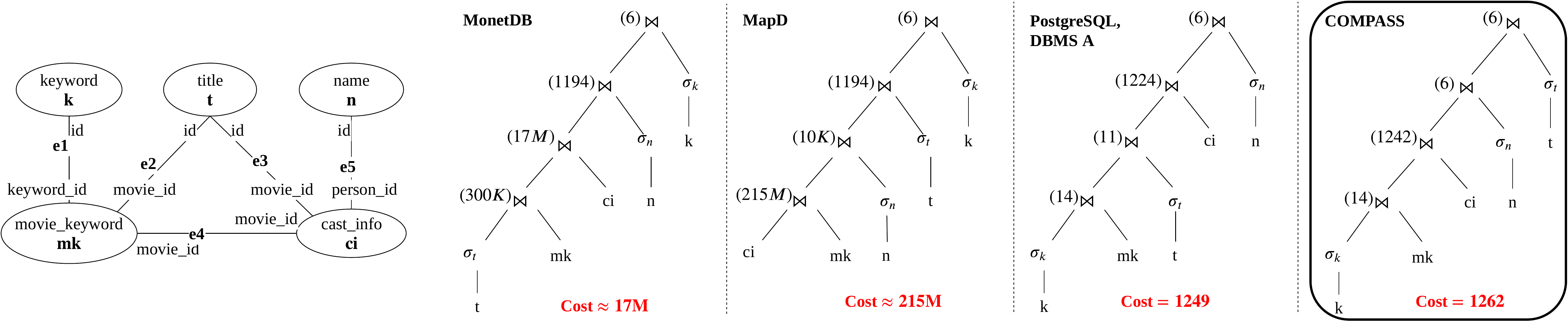}
  \caption{Join graph and corresponding execution plans for query JOB 6a. The numbers represent cardinality.}
  \label{fig:query_6a_plans}
\end{figure*}

Consider query 6a from the JOB benchmark~\cite{Leis:JOB:vldb-2018}:
\begin{sql}
\textbf{SELECT} MIN(k.keyword), MIN(n.name), MIN(t.title)
\textbf{FROM} cast\_info \textbf{ci}, keyword \textbf{k}, movie\_keyword \textbf{mk}, name \textbf{n}, title \textbf{t}
\textbf{WHERE}
  $\triangleright$ \textit{selection predicates}
    k.keyword = 'marvel-cinematic-universe' \textbf{AND}
    n.name LIKE '\%Downey\%Robert\%' \textbf{AND} t.production\_year > 2010 \textbf{AND}
  $\triangleright$ \textit{join predicates}
    k.id = mk.keyword\_id \textbf{AND} t.id = mk.movie\_id \textbf{AND} t.id = ci.movie\_id \textbf{AND}
    ci.movie\_id = mk.movie\_id \textbf{AND} n.id = ci.person\_id
\end{sql}
The query has 3 selection predicates -- point, subset, and range -- and joins 5 tables with 5 join predicates---there is a triangle subquery between tables \textit{t}, \textit{mk}, and \textit{ci}. The corresponding join graph is depicted in Figure~\ref{fig:query_6a_plans}. For each join, the graph contains a named edge $e{1}$--$e{5}$ that connects the tables involved in the join predicate. For example, edge $e{1}$ represents the join predicate \textit{k.id = mk.keyword\_id}.

%%%%%%%%%%%%%%%%%%%%%%%%%%%%%%%%%%%%%%%%%%%%%%%%%%%%%%%%%%%%%%%%%
%%%%%%%%%%%%%%%%%%%%%%%%%%%%%%%%%%%%%%%%%%%%%%%%%%%%%%%%%%%%%%%%%

Figure~\ref{fig:query_6a_plans} also includes the execution plans together with their cost -- the total cardinality of the intermediate results -- for COMPASS and the four other databases considered in the paper. Although all the plans are left-deep trees, their cost ranges from $1,249$ to $215$ millions tuples. This is entirely due to the statistics used for cardinality estimation. MapD~\cite{mapd} does not use any statistics, thus its cost is orders of magnitude higher. The plan is determined by sorting the tables in decreasing order of their size---number of tuples. MonetDB~\cite{monetdb-web} has a rule-based optimizer with minimum support for statistics~\cite{Stratos:MONETDB:deb-2012} which generates a better plan. The reason why both of these systems have primitive optimizers is because they are relatively ``young'' and are targeted at modern architectures. They try to compensate bad plans with highly-optimized execution engines that make use of extensive in-memory processing supported by massive multithread parallelism and vectorized instructions. However, this approach is clearly limited.

PostgreSQL~\cite{postgres} and the industrial-grade DBMS A -- name anonymized for legal reasons -- are ``mature'' databases with advanced query optimizers. In order to find the much better plan, they use a large variety of statistics. Histograms, most frequent values, and number of distincts are used to estimate the selectivity of the point predicate on attribute \textit{k.keyword} and of the range predicate on \textit{t.production\_year}. The subset LIKE predicate on \textit{n.name} is estimated with table-level samples. Estimating join cardinality requires correlated statistics on the join attributes. While such statistics exist, e.g., correlated samples~\cite{Kader:ROX:sigmod-2009,Yu:CS2:sigmod-2013,Leis:index-join-sample:cidr-2017}, they require the existence of indexes on every join attribute combination, which severely limits their applicability in the case of multi-way joins. As a result, even advanced optimizers rely on crude formulas that assume uniformity, inclusion, and independence---which are likely to produce highly sub-optimal execution plans~\cite{Leis:QOREALLY:pvldb-2015}. Since implementing and maintaining these many statistics requires considerable effort, it is completely understandable that only mature systems implement them.

%%%%%%%%%%%%%%%%%%%%%%%%%%%%%%%%%%%%%%%%%%%%%%%%%%%%%%%%%%%%%%%%%
%\textbf{Problem.}
\paragraph*{Problem.}
We investigate how to design a lightweight -- yet effective -- query optimizer for modern in-memory databases. We have two design principles. First, we aim to capitalize on the highly-parallel execution engine in the query optimization process. Since query execution is already fast, it is challenging to minimize the overhead incurred by the additional optimization. Second, the type and number of synopses included in the optimizer has to be minimal. Our goal is to employ a single type of synopsis built exclusively for single-attributes and without the requirement of additional data structures such as indexes. The challenge is to design a composable -- and consistent -- synopsis that provides incremental cardinality estimates for the sub-plans generated in plan enumeration.

%%%%%%%%%%%%%%%%%%%%%%%%%%%%%%%%%%%%%%%%%%%%%%%%%%%%%%%%%%%%%%%%%
%\textbf{The COMPASS query optimizer.}
\paragraph*{COMPASS query optimizer.}
We introduce the online sketch-based COMPASS query optimizer. Fast-AGMS sketches~\cite{Cormode:FAGMS:vldb-2005} are the only statistics present in COMPASS. These sketches are a type of correlated synopses for join cardinality estimation~\cite{Rusu:SAS:sigmod-2007,Rusu:SJSE:tods-2008} that use small space, can be computed efficiently in a single scan over the data, are linearly composable, and -- more importantly -- have statistically high accuracy. These properties allow for Fast-AGMS sketches to be computed online in COMPASS by leveraging the optimized parallel execution engine in modern databases. This is realized by decomposing query processing into two stages performed before and after the optimization. In the first stage, selection predicates are pushed-down and Fast-AGMS sketches are built concurrently only over the relevant tuples. Sketches are built for each two-way join independently---not for every combination of tables. In the query optimization stage, plan enumeration is performed over the join graph by incrementally composing the corresponding two-way join sketches in order to estimate the cardinality of multi-way joins. The optimal join ordering is finally passed to the execution engine to finalize the query. As shown in Figure~\ref{fig:query_6a_plans}, COMPASS identifies a plan as good as PostgreSQL and DBMS A, while relying exclusively on a single synopsis---Fast-AGMS sketches. In addition to the novel query optimization paradigm, we make the following technical contributions:

\begin{itemize}[leftmargin=*,noitemsep,nolistsep]
\item We present a systematic approach of using sketches for join cardinality estimation in a query optimizer. This includes two-way and multi-way joins. We do this for two types of sketches---AGMS~\cite{Alon:AGMS:pods-1999} and Fast-AGMS~\cite{Cormode:FAGMS:vldb-2005}.

\item We introduce two novel strategies to extend Fast-AGMS sketches to multi-way join cardinality estimation. The first strategy -- sketch partitioning -- is a theoretically sound estimator for a given multi-way join. Since it does not support composition, sketch partitioning is not scalable for join order enumeration. The second strategy -- sketch merging -- addresses scalability by incrementally creating multi-way sketches from two-way sketches. Although this is done heuristically for a certain multi-way join taken separately, all the multi-way joins with a given size are equally impacted. This property guarantees estimation consistency in plan enumeration.

\item We prototype COMPASS in MapD and perform extensive experiments over the complete JOB benchmark---113 queries. The results prove the reduced overhead COMPASS incurs -- below 500 milliseconds  -- while generating similar or better execution plans compared to the four databases systems included in Figure~\ref{fig:query_6a_plans}. COMPASS outperforms the other databases both in terms of the number of queries it obtains the best result on, as well as on the cumulative workload execution time.
\end{itemize}

%%%%%%%%%%%%%%%%%%%%%%%%%%%%%%%%%%%%%%%%%%%%%%%%%%%%%%%%%%%%%%%%%
%\textbf{Outline.}
\paragraph*{Outline.}
The paper is organized as follows. Background information on cost-based query optimization and sketches is given in Section~\ref{sec:prelims}. A high-level overview of COMPASS is presented in Section~\ref{sec:high-level}, followed by the technical details of sketch-based cardinality estimation in Section~\ref{sec:sketches}. The novel Fast-AGMS sketches for multi-way joins are introduced in Section~\ref{sec:fagms-join-order}. In Section~\ref{sec:enumeration}, we show how the sketches are integrated in a typical enumeration algorithm. The empirical evaluation of COMPASS is detailed in Section~\ref{sec:experiments}. We discuss related work in Section~\ref{sec:rel-work} and conclude with future work directions in Section~\ref{sec:conclusions}.

%%%%%%%%%%%%%%%%%%%%%%%%%%%%%%%%%%%%%%%%%%%%%%%%%%%%%%%%%%%%%%%%%
%\input{preliminaries}
\section{PRELIMINARIES}\label{sec:prelims}

%%%%%%%%%%%%%%%%%%%%%%%%%%%%%%%%%%%%%%%%%%%%%%%%%%%%%%%%%%%%%%%%%
%\subsection{Cost-based Query Optimization}
%\textbf{Cost-based query optimization.}
\paragraph*{Cost-based query optimization.}
The query optimization problem~\cite{Lohman:QOSP:2014,Leis:JOB:vldb-2018,Leis:QOREALLY:pvldb-2015,Chaudhuri:OQO:pods-1998} consists in finding the best execution plan -- which typically corresponds to the one with the fastest execution time -- for a given query. The search space is defined over all the valid plans -- combinations of relational algebra operators -- which can answer the query correctly. The number of potential plans is exponentially factorial in the number of tables. Thus, inspecting all of them is not practical for a large number of tables. \textit{Plan enumeration} is the procedure that defines the plans in the search space. Since the execution time of a plan cannot be determined without running it -- which defeats the purpose -- alternative cost functions are defined. The most common \textit{cost function is the total size -- or cardinality -- of the intermediate results produced by all the operators in the plan}. This function captures the correlation between the amount of accessed data and execution time---which is true in general. Computing the cardinality of a relational algebra operator is itself a difficult problem and requires knowledge about the data on which the operator is performed. This knowledge is captured by incomplete statistics -- or synopses -- about the data. Different classes of statistics~\cite{Cormode:SMD:fnt-2012} are useful for different relational operators. For example, attribute histograms and number of distinct values are optimal for selection predicates, while correlated samples are better for join predicates. With statistics, the cardinality can only be estimated---it is not exact. While accurate for simple predicates over a small number of attributes, \textit{cardinality estimation} becomes harder for correlated predicates and multi-way joins. This is not necessarily a problem if all the plans are equally impacted. However, estimation errors vary widely across sub-plans and this can potentially lead to a highly suboptimal plan. The COMPASS query optimizer includes solutions both for effective plan enumeration as well as incremental cardinality estimation for the enumerated sub-plans.

%%%%%%%%%%%%%%%%%%%%%%%%%%%%%%%%%%%%%%%%%%%%%%%%%%%%%%%%%%%%%%%%%
%\subsection{Parallel In-Memory Databases}
%\textbf{Parallel in-memory databases.}
\paragraph*{Parallel in-memory databases.}
Database systems for modern computing architectures rely on extensive in-memory processing supported by massive multithread parallelism and vectorized instructions. GPUs represent the pinnacle of such architectures, harboring thousands of SMT threads which execute tens of vectorized SIMD instructions simultaneously. MapD, Ocelot~\cite{ocelot}, CoGaDB~\cite{cogadb}, Kinetica~\cite{kinetica}, and Brytlyt~\cite{brytlyt} are a few examples of modern in-memory databases with GPU support. They provide relational algebra operators and pipelines for GPU architectures~\cite{He:RQC:tods-2009,Bress:GPUADS:tlkds-2014,Funke:PQP:sigmod-2018} that optimize memory access and bandwidth. This results in considerable performance improvement for certain classes of queries. However, these databases provide only primitive rule-based query optimization---if at all. This limits drastically their applicability to general workloads. In COMPASS, we leverage the optimized execution engine of MapD to build a lightweight -- yet accurate and general -- query optimizer based on a single type of synopsis.

%%%%%%%%%%%%%%%%%%%%%%%%%%%%%%%%%%%%%%%%%%%%%%%%%%%%%%%%%%%%%%%%%
%\subsection{Sketches}
%\textbf{Sketches.}
\paragraph*{Sketches.}
Sketch synopses~\cite{Cormode:SMD:fnt-2012} summarize the tuples of a relation as a set of random values. This is accomplished by projecting the domain of the relation on a significantly smaller domain using random functions or seeds. In the case of join attributes, correlation between attributes is maintained by using the same random function. While sketches compute only approximate results with probabilistic guarantees, they satisfy several major requirements of a query optimizer for in-memory databases---single-pass computation, small space, fast update and query time, and linearity:
\begin{itemize}[leftmargin=*,noitemsep,nolistsep]
\item A sketch is built by streaming over the input data and considers each tuple at most once.
\item A basic sketch is composed of a single counter and one or more random seeds---a few bytes. In order to improve accuracy, a standard method is to use multiple independent basic sketch instances. The number of instances is derived from the desired accuracy and confidence levels. In practice, very good accuracy can be achieved with sketches having size in kilobytes.
\item The update of a sketch with a new tuple consists in generating one or more random numbers and adding them to the sketch counter. The answer to a query involves simple arithmetic operations on the sketch. In the case of multiple sketches, both the update and query are applied to all the instances. Overall, update and query time are linearly proportional with the sketch size.
\item A sketch can be computed by partitioning the input relation into multiple parts, building a sketch for every part, and then merging the partial sketches. This mergeable property makes sketches amenable for parallel processing on modern hardware and can result in linear speedups in update and query time~\cite{Alonso:AugSketch:sigmod-2016,Charalampos:DelSketch:eurosys-2020}.
\end{itemize}

\noindent
While previous work addresses how to apply sketches to certain cardinality estimation problems that occur in query optimization, COMPASS is a complete query optimizer based exclusively on sketches. In addition to cardinality estimation, we show how to integrate the sketch estimations in plan enumeration. We are not aware of any work that integrates sketches effectively with plan enumeration. This is the main reason why sketches have not been integrated in a query optimizer before. COMPASS solves this problem.

%%%%%%%%%%%%%%%%%%%%%%%%%%%%%%%%%%%%%%%%%%%%%%%%%%%%%%%%%%%%%%%%%
%\input{new_paradigm}
\section{COMPASS: ONLINE SKETCH-BASED QUERY OPTIMIZATION}\label{sec:high-level}

In this section, we provide a high-level description of the COMPASS query optimization paradigm, while the technical details of cardinality estimation, join ordering, and plan enumeration are presented in Section~\ref{sec:sketches}, \ref{sec:fagms-join-order}, and \ref{sec:enumeration}, respectively.

%%%%%%%%%%%%%%%%%%%%%%%%%%%%%%%%%%%%%%%%%%%%%%%%%%%%%%%%%%%%%%%%%
%\subsection{Workflow}
%\textbf{Workflow.}

\begin{figure*}[htbp]
 \centering
 \includegraphics[width=\textwidth]{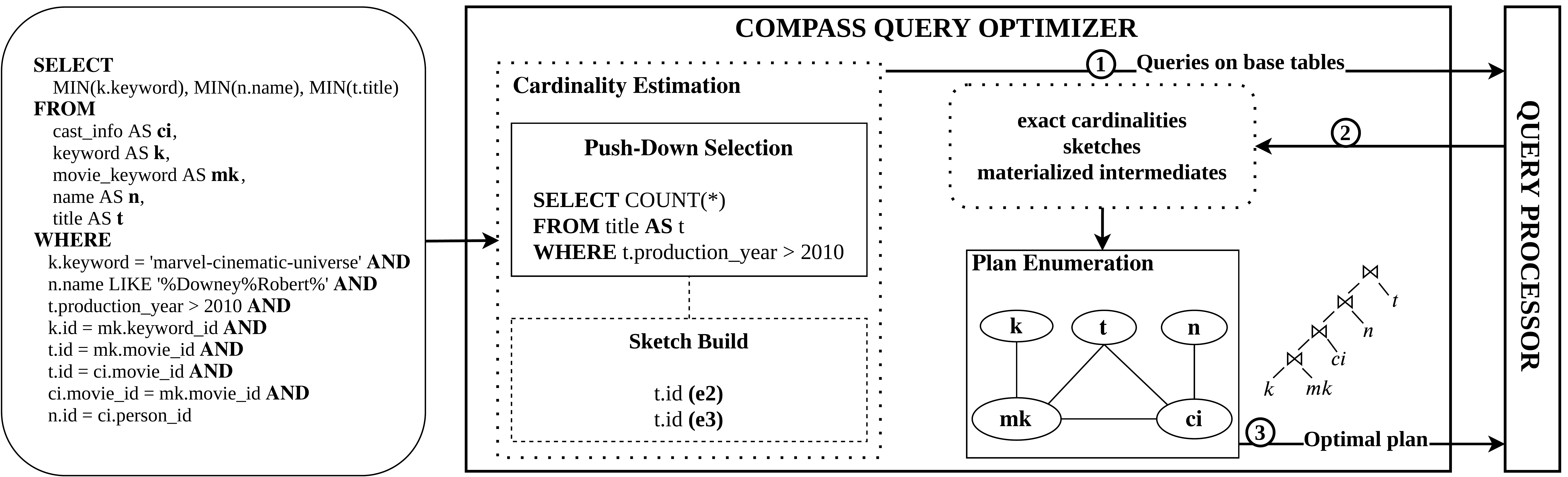}
 \caption{COMPASS workflow: online sketch-based query optimization for in-memory databases.}
 \label{fig:workflow}
\end{figure*}

\paragraph*{Workflow.}
The workflow performed by the COMPASS query optimizer is depicted in Figure~\ref{fig:workflow}. It consists of a two-step process that requires interaction with the query processor. First, the optimizer extracts the selection predicates and join attributes for every table. A sketch is built for every join attribute while performing the selection query on the base table, and only over the tuples that satisfy the predicate. Figure~\ref{fig:workflow} shows the procedure for table \textit{title} which has a range predicate and two join conditions---although both join predicates involve the same attribute \textit{t.id}, two independent sketches have to be built. COMPASS leverages the high-parallelism of in-memory databases and the mergeable property of sketches to execute this process with minimal overhead. Two additional optimizations can be applied to further reduce the overhead. Sketches for join attributes from tables without selection predicates can be built offline and plugged-in directly. Sketches can be built only over a sample~\cite{Rusu:SSD:icde-2009}, which, however, incurs a decrease in accuracy. In the second step of the workflow, plan enumeration is performed by estimating the cardinality of all the sub-plans using the sketches built in the first step. This is possible only because the attribute-level sketches we design are incrementally composable. Otherwise, separate sketches have to be built for every enumerated sub-plan. In our example, there are two sketches on attribute \textit{t.id}, one for join \textit{e2} and one for join \textit{e3} in the join graph (Figure~\ref{fig:query_6a_plans}). The sketch for \textit{e2} is included in all the sub-plans that contain this join attribute---similar for \textit{e3}. In a sub-plan that includes both \textit{e2} and \textit{e3}, these two sketches are first merged and then used in estimation as before. This process is performed incrementally during plan enumeration. Finally, the optimal plan is submitted for execution together with any materialized intermediates from step one.
% This would increase the number of sketches that have to be built in step one by a considerable factor.

%%%%%%%%%%%%%%%%%%%%%%%%%%%%%%%%%%%%%%%%%%%%%%%%%%%%%%%%%%%%%%%%%
%%%%%%%%%%%%%%%%%%%%%%%%%%%%%%%%%%%%%%%%%%%%%%%%%%%%%%%%%%%%%%%%%

%%%%%%%%%%%%%%%%%%%%%%%%%%%%%%%%%%%%%%%%%%%%%%%%%%%%%%%%%%%%%%%%%
%\subsection{Partitioned Query Execution}
%\textbf{Partitioned query execution.}

\paragraph*{Partitioned query execution.}
As shown in Figure~\ref{fig:workflow}, COMPASS intertwines query optimization and evaluation by partitioning execution into push-down selection (step 1) and join computation (step 3). Query optimization, i.e., join ordering plan enumeration, is performed in-between these two stages. Since plan enumeration and join computation are standard, we focus on push-down selection, where online sketch building is performed.  Push-down selection computes the exact selectivity cardinalities for all the base tables that have selections. This is similar to the ESC approach introduced in~\cite{Shin:ESC:arxiv-2019}. However, in addition to predicate evaluation, COMPASS also builds sketches for every join attribute in the table by piggybacking on the same traversal---sketch building is performed during the selection. Notice that this works both for sequential and index scans. It is important to emphasize that only the tuples that satisfy the predicate are included in the sketch, which increases their accuracy significantly. Moreover, the sketch update overhead is kept to the minimum necessary. While the exact cardinalities and sketches are always materialized due to their reduced size and role in optimization, the decision to materialize the selection output -- the intermediate result -- depends on its size. COMPASS follows the same approach as in~\cite{Shin:ESC:arxiv-2019}. If the intermediate size is smaller than a threshold, it is materialized. Otherwise, it is not, since the space reduction does not compensate for the access time reduction. Notice, though, that, even when intermediates are not materialized, sketches still contain only the relevant tuples for join cardinality estimation.

While the idea of partitioned query execution for XML processing is introduced in ROX~\cite{Kader:ROX:sigmod-2009}, the COMPASS approach is different in several aspects. First, similar to adaptive query processing~\cite{Deshpande:AQP:fnt-2007}, COMPASS works for relational data and operators. However, COMPASS does not change the plan while the query is executing. This is not necessary because the sketch-based optimization strategy finds better plans in the first place. ROX can decompose a join graph into an arbitrary number of stages, each of which requiring materialization. COMPASS, on the other hand, splits execution in exactly two stages and intermediate result materialization is only optional. The reason ROX requires materialization is because it uses chain sampling to estimate cardinalities. In order to provide acceptable accuracy, samples have to be extracted from the most recent intermediate results---not the base tables. Moreover, ROX chain sampling requires indexes on all the join attributes to guarantee a minimum sample size. This is a stringent constraint hardly satisfied in most real-world databases. Sketches, on the other hand, do not impose any constraints. Lastly, due to its incremental greedy exploration of the join order space, ROX considers only a limited number of plans---possibly sub-optimal. In COMPASS, plan enumeration is performed at once after push-down selection and can cover any portion of the join space. This can be achieved with the base table sketches which can be composed without the risk to become empty---the case for chain sampling.

%%%%%%%%%%%%%%%%%%%%%%%%%%%%%%%%%%%%%%%%%%%%%%%%%%%%%%%%%%%%%%%%%
%\subsection{Plan Enumeration}
%\textbf{Plan enumeration.}

\paragraph*{Plan enumeration.}
The join attribute-level sketches computed during push-down selection can be composed to estimate the cardinality of any valid join order -- excluding cross products -- generated during plan enumeration. In most cases, cross products are ignored by join enumeration algorithms anyway~\cite{Leis:index-join-sample:cidr-2017}. As shown in Section~\ref{sec:fagms-join-order}, sketch composition consists of two stages. First, the sketches of all the relevant join attributes in a table are merged together. An attribute is relevant for a partial join order if its join is part of the order. Second, the sketches across tables are combined to estimate the cardinality of the join order. Since the overall composition consists only of arithmetic operations, sketches can be integrated into any enumeration algorithm---exhaustive, bushy, or left-deep. Essentially, sketches can readily replace the standard join cardinality estimation formula based on table and join attribute distinct cardinality~\cite{Ullman:db-book}. However, since sketches capture the correlation between join attributes and do not make the independence and containment assumptions, their accuracy is expected to be better.

%%%%%%%%%%%%%%%%%%%%%%%%%%%%%%%%%%%%%%%%%%%%%%%%%%%%%%%%%%%%%%%%%
%\subsection{Sketches vs. Other Synopses}
%\textbf{Sketches vs. other synopses.}

\paragraph*{Sketches vs. other synopses.}
The decision to exclusively use sketches in COMPASS may seem questionable given that sketches are designed for specific stream processing tasks, while traditional databases support generic batch-oriented execution. To put it differently, there is a specific sketch for every streaming query, while synopses are for the entire database. To achieve generality, COMPASS has to build a set of sketches for every query---except base tables without predicates. However, this is done concurrently with push-down selection and is highly-parallel, resulting in low overhead (Section~\ref{sec:experiments}). As a result, sketches do not require any maintenance under modification operations since they are built on the current data. This is not possible for any of the other database synopses. The benefit of having query-specific synopses is also exploited in~\cite{Leis:index-join-sample:cidr-2017}, where index-based join sampling -- a variation of ROX chain sampling~\cite{Kader:ROX:sigmod-2009} -- is introduced. Index-based join sampling is performed during the plan enumeration of every query under the corresponding selection predicates. Since the sample size -- both minimum and maximum -- is carefully controlled, index-based join sampling has improved memory usage and accuracy because it avoids empty results. Compared to sketches, though, this sampling strategy has two serious shortcomings. First, it requires the existence of an index and complete frequency distribution on every join attribute. Sketches require nothing beyond the data. Second, the estimation of every join cardinality requires separate sampling from each of the involved tables. Since this process is time-consuming, plan enumeration is performed bottom-up -- or breadth-first -- in a limited time budget. Sketches can be composed incrementally in any order, without the need to access the data.

The other types of synopses -- histograms and distinct cardinality -- are not query-specific. Thus, they do not incur any creation overhead during optimization. To estimate join cardinality, the attribute-level instances of these synopses are composed by simple arithmetic operations~\cite{Selinger:APSRDMS:sigmod-1979,Ullman:db-book}. However, due to the strong assumptions -- uniformity, independence, inclusion, ad-hoc constants -- made by these operations, the estimates can be highly-inaccurate. Sketches do not make any of these assumptions because they capture correlations by design.

%%%%%%%%%%%%%%%%%%%%%%%%%%%%%%%%%%%%%%%%%%%%%%%%%%%%%%%%%%%%%%%%%
\begin{figure*}[htbp]
 \centering
 \includegraphics[width=\textwidth]{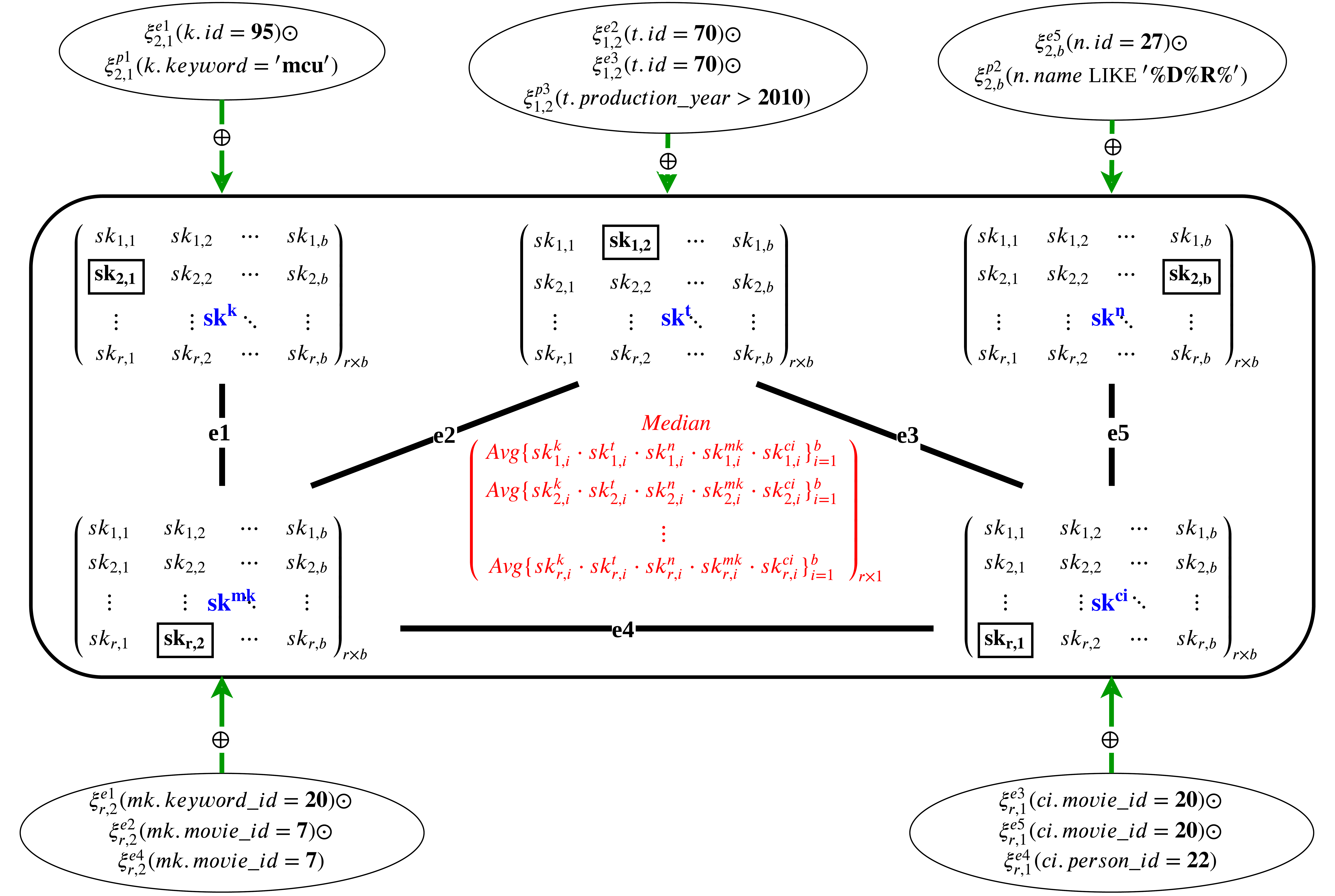}
 \caption{Cardinality estimation for query JOB 6a with AGMS sketches.}
 \label{fig:agms_6a}
\end{figure*}
%%%%%%%%%%%%%%%%%%%%%%%%%%%%%%%%%%%%%%%%%%%%%%%%%%%%%%%%%%%%%%%%%

%%%%%%%%%%%%%%%%%%%%%%%%%%%%%%%%%%%%%%%%%%%%%%%%%%%%%%%%%%%%%%%%%
%\input{sketches}
\section{SKETCH CARDINALITY ESTIMATION}\label{sec:sketches}

In this section, we present how the class of AGMS sketches are applied for estimating the cardinality of complex queries involving selection predicates and multi-way joins. We organize the presentation around the original AGMS sketches~\cite{Alon:AGMS:pods-1999} which have known solutions to these problems. However, AGMS sketches are too inefficient to be accurate and cannot be integrated in query plan enumeration. This leads us to the Fast-AGMS sketches~\cite{Cormode:FAGMS:vldb-2005} which are asymptotically more efficient and have been shown to be statistically more accurate~\cite{Rusu:SAS:sigmod-2007,Rusu:SJSE:tods-2008}. However, Fast-AGMS sketches are limited to estimating two-way join cardinality. Our main contributions are to extend Fast-AGMS sketches to multi-way joins and to effectively integrate them in query plan enumeration.

%%%%%%%%%%%%%%%%%%%%%%%%%%%%%%%%%%%%%%%%%%%%%%%%%%%%%%%%%%%%%%%%%
\subsection{AGMS Sketches}\label{ssec:sketches:agms}

The basic AGMS sketch of an attribute from a relation consists of a single random value \(sk\) that summarizes the values of the attribute across all the tuples in the relation. For example, all the values of attribute \textit{id} from table \textit{keyword} can be summarized by a sketch \(sk(k.id)\) computed as $\textit{sk(k.id)} = \sum_{t\in k}\xi \left(\textit{t.id}\right)$, where $\xi$ is a family of $\{+1,-1\}$ random variables that are 4-wise independent. Essentially, a random value of either $+1$ or $-1$ is associated to each point in the domain of attribute \textit{k.id}. Then, the corresponding random value is added to the sketch \(sk(k.id)\) -- initialized to $0$ -- for each tuple $t$ in table \textit{keyword}. Intuitively, the more frequent a value is, the more is ``pulling'' the sketch to its frequency. Since all the tuples are combined in the same sketch \(sk(k.id)\), they are conflicting and the output can be far away from the frequency of each single tuple. This is where the 4-wise independence property of $\xi$ is important. It guarantees that for any group of at most 4 different values of attribute \textit{k.id}, the product of their corresponding $\xi$ values is $0$ on expectation---they cancel out. This, in turn, allows for each individual attribute value frequency to be unbiasedly estimated by multiplying the sketch with the corresponding $\xi$ random value. For example, the frequency of \(k.id=5\) is estimated by the product \(sk(k.id) \cdot \xi(5)\).

%%%%%%%%%%%%%%%%%%%%%%%%%%%%%%%%%%%%%%%%%%%%%%%%%%%%%%%%%%%%%%%%%
\subsubsection{Two-Way Join Cardinality Estimation}

Consider the join $e{1}$ between tables \textit{keyword} and \textit{movie\_keyword} with predicate \textit{k.id = mk.keyword\_id} (Figure~\ref{fig:query_6a_plans}). The cardinality of this join operator can be estimated with two AGMS sketches \(sk(k.id)\) and \(sk(mk.keyword\_id)\) built on the join attributes. The requirement is that these sketches share the same family $\xi$ of random variables---$\xi^{e1}$ is associated with edge $e{1}$. $\xi^{e1}$ guarantees that join keys with the same value are assigned the same $\{+1,-1\}$ random value---they are correlated. The basic AGMS estimator is the product of \(sk(k.id)\) and \(sk(mk.keyword\_id)\):
\begin{equation*}\label{def:sketch-est:2-way}
	\textit{Est}\left(\left|e{1}\right|\right) = \textit{sk(k.id)} \cdot \textit{sk(mk.keyword\_id)} = \sum_{x\in k} \sum_{y\in \textit{mk}} {\xi^{e1} \left(\textit{x.id} \right) \cdot \xi^{e1} \left(\textit{y.keyword\_id} \right)}
\end{equation*}
Due to the 4-wise independence property of $\xi^{e1}$, this estimator is unbiased---its expectation equals the true $\left|e{1}\right|$ cardinality. However, its variance is high---it has poor accuracy. This is expected since a full table with any number of tuples is summarized as a single number. The standard technique to improve accuracy is to build multiple independent basic sketch estimators. This is achieved by using independent families of random variables $\xi^{e1}$. It is theoretically proven that, in order to obtain an estimator with relative error at most $\epsilon$ with confidence $\delta$, $\mathcal{O} \left( 1/\epsilon^{2} \log{(1/\delta)} \right)$ basic sketches are necessary. As shown in Figure~\ref{fig:agms_6a}, they are grouped into a matrix of $r$ rows and $b$ columns. Then, the final AGMS estimator is obtained by averaging the $b$ instances in each row and taking the median over the resulting $r$ averages. In summary, an AGMS sketch has $\Omega(r\cdot b)$ update and query time, and its space usage is also $\Omega(r\cdot b)$. This assumes that the random number generators $\xi$ have small seeds and produce their values fast---aspects that require careful implementation.

%%%%%%%%%%%%%%%%%%%%%%%%%%%%%%%%%%%%%%%%%%%%%%%%%%%%%%%%%%%%%%%%%
\subsubsection{Multi-Way Join Cardinality Estimation} \label{agms-multi-way-join-estimation}

We show how to extend AGMS sketches to multi-way join cardinality estimation. For this we add the join $e{2}$ between \textit{movie\_keyword} and \textit{title} to $e{1}$ and aim to estimate the cardinality of this 3-table query. Following the approach for two-way joins, a family of sketches is built for edge $e{2}$ on attributes \textit{mk.movie\_id} and \textit{t.id}, respectively. These sketches share their own family $\xi^{e{2}}$ of random variables. Since two attributes from \textit{mk} -- \textit{keyword\_id} and \textit{movie\_id} -- participate in join operators with other tables, we have to preserve their tuple connection. This is achieved by creating a single composed sketch \(sk(mk.k\_id,mk.m\_id)\) instead of separate sketches for each attribute~\cite{Dobra:SBMQPDS:edbt-2004}. The value of \(sk(mk.k\_id,mk.m\_id)\) is computed as:
\begin{equation*}\label{def:sketch:3-way}
	\textit{sk(mk.k\_id,mk.m\_id)} = \sum_{t\in \textit{mk}}\xi^{e{1}} \left(\textit{t.k\_id}\right) \cdot \xi^{e{2}} \left(\textit{t.m\_id}\right)
\end{equation*}
where the product of the two random variables is added to the sketch. The cardinality estimator is defined as the product of three sketches in this case:
\begin{equation*}\label{def:sketch-est:3-way}
	\textit{Est}\left(\left|e{1}\cup e{2}\right|\right) = \textit{sk(k.id)} \cdot \textit{sk(mk.k\_id,mk.m\_id)} \cdot \textit{sk(t.id)} =
	\hspace*{.05cm} \sum_{x\in k} \sum_{y\in \textit{mk}} \sum_{z\in \textit{t}} { \xi^{e1} \left(\textit{x.id} \right) \cdot \xi^{e1} \left(\textit{y.k\_id} \right) \cdot \xi^{e2} \left(\textit{y.m\_id} \right) \cdot \xi^{e2} \left(\textit{z.id} \right) }
\end{equation*}
As long as the families $\xi^{e{1}}$ and $\xi^{e{2}}$ are independent, this estimator is unbiased. However, its variance can be exponentially worse than that of the two-way join estimator---which makes sense, given the additional degree of randomness. Thus, to achieve the same accuracy, a considerably larger number of basic sketches are required.

This strategy can be generalized to complex queries involving any number of tables and join predicates. A sketch is built for every table. Independent random families $\xi$ are used for every join predicate. The sketch corresponding to a table is updated with the product of all the $\xi$ families incident to it, applied to the corresponding join attribute. In the case of our example query JOB 6a with 5 tables and 5 join predicates (Figure~\ref{fig:agms_6a}), there are 5 sketches and 5 families $\xi$. The sketch $\textit{sk}^{\textit{mk}}$ for table \textit{mk} is updated with the product $\xi^{e{1}} \left(\textit{k\_id}\right) \cdot \xi^{e{2}} \left(\textit{m\_id}\right) \cdot \xi^{e{4}} \left(\textit{m\_id}\right)$ which includes a factor for each of the three join predicates. The unbiased cardinality estimator is the product of the 5 sketches $\textit{sk}^{\textit{k}} \cdot \textit{sk}^{\textit{mk}} \cdot \textit{sk}^{\textit{t}} \cdot \textit{sk}^{\textit{ci}} \cdot \textit{sk}^{\textit{n}}$. For the same number of basic sketches $r\cdot b$ as in the case of the $|e1|$ join, the accuracy of the $|e1\cup e2\cup e3\cup e4\cup e5|$ join can be exponentially worse.

%%%%%%%%%%%%%%%%%%%%%%%%%%%%%%%%%%%%%%%%%%%%%%%%%%%%%%%%%%%%%%%%%
\subsubsection{Selection Cardinality Estimation}

Query JOB 6a contains 3 selection predicates---point on \textit{k}, subset on \textit{n}, and range on \textit{t}. These have to be accounted for when estimating the overall query cardinality. AGMS sketches can handle selection predicates as long as the domain of the attribute is discrete---which is the case for the fixed-size data types in databases. The idea is to express the selection as a join predicate between the table and the domain of the selection attribute~\cite{Rusu:FRSRV:sigmod-2006,Rusu:PRNG:tods-2007}. Following the two-way join approach, a sketch is built on the selection attribute over all the tuples in the table. The sketch over the domain -- which shares the same random family $\xi$ -- summarizes the values in the domain which satisfy the predicate by adding an entry for each of them to the sketch---for a point predicate, the sketch includes only the $\xi$ value corresponding to the constant in the predicate; for a range, the $\xi$ values for all points in the range; for a subset, the $\xi$ values for the points in the subset. As long as the number of points is small, these sketches can be computed fast. Moreover, even for ranges, there is a specific fast range-summable random family $\xi$ for which the sketch can be computed in constant time, independent of the range size~\cite{Rusu:FRSRV:sigmod-2006,Rusu:PRNG:tods-2007}. In the JOB 6a query depicted in Figure~\ref{fig:agms_6a}, the sketch update procedure for tables with predicates includes an additional factor corresponding to the selection attribute. For example, the sketch $\textit{sk}^{\textit{k}}$ for table \textit{keyword} is updated with the product $\xi^{e{1}} \left(\textit{id}\right) \cdot \xi^{p{1}} \left(\textit{keyword}\right)$. Overall, 8 families $\xi$ and 8 sketches are required---the sketches over the domain of the selection attributes are not included in Figure~\ref{fig:agms_6a}. The final estimator is the product of these 8 sketches. Since this estimator is a multi-way join with a larger number of sketches, its accuracy becomes worse than that of the join sketches only.

%%%%%%%%%%%%%%%%%%%%%%%%%%%%%%%%%%%%%%%%%%%%%%%%%%%%%%%%%%%%%%%%%
\subsubsection{Why AGMS Sketches Are Not Practical for Query Optimization?}

As shown, AGMS sketches can be theoretically used to estimate the cardinality of arbitrary complex queries with join and selection predicates. While all the sketches for a table can be built in a single scan, since the update time per AGMS sketch is linear in the sketch size, updating an exponential number of sketches becomes dominant. Moreover, the space requirement for all the sketches is also a problem. These scalability issues hinder the application of AGMS sketches to join order enumeration. However, AGMS sketches suffer from a more serious problem in query optimization---they cannot be incrementally composed. What this means is that a sketch used to estimate a two-way join between two relations cannot be used to estimate a three-way join that includes another relation. The addition of join $e2$ to $e1$ in our example illustrates this well. It is not possible to compute the sketch \(sk(mk.k\_id,mk.m\_id)\) from sketch \(sk(mk.keyword\_id)\). It is not even possible to compute \(sk(mk.k\_id,mk.m\_id)\) from \(sk(mk.keyword\_id)\) and \(sk(mk.movie\_id)\). The reason is the order of multiplication and addition. The other direction -- use \(sk(mk.k\_id,mk.m\_id)\) instead of \(sk(mk.keyword\_id)\) or \(sk(mk.movie\_id)\) -- is also not possible. Thus, in order to support plan enumeration, a separate sketch has to be built for every combination of the join attributes---which is an exponential number. For example, 7 sketches have to be built for both tables \textit{mk} and \textit{ci} which participate in 3 join predicates. If we include the attributes that can appear in selection predicates, the number of sketches that has to be built for a table can become exponential in the number of attributes in the table. While workload information can be used to reduce this number, there is little that can be done for tables that join with several other tables on different attributes. Practically, AGMS sketches cannot achieve the goal of having synopses only for single attributes.

\medskip
\medskip
%%%%%%%%%%%%%%%%%%%%%%%%%%%%%%%%%%%%%%%%%%%%%%%%%%%%%%%%%%%%%%%%%
\begin{figure*}[htbp]
 \centering
 \includegraphics[width=\textwidth]{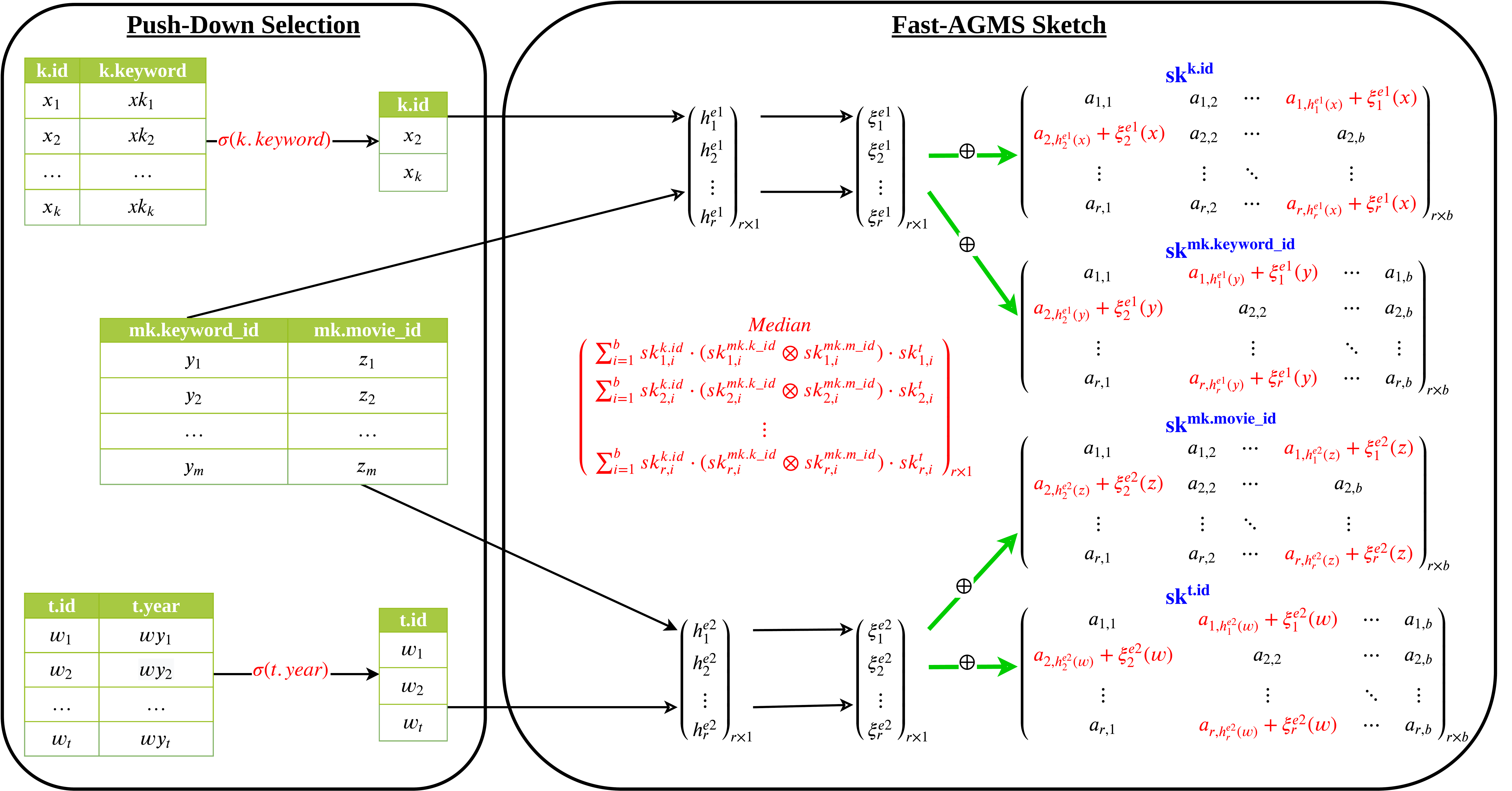}
 \caption{Cardinality estimation for query JOB 6a with Fast-AGMS sketches.}
 \label{fig:fagms_6a}
\end{figure*}
%%%%%%%%%%%%%%%%%%%%%%%%%%%%%%%%%%%%%%%%%%%%%%%%%%%%%%%%%%%%%%%%%

%%%%%%%%%%%%%%%%%%%%%%%%%%%%%%%%%%%%%%%%%%%%%%%%%%%%%%%%%%%%%%%%%
\subsection{Fast-AGMS Sketches}\label{ssec:sketches:f-agms}

Fast-AGMS sketches preserve the $(r\times b)$ matrix structure of AGMS sketches. However, they define a complete row of $b$ counters as a basic sketch element (Figure~\ref{fig:fagms_6a}). Only one of these counters is updated for every tuple, thus, a factor $b$ reduction in update time is obtained. The updated counter is chosen by a random hash function $h$ associated with the row. The purpose of $h$ is to spread tuples with different values as evenly as possible---tuples with the same key still end up in the same bucket. On average, a factor $b$ less tuples collide on the same counter, which preserves the frequency of each of them better. Since a full row is a sketch element, a single $\xi$ family of random variables is associated with every row. Thus, a Fast-AGMS sketch with $r$ rows requires only $r$ hash and $\xi$ random functions. The value of a counter $j$ is $\textit{sk(k.id)}_{j} = \sum_{t\in k, h(\textit{t.id})=j}\xi \left(\textit{t.id}\right)$.

%%%%%%%%%%%%%%%%%%%%%%%%%%%%%%%%%%%%%%%%%%%%%%%%%%%%%%%%%%%%%%%%%
\subsubsection{Two-Way Join Cardinality Estimation} \label{fagms-two-way-join-estimation}

In order to estimate join cardinality, the same principle applies---Fast-AGMS sketches are built over the join attributes using the same random functions $h$ and $\xi$. The hash function $h$ lands identical keys to the same bucket, while $\xi$ gives the same sign. The unbiased estimator for a basic sketch sums up the product of the corresponding buckets:
\begin{equation*}\label{def:f-sketch-est:2-way}
	\textit{Est}\left(\left|e{1}\right|\right) = \sum_{j=1}^{b}\textit{sk(k.id)}_{j} \cdot \textit{sk(mk.keyword\_id)}_{j}
\end{equation*}
Summation is necessary because $h$ partitions the tuples. As for AGMS sketches, the final estimate is obtained by taking the median of the $r$ independent basic sketches. Although the accuracy of Fast-AGMS sketches is asymptotically equal to that of AGMS sketches~\cite{Cormode:FAGMS:vldb-2005} in the worst case, it has been shown statistically that Fast-AGMS sketches have considerably better accuracy than any other sketching technique on average~\cite{Rusu:SAS:sigmod-2007}. The combined accuracy and fast update time make Fast-AGMS sketches suitable for query optimization.

%%%%%%%%%%%%%%%%%%%%%%%%%%%%%%%%%%%%%%%%%%%%%%%%%%%%%%%%%%%%%%%%%
\subsubsection{Why Fast-AGMS Sketches Are Not Applicable to Query Optimization?}

As far as we know, there is no work that extends Fast-AGMS sketches to multi-way join estimation. The main problem is the requirement to have independent hash functions $h^{e1}$ and $h^{e2}$ for the two join attributes. These functions allocate the attributes to different buckets, which means that the tuple is added to the sketch twice. Moreover, the relationship between attributes is lost. Since sketch-based selectivity estimation is also reduced to a join between the selection attribute and its domain, this implies that Fast-AGMS sketches cannot be used to estimate the cardinality of two-way joins with predicates. In fact, computing optimally the Fast-AGMS sketch of the domain of a range predicate does not have a solution. This is because there is no order relationship between the hash values of adjacent points in the domain. Due to these limitations, Fast-AGMS sketches have not been used in query optimization before. COMPASS introduces Fast-AGMS extensions for multi-way joins and solves the selectivity issue by pushing-down predicates during query optimization, and adding only the relevant tuples to the sketch.

%%%%%%%%%%%%%%%%%%%%%%%%%%%%%%%%%%%%%%%%%%%%%%%%%%%%%%%%%%%%%%%%%
%\input{fagms-ordering}
\section{FAST-AGMS SKETCH MULTI-WAY JOIN ESTIMATION}\label{sec:fagms-join-order}

We present two strategies to extend Fast-AGMS sketches to multi-way join cardinality estimation. The first strategy -- sketch partitioning -- is a theoretically sound estimator for a given multi-way join. Its limitation is that it cannot be composed/decomposed, thus, it is not scalable for plan enumeration. The second strategy -- sketch merging -- addresses the scalability issue by incrementally creating multi-way sketches from two-way sketches. Although this is done heuristically for a certain multi-way join taken separately, all the multi-way joins with a given size are equally impacted. We show empirically that this property is a good surrogate for accuracy -- which is much harder to consistently achieve -- in join order enumeration.
\begin{figure*}[htbp]
 \centering
 \includegraphics[width=\textwidth]{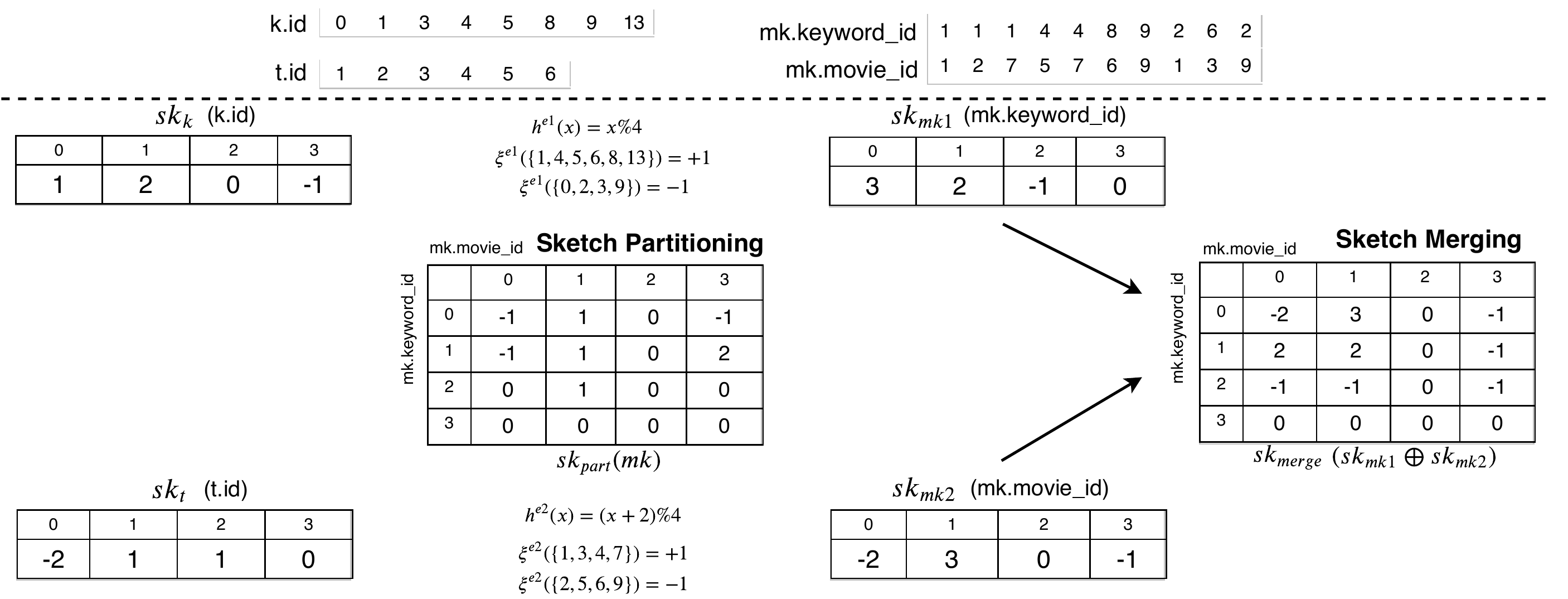}
 \caption{Fast-AGMS sketches for multi-way join cardinality estimation on query JOB 6a.}
 \label{fig:fagms_6a-multiway}
\end{figure*}
%%%%%%%%%%%%%%%%%%%%%%%%%%%%%%%%%%%%%%%%%%%%%%%%%%%%%%%%%%%%%%%%%

%%%%%%%%%%%%%%%%%%%%%%%%%%%%%%%%%%%%%%%%%%%%%%%%%%%%%%%%%%%%%%%%%
\subsection{Sketch Partitioning}

The idea of sketch partitioning is to reorganize the $b$ buckets of the elementary sketch into a $(b_{1}\times b_{2})$ 2-D matrix---as done in~\cite{Cai:PCETUB:sigmod-2019} for frequency sketches. $h^{e1}$ hashes a tuple \(mk(k\_id,m\_id)\) to one of the $b_{1}$ rows, while $h^{e2}$ hashes to one of the $b_{2}$ columns. Then, only the counter at indices $\left[h^{e1}(k\_id),h^{e2}(m\_id)\right]$ is updated with the product $\xi^{e{1}}(k\_id) \cdot \xi^{e{2}}(m\_id)$. This process is depicted in Figure~\ref{fig:fagms_6a-multiway}. For example, tuple (6,3) in \(mk\) adds 1 to the counter [2,1]. $h^{e1}$ guarantees that all the tuples with \(k\_id=6\) are hashed to row 2, while $h^{e2}$ sends tuples with \(m\_id=3\) to column 1. Conflicts happen only when the output of both hash functions is identical. Given the quadratic number of buckets compared to the sketch for a single attribute -- while the number of tuples is the same -- conflicts are less frequent. The cardinality estimate for the 3-table join \(k \bowtie mk \bowtie t\) is obtained by summing up all the entries in the matrix resulted after the scalar multiplication between \(sk(k.id)\) and every row in \(sk_{part}(mk)\), followed by the scalar multiplication between the transpose of \(sk(t.id)\) and every column in \(sk_{part}(mk)\). This can be written as:
\begin{equation*}\label{def:fagms-est:3-way}
\textit{Est}\left(\left|e{1}\cup e{2}\right|\right) = \sum_{0\leq i < b_{1}} \sum_{0\leq j < b_{2}} \textit{sk}_{k}[i] \cdot \textit{sk}_{\textit{part}}\textit{(mk)}[i,j] \cdot \textit{sk}_{t}[j]
\end{equation*}
It can be shown theoretically that this estimator is unbiased following the same proof as for AGMS sketches in~\cite{Dobra:SBMQPDS:edbt-2004}. Moreover, given the larger size of sketch \(sk_{part}(mk)\), its accuracy is expected to be better. This procedure can be generalized to any number of join attributes by partitioning -- or replicating -- $b$ into the corresponding number of dimensions. For example, a table with 3 joins has a 3-D tensor as its sketch, with one dimension for every join attribute. Thus, there is a polynomial factor increase in the size of the sketch and the estimate computation. This has to be carefully accounted for in the overall memory budget since the likelihood of conflicts varies with the dimensionality of the sketch tensor. The constraint to have the same number of buckets for a join predicate, e.g., \(sk(k.id)\) has as many buckets as the number of rows in \(sk_{part}(mk)\), makes memory allocation among sketches more complicated than for the 1-D AGMS sketch vectors.

Partitioned Fast-AGMS sketches are not scalable for join order enumeration. This is because separate sketches are required for every join. For example, in Figure~\ref{fig:fagms_6a-multiway}, the 2-D sketch \(sk_{part}(mk)\) is used for the 3-way join \(k \bowtie mk \bowtie t\), while the 1-D sketches \(sk_{mk1}\) and \(sk_{mk2}\) are used for the 2-way joins \(k \bowtie mk\) and \(t \bowtie mk\), respectively. Building and storing these many sketches is impractical in query optimization. One alternative is to build only the sketches for up to k-way joins and use other methods to estimate higher-order join cardinality. This strategy is applied for run-time join samples in~\cite{Leis:index-join-sample:cidr-2017}. The drawback is that other statistics are required for the higher-order joins and the interaction between the estimates produced by these statistics and the sketch estimates has to be carefully controlled.

Our goal is to exclusively use sketches. Intuitively, we want to be able to either generate the 2-D sketch from the 1-D sketches or extract the 1-D sketches from the 2-D sketch. Unfortunately, none of these have a clear solution for Fast-AGMS sketches. The composition of \(sk_{mk1}\) and \(sk_{mk2}\) requires to determine how to combine all the pairs of buckets in the 1-D sketches in order to compute the quadratic number of entries in the 2-D sketch. Since the identity of tuples is lost when they are inserted in the 1-D sketch, it is not possible to recreate the tuple and determine its corresponding 2-D bucket. Moreover, due to conflicts in the $\xi$ random functions, we do not even know how many tuples belong to a 1-D bucket. For example, bucket 1 in \(sk_{mk1}\) is 2 even though 4 tuples are hashed to it. The extraction of a 1-D sketch from the 2-D sketch also does not work because of the $\xi$ variables. Specifically, the update by the product $\xi^{e{1}} \cdot \xi^{e{2}}$ makes it impossible to retrieve the value of a 1-D bucket by summing up the corresponding 2-D buckets. For example, the value of bucket 0 in \(sk_{mk1}\) is not the sum of the buckets in row 0 of sketch \(sk_{part}(mk)\). This property is true only for hash-based sketches~\cite{Cai:PCETUB:sigmod-2019}.

%%%%%%%%%%%%%%%%%%%%%%%%%%%%%%%%%%%%%%%%%%%%%%%%%%%%%%%%%%%%%%%%%
\subsection{Sketch Merging}\label{sec:sk-order}

We introduce the sketch merging heuristic as a lightweight method to compose two-way join Fast-AGMS sketches in order to estimate the cardinality of multi-way joins. The procedure works as follows. We build sketches for every two-way join predicate independently, as shown in Figure~\ref{fig:fagms_6a-multiway}. The number of sketches corresponding to a table is equal to the number of joins it participates in. For example, tables \(k\) and \(t\) have one sketch, while \(mk\) has two sketches. We estimate any join combination generated during plan enumeration using only these sketches. The two-way joins \(k \bowtie mk\) and \(t \bowtie mk\) are estimated optimally with the sketch pairs \((sk_{k},sk_{mk1})\) and \((sk_{t},sk_{mk2})\), respectively. These are the most accurate sketch estimates we can get. For the 3-way join \(k \bowtie mk \bowtie t\), we create a merged sketch \(sk_{merge}(mk)=sk_{mk1} \oplus sk_{mk2}\) from the two 2-way join sketches on demand during plan enumeration. This merged sketch approximates the partitioned sketch \(sk_{part}(mk)\) computed with the same random functions, without accessing the tuples. A bucket \([i,j]\) in \(sk_{merge}\) is set to the value having the minimum absolute magnitude among the corresponding \([i]\) and \([j]\) buckets in the two basic sketches:
\begin{equation*}\label{def:fagms-merge}
\textit{sk}_{\textit{merge}}[i,j] = \left\lbrace \begin{array}{ll} \textit{sk}_{\textit{mk1}}[i], & \textit{if} \hspace*{.1cm} \left|\textit{sk}_{\textit{mk1}}[i]\right| \leq \left|\textit{sk}_{\textit{mk2}}[j]\right| \\ \textit{sk}_{\textit{mk2}}[j], & \textit{if} \hspace*{.1cm} \left|\textit{sk}_{\textit{mk1}}[i]\right| > \left|\textit{sk}_{\textit{mk2}}[j]\right| \end{array} \right.
\end{equation*}
For the example in Figure~\ref{fig:fagms_6a-multiway}, bucket $[0,0]$ is set to $-2$ because $|3|>|-2|$, while bucket $[0,2]$ to $0$ because $|0|<|3|$. The reason for this merge procedure is multifolded. The interaction between the random functions \(\xi\) is considered -- albeit not through a direct multiplication -- by preserving the sign of the value in the basic sketch. The absolute magnitude corresponds to the maximum number of tuples with a given join key that are hashed to the bucket---assuming no conflicts. These tuples are partitioned across the buckets of the other join key. The minimum is chosen because this is the maximum number of tuples that can have identical values for both join keys when considered together. However, this is an overestimate because the exact tuple pairing is lost. This can be seen when comparing the magnitude of the values in the two 2-D sketches in Figure~\ref{fig:fagms_6a-multiway}. In fact, sketch merging is likely to always overestimate join cardinality. The only caveat is the interaction between the \(\xi\) functions.

Sketch merging can be generalized to any number of joins by applying the procedure iteratively. Moreover, (n+1)-D sketches can be derived incrementally from n-D sketches in a single step---without the need to always start from the basic sketches. This property can be exploited to speed up the computation and reduce memory usage in bottom-up plan enumeration since only the highest dimensional sketches have to be maintained. An even more important property of sketch merging is that it is consistent in how it handles the multi-way joins with the same number of predicates. Specifically, all these joins rely on the same basic sketches and the same assumptions for merging. Thus, it is likely that these estimates exhibit similar accuracy behavior---same type of errors for equal join size.

%%%%%%%%%%%%%%%%%%%%%%%%%%%%%%%%%%%%%%%%%%%%%%%%%%%%%%%%%%%%%%%%%
\begin{figure}[htbp]
\centering
\begin{subfigure}{0.49\textwidth}
\includegraphics[width=\linewidth,height=\linewidth]{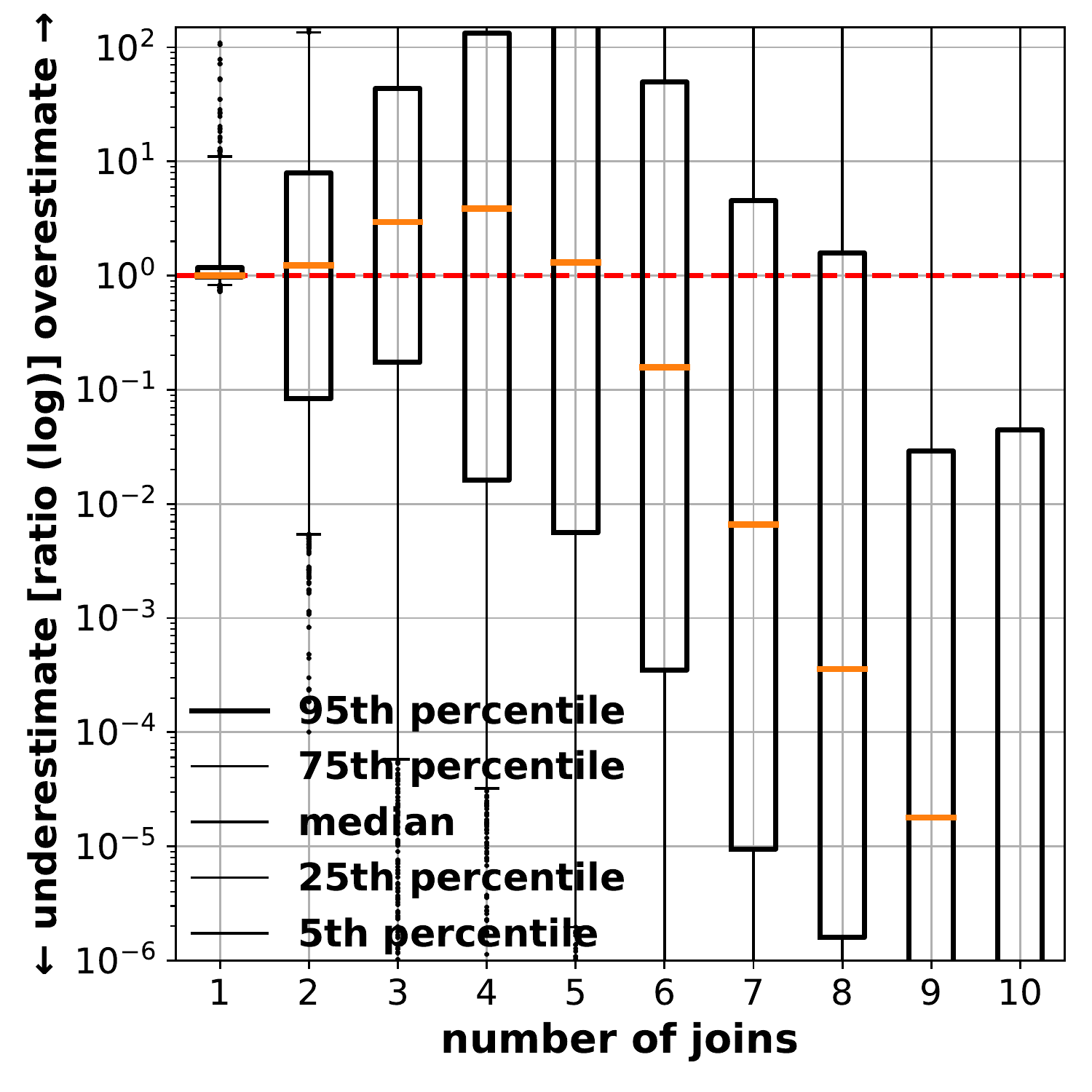}
\caption{estimate/true}
\label{fig:factor_error_comparison}
\end{subfigure}
\hfill
\begin{subfigure}{0.49\textwidth}
\includegraphics[width=\linewidth,height=\linewidth]{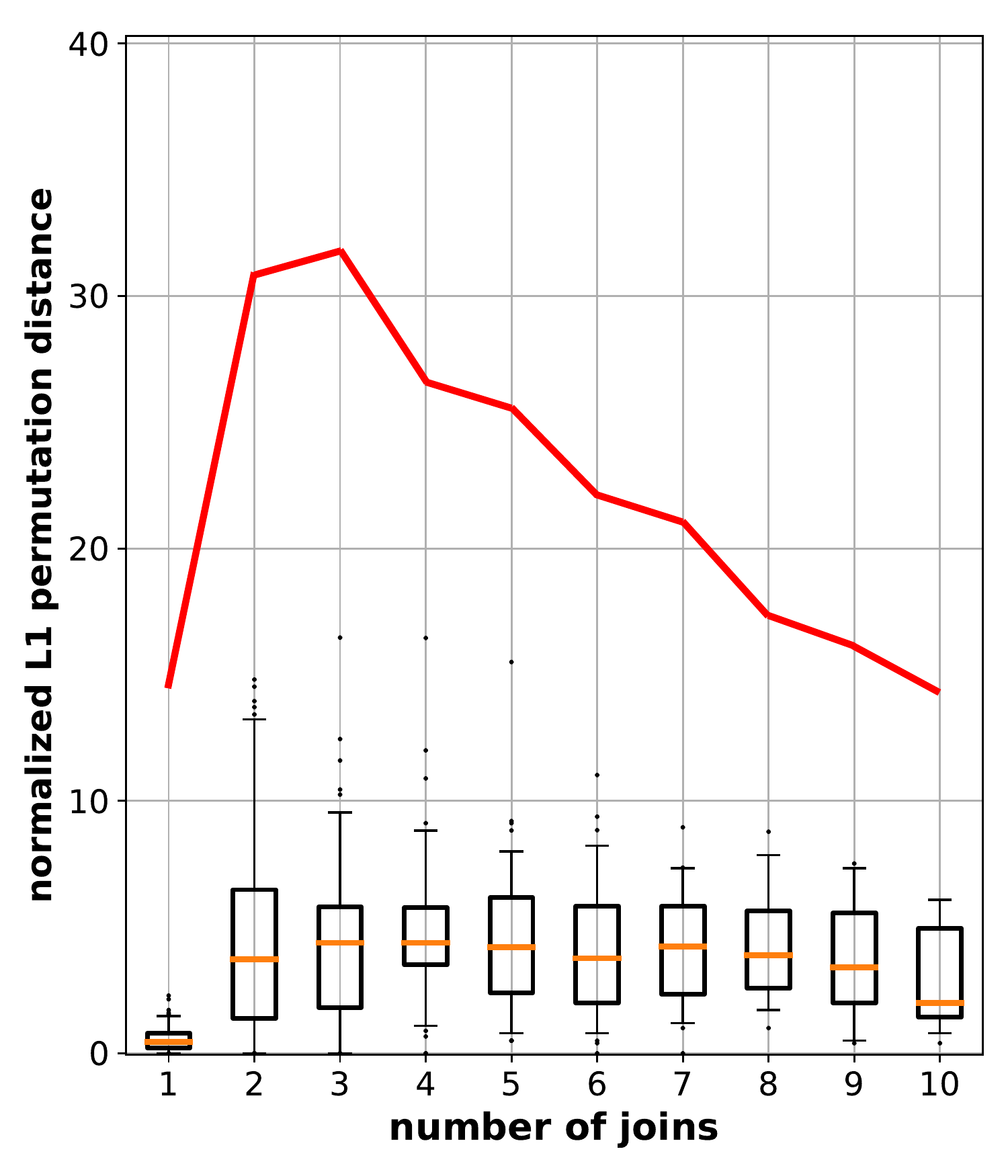}
\caption{L1-distance}
\label{fig:l1-distance}
\end{subfigure}
\caption{Accuracy ratio (a) and L1-distance between the estimated sketch permutation and the correct join order (b).}
\label{fig:order_error}
\end{figure}
%%%%%%%%%%%%%%%%%%%%%%%%%%%%%%%%%%%%%%%%%%%%%%%%%%%%%%%%%%%%%%%%

In order to verify this claim, we depict the accuracy of sketch merging for the JOB queries in Figure~\ref{fig:order_error}. We use two measures to quantify accuracy. The first is the ratio between the sketch estimate and the true cardinality for all the enumerated sub-plans having at most ten joins (Figure~\ref{fig:factor_error_comparison}). We observe that the median ratio is within a factor of 10 for up to six joins, which is better than any previous practical results~\cite{Leis:index-join-sample:cidr-2017}. For a larger number of joins, sketch merging generates underestimates systematically. In previous results~\cite{Leis:JOB:vldb-2018}, this behavior occurs starting from 3-way joins.

%%%%%%%%%%%%%%%%%%%%%%%%%%%%%%%%%%%%%%%%%%%%%%%%%%%%%%%%%%%%%%%%%
\begin{table}[htbp]
  \centering

  \begin{tabular}{|c|*{6}{c|}} \hline
   2-join & \(mk \bowtie k\) & \(ci \bowtie n\) & \(mk \bowtie t\) & \(ci \bowtie t\) & \(mk \bowtie ci\) \\ \hline
   True cardinality & 14 & 486 & 0.3M & 6M & 215M \\ \hline
   Sketch merging & 3K & 7.5K & 6.4M & 14M & 397M \\ \hline
  \end{tabular}

  \par\medskip

  \begin{tabular}{|c|*{6}{c|}} \hline
   3-join & \(mk \bowtie k \bowtie t\) & \(t \bowtie ci \bowtie n\) & \(mk \bowtie ci \bowtie k\) & \(mk \bowtie ci \bowtie n\) & \(mk \bowtie ci \bowtie t\) \\ \hline
   True cardinality & 11 & 61 & 1242 & 10K & 17M \\ \hline
   Sketch merging & 6.7K & 35K & 0.7M & 1.5M & 8.7B \\ \hline
  \end{tabular}

  \par\medskip

  \begin{tabular}{|c|*{4}{c|}} \hline
   4-join & \(mk \bowtie ci \bowtie n \bowtie k\) & \(mk \bowtie ci \bowtie n \bowtie t\) & \(mk \bowtie ci \bowtie k \bowtie t\) \\ \hline
   True cardinality & 6 & \underline{1194} & \underline{1224} \\ \hline
   Sketch merging & 198 & \underline{\textbf{18M}} & \underline{\textbf{6.5M}} \\ \hline
  \end{tabular}
  
  \caption{2-, 3-, and 4-way join L1 permutation distance for JOB 6a.}
  \label{table:agms_vs_fagms}
\end{table}
%%%%%%%%%%%%%%%%%%%%%%%%%%%%%%%%%%%%%%%%%%%%%%%%%%%%%%%%%%%%%%%%%

The second measure is the normalized L1 distance~\cite{stack:perm-l1:2017} between the permutation generated by sketch merging and the correct join order. Given $n$ sub-plans of the same size, the correct order $\mathcal{C}$ is obtained by sorting them in increasing order of their cardinality. The permutation corresponding to sketch merging $\mathcal{S}$ is obtained by sorting the sub-plans based on the sketch estimates. The L1 distance is defined as $\sum_{i=1}^{n}{|\mathcal{S}_{i}-\mathcal{C}_{i}|}$, the sum of the differences between the position in the permutation and the correct order. For example, the L1 distance for the 4-way joins in query JOB 6a (Table~\ref{table:agms_vs_fagms}) is $0+1+1=2$. The normalized L1 distance -- we divide the distance by the number of sub-plans in the query -- is depicted in Figure~\ref{fig:l1-distance}. The closer the distance is to zero, the more similar is the permutation to the correct order. For reference, we plot the line corresponding to the maximum L1 distance. The join orders generated by sketch merging have an L1 distance that is significantly below the maximum. In particular, for 2-way join sub-plans, the distance is almost zero, while for sub-plans with more joins, the distance is constantly below 10. This confirms that sketch merging selects orders that are close to optimal most of the time.

%%%%%%%%%%%%%%%%%%%%%%%%%%%%%%%%%%%%%%%%%%%%%%%%%%%%%%%%%%%%%%%%%
%\input{plan-enumeration}
\section{FAST-AGMS SKETCH JOIN ORDER ENUMERATION}\label{sec:enumeration}

In this section, we present how merged Fast-AGMS sketches are integrated into a novel plan enumeration algorithm we introduce in this work. It is important to emphasize that -- due to the proposed generalization to multi-way joins -- Fast-AGMS sketches can be embedded into any enumeration algorithm. According to the investigation of different join order enumeration algorithms performed in~\cite{Leis:JOB:vldb-2018,Leis:QOREALLY:pvldb-2015}, plan enumeration is not the most critical component of a query optimizer---it has a relatively small impact on plan quality compared to cardinality estimation accuracy. This confirms the approach taken by many query optimizers~\cite{postgres,monetdb-web} that consider only a plan subset ---left- or right-deep plans, which are a permutation of the query tables. We follow a similar approach and introduce an heuristic plan enumeration algorithm that allows us to explore the overall impact of merged Fast-AGMS sketches. Although this algorithm borrows ideas from previous work~\cite{Steinbrunn:HRO:jvldb-1997}, we argue that it is original in the presented form.

%%%%%%%%%%%%%%%%%%%%%%%%%%%%%%%%%%%%%%%%%%%%%%%%%%%%%%%%%%%%%%%%%
\begin{algorithm}[htbp]
\caption{Fast-AGMS Sketch Join Order Enumeration}\label{alg:dfs}
% Can be bounded by time or number of traversals
% Start DFS from each node starting with smallest relation
\begin{multicols}{2}
\begin{algorithmic}[1]

%\State \textbf{Input:}
\State Let $G=(V,E)$ be the join graph
\State Let $C$ be the join order set, initially $\emptyset$
\State Let $p$ be the number of complete plans, initially $0$
\State Let $\textit{min\_cost}$ be the minimum cost, initially $\infty$
\State Let $\textit{max\_plans}$ be the maximum threshold of plans enumerated from a source vertex

\Procedure{Plan Enumeration}{$G$}
	\State Let $S$ be the set of vertices $V$ sorted in increasing order of function $f$ that combines table cardinality with the degree of $v$ in $G$: $f(v) = \alpha \cdot \frac{cardinality(v)}{\max\{cardinality(u), \forall u \in V\}} + \beta \cdot \frac{deg(v)}{max\{deg(u), \forall u \in V\}}$, where $\alpha$ and $\beta$ are user-defined constants
    \For{each vertex $v \in S$}
    	\State $p \leftarrow 0$
		\State \Call{DFS Traversal}{$G$, $\{v\}$, $0$}
    \EndFor
\EndProcedure

\Procedure{DFS Traversal}{$G$, $C$, $\textit{cost}$}

	\State $\textit{curr\_est} \leftarrow$ \Call{Fast-AGMS Estimate}{$C$}
   	\State $\textit{cost} \leftarrow \textit{cost} + \textit{curr\_est}$

	\LineComment{Early pruning}
   	\State \textbf{if} $\textit{cost} > \textit{min\_cost}$ \textbf{then} \Return
	
	\LineComment{Evaluate complete plan}
    \If{$|C| = |V|$}
    	\If{$\textit{cost} < \textit{min\_cost}$}
        	\State $\textit{min\_cost} \leftarrow \textit{cost}$
        	\State $\textit{opt\_path} \leftarrow C$
        \EndIf
        \State $p \leftarrow p+1$
    	\State \textbf{if} $p = \textit{max\_plans}$ \textbf{then} \textbf{abort}
    	\State \Return
    \EndIf

	\LineComment{Recursive enumeration in increasing order of the join cardinality estimates}
	\State Let $L$ be the set of vertices $v \notin C$ that are adjacent to a vertex $u \in C$
    \For{each vertex $v \in L$}
		\State $e[v] \leftarrow $ \Call{Fast-AGMS Estimate}{$C \cup \{v\}$}
    \EndFor

	\State Let $L'$ be set $L$ sorted in increasing order of $e[v]$
    \For{each vertex $v \in L'$}
		\State \Call{DFS Traversal}{$G$, $C \cup \{v\}$, $\textit{cost}$}
    \EndFor

\EndProcedure

\end{algorithmic}
\end{multicols}
\end{algorithm}
%%%%%%%%%%%%%%%%%%%%%%%%%%%%%%%%%%%%%%%%%%%%%%%%%%%%%%%%%%%%%%%%%

COMPASS uses the join graph (Figure~\ref{fig:query_6a_plans}) in plan enumeration. This guarantees that only valid join order plans are considered and cross products are ignored. Plan enumeration becomes a graph traversal problem. We design a depth-first search (DFS) traversal algorithm (Algorithm~\ref{alg:dfs}) that enumerates left-deep plans following the edges in the join graph. This is achieved by considering all the vertices as the source of DFS and backtracking whenever a complete plan is reached (line 26). The number of plans explored from a source vertex is controlled by a user-defined parameter $max\_plans$ (line 25) that plays a similar role to the timeout used to confine the plan search space in~\cite{Leis:index-join-sample:cidr-2017}. However, our algorithm does not require bottom-up plan enumeration since sketches can be merged and combined in any order. The $max\_plans$ parameter also allows the plan search to restart from other sources in the presence of high-degree vertices and cycles---instead of getting locked on the initial selections. On a continuum spectrum that has left-deep greedy search at one extreme and exhaustive enumeration at the other~\cite{Ullman:db-book}, the proposed algorithm can be configured anywhere in-between by controlling the value of parameter $max\_plans$. When a single plan is enumerated from every vertex, we obtain left-deep ordering. This is achieved by selecting the vertex $v$ that has the smallest cardinality when appended to the current plan in line 35. When all the plans are enumerated from every vertex -- set $max\_plans=\infty$ -- we obtain exhaustive enumeration. The tradeoff between these two alternatives is evident---number of explored plans vs. enumeration overhead. Since sketch merging and estimation are fast operations amenable to parallelization, the overhead is small---see the experiments in Section~\ref{sec:experiments}. Thus, $max\_plans$ values larger than $1$ are amenable---$max\_plans$ is set by default to $10$. This allows for a more comprehensive exploration of the join order space compared to alternative synopses that incur higher overhead~\cite{Kader:ROX:sigmod-2009,Leis:index-join-sample:cidr-2017}.

We design two heuristics that control the order in which the plan space is explored and the depth of exploring sub-optimal plans. First, we sort the vertices according to a normalized cost function $f(v)$ that combines vertex cardinality and the number of join predicates the vertex participates in (line 7). The configurable weights $\alpha$ and $\beta$ control the relative importance of these factors. They are set by default to $0.5$, which assigns equal weight to each factor. \textit{DFS Traversal} is invoked from the source vertices in increasing order of cost $f(v)$. The intuition is to generate sub-plans with small cardinalities and limited orders as early as possible in the enumeration---get the left-deep plan corresponding to $max\_plans=1$ first. The second heuristic is early pruning of sub-optimal plans (line 17). Whenever the cost of a sub-plan exceeds the minimum cost, we backtrack to a sub-plan that can still become optimal.

Fast-AGMS sketch cardinality estimation is invoked for every enumerated sub-plan (line 14) and to decide the order in which vertices are explored (line 31)---while the call on line 14 can be eliminated, we keep it for clarity. The \textit{Fast-AGMS Estimate} function performs sketch merging and estimation only for the joins included in the sub-plan. In order to avoid recomputation, the estimates are cached. Alternatively, the merged sketches corresponding to join subsets a table is involved in and the product of the basic sketches corresponding to a sub-plan can also be cached. They provide different levels of reuse and computation to generate the estimate. Caching the estimate provides the least reuse, while caching merged sketches and products allows for incremental estimate evaluation. In our implementation, we settle for a combined solution in which all the estimates, and merged sketches and products of up to three sketches are cached. This insures that the estimate corresponding to any sub-plan is computed only once and allows for incremental extension of the core sub-plans. It is important to emphasize that the input to \textit{Fast-AGMS Estimate} is always represented only by the sketches corresponding to the two-way joins.

We illustrate how the plan enumeration algorithm works for query JOB 6a based on the join graph in Figure~\ref{fig:query_6a_plans} and the join cardinality estimates in Table~\ref{table:agms_vs_fagms}. The five vertices are sorted in the order $n-k-t-mk-ci$ based on function $f(v)$ applied on the tables resulted after selection push-down. $mk$ and $ci$ are the last two because they have the largest degree and cardinality---and have no selection predicates. $k$ and $n$ come before $t$ because they have a smaller degree. Although $n$ has an order of magnitude more tuples than $k$, the predicate on $n$ is very selective and outputs a smaller cardinality. The degree being the same, $n$ is first in the order. Thus, \textit{DFS Traversal} is performed with $n$ as the first source. The first enumerated plan is $n-ci-t-mk-k$ and its estimated cost is $\approx 18M$. While a left-deep plan search finishes the enumeration at this point, our algorithm backtracks and explores alternative plans---we set $max\_plans=2$ in this case. The other plan enumerated from $n$ is $n-ci-mk-k-t$ which has an estimated cost of $\approx 1.5M$. This is the optimal plan from $n$. Since the number $p$ of explored plans from $n$ reaches the value of $max\_plans$, in the next step, \textit{DFS Traversal} is performed from $k$. The first plan enumerated from $k$ is $k-mk-t-ci$ which is pruned early because its partial cost of $\approx 6.5M$ is larger than the complete minimum cost of $\approx 1.5M$. The next plan $k-mk-ci-n-t$ has an estimated cost of only $\approx 0.7M$ and becomes the optimal plan up to this point---in fact, it is the optimal plan identified by COMPASS (Figure~\ref{fig:query_6a_plans}). The third -- and final -- plan considered from $k$ is $k-mk-ci-t$. This plan is pruned early. Notice that the enumeration terminates without reaching the maximum number of allowed plans. The enumeration starting from $t$ does not proceed beyond its immediate neighbors because of the large cardinality estimates. The only complete plan enumerated from $mk$ is $mk-k-ci-n-t$ which has the same cost as the minimal cost plan---they are equivalent. All the plans starting from $ci$ are pruned early. The left-deep plan identified by PostgreSQL and DBMS A is $k-mk-t-ci-n$ (Figure~\ref{fig:query_6a_plans}). Although this plan has a slightly lower cost, it is not enumerated by our algorithm. The reason is the large sketch estimate for the join $k \bowtie mk \bowtie t \bowtie ci$ which stops the enumeration early. While this estimate is inaccurate, the more thorough plan space exploration allows our algorithm to identify an alternative plan with a cost almost identical. We point out that if we apply the same greedy strategy as in left-deep plan search using the sketch estimates in Table~\ref{table:agms_vs_fagms}, we would get the same optimal plan as PostgreSQL and DBMS A. This is because, once we reach the sub-plan $k-mk-t$, the only alternative is to choose $ci$---there is no backtracking. Thus, left-deep search identifies the plan $k-mk-t-ci-n$ by chance rather than by considering estimates for four-way joins---known to be unreliable for any type of synopses, not only sketches.

%%%%%%%%%%%%%%%%%%%%%%%%%%%%%%%%%%%%%%%%%%%%%%%%%%%%%%%%%%%%%%%%%
%\input{experiment}
\section{EMPIRICAL EVALUATION}\label{sec:experiments}

We perform an extensive experimental study over the complete JOB benchmark~\cite{Leis:JOB:vldb-2018} in order to evaluate the performance of COMPASS and compare it against four other database query optimizers (Figure~\ref{fig:query_6a_plans}). While our main goal is to determine whether COMPASS is a complete optimizer -- which requires an effective integration of cardinality estimation in plan enumeration -- we also perform a detailed comparison between Fast-AGMS sketches and several state-of-the-art methods for multi-way join cardinality estimation. Moreover, we assess the impact of the proposed plan enumeration algorithm. To this end, our evaluation investigates the following questions:
\begin{itemize}[leftmargin=*,noitemsep,nolistsep]
\item What is the quality of the query execution plans generated by COMPASS? We measure plan quality as the total cardinality of the intermediate results since this is independent from specific execution engine optimizations. Moreover, logical optimizers use cardinality information as the main criterion to rank plans.

\item What is the execution time -- or runtime -- for the COMPASS plans? Since this is highly dependent on the underlying query processing engine, we execute the plans in MapD, PostgreSQL and DBMS A. This allows us to identify the correlation -- if there is one -- between plan quality and execution time.

\item What is the overall JOB workload runtime? While individual queries allow for localized analysis, the workload execution time measures the reliability of COMPASS. However, due to the high variance in JOB query complexity, this measure alone is not an absolute indicator of the quality of an optimizer.

\item How does COMPASS compare against the pessimistic optimizers that minimize upper bound cardinality, i.e., over-estimates? Since the highly-optimized pessimistic plans are shown to be considerably faster than the default PostgreSQL plans~\cite{Cai:PCETUB:sigmod-2019,Hertzschuch:simplicity:cidr-2021}, we are interested where COMPASS stands on this scale.

\item What is the optimization overhead incurred by sketch merging in plan enumeration? While significantly improving upon sketch partitioning, it is not clear if online sketch merging during push-down selection is fast enough to be practical. We deem COMPASS to be a practical optimizer if it manages to consistently outperform the other databases and also incurs a reduced overhead.

\item How does the accuracy of Fast-AGMS sketch merging compare against state-of-the-art methods for multi-way join cardinality estimation? Previous studies~\cite{Vengerov:JSEFC:pvldb-2015,Kiefer:KDE:pvldb-2017} include only AGMS sketches, which are known to be considerably worse than Fast-AGMS for two-way join estimation~\cite{Rusu:SAS:sigmod-2007,Rusu:SJSE:tods-2008}.

\item How does the plan enumeration algorithm driven by sketch merging compare against standard algorithms? Does the larger search space improve the plan quality compared to the greedy left-deep enumeration? Alternatively, how close (far) is the proposed algorithm to exhaustive enumeration?
\end{itemize}

%%%%%%%%%%%%%%%%%%%%%%%%%%%%%%%%%%%%%%%%%%%%%%%%%%%%%%%%%%%%%%%%%
\subsection{Experimental Setup}

%%%%%%%%%%%%%%%%%%%%%%%%%%%%%%%%%%%%
\paragraph*{Implementation.}
We implement COMPASS in MapD (version 3.6.1)~\cite{mapd}. The source code is publicly available in Github~\cite{Izenov:compass-github:2020}. MapD has a highly-parallel GPU-accelerated query execution engine. Relational operators are compiled into CUDA kernels that are executed concurrently across the SIMD GPU  architecture. In order to reduce data movement, MapD coalesces multiple relational operators into a single CUDA kernel. For joins, this corresponds to a worst-case optimal join algorithm~\cite{Ngo:WORST:pods-2012}. The MapD query optimizer, however, is not as sophisticated as its execution engine. It relies on the Calcite SQL compiler~\cite{calcite}
to get a lexicographic -- in the order in which the query is written -- query execution plan. The join order is computed based on a primitive heuristic that sorts the tables in decreasing order of their cardinality. Moreover, selection predicates are not considered in the optimization. COMPASS brings a principled cost-based optimization procedure to the MapD query optimizer.

The COMPASS implementation consists of two modules---a scan operator that integrates Fast-AGMS sketch construction with push-down selection and a lightweight join order enumeration algorithm. For sketch construction, we adapt a publicly available two-way join Fast-AGMS sketch implementation~\cite{Rusu:sketch-library} to the MapD CUDA kernel API. This requires parallelizing both the update and the estimation functions. The scan operator filters only the relevant tuples to be passed to the sketch update. Since separate sketch instances are created for every GPU block warp, this requires an additional merge stage---currently performed on the CPU. The sketches used throughout the experiments have 11 rows of 1023 buckets, for a total of roughly $\approx$ 11K integers. Assuming 4-byte integers, the memory usage of a sketch is $\approx$ 45KB. The largest query has 28 joins. With two sketches per join, the maximum memory usage for a query is $\approx$ 2.5MB ($2\cdot 28\cdot 45$), which is quite small. Depending on the parallelization approach, there can be a sketch instance on every GPU block warp. In our case, the NVIDIA Tesla K80 has 26 block warps, resulting in a total of $\approx$ 65MB memory usage for the most complex JOB query. The COMPASS plan enumeration algorithm depicted in Algorithm~\ref{alg:dfs} replaces the primitive sorting heuristic from MapD. It determines the optimal plan based on the sketches computed by the scan operator and the configurable enumeration logic. Greedy join enumeration is the default algorithm used throughout experiments.

%%%%%%%%%%%%%%%%%%%%%%%%%%%%%%%%%%%%
\paragraph*{Database systems \& hardware.}
The other three databases we use in addition to MapD are PostgreSQL (v.11.5), MonetDB (v.11.33.11), and the commercial DBMS A. PostgreSQL and DBMS A are used as the common ground in all the experiments because of their extensibility. Both of them allow us to inject and execute the join orders computed by the other databases---the \texttt{CROSS JOIN} statement in PostgreSQL and the hints in DBMS A, respectively. We configure PostgreSQL with 2GB memory per operator, 32GB buffer cache size, and we force the optimizer to use dynamic programming in plan enumeration for queries with no more than 18 join predicates. These settings follow prior art~\cite{Leis:JOB:vldb-2018}. We use an optimized docker image publicly available for DBMS A, while for MonetDB we keep the default configuration. All the systems run on a Ubuntu 16.04 LTS machine with 56 CPU cores (Intel Xeon E5-2660), 256GB RAM, HDD storage, and an NVIDIA Tesla K80 GPU.

%%%%%%%%%%%%%%%%%%%%%%%%%%%%%%%%%%%%%%%%%%%%%%%%%%%%%%%%%%%%%%%%%
\begin{figure*}[htbp]
  \centering
  \includegraphics[width=\textwidth]{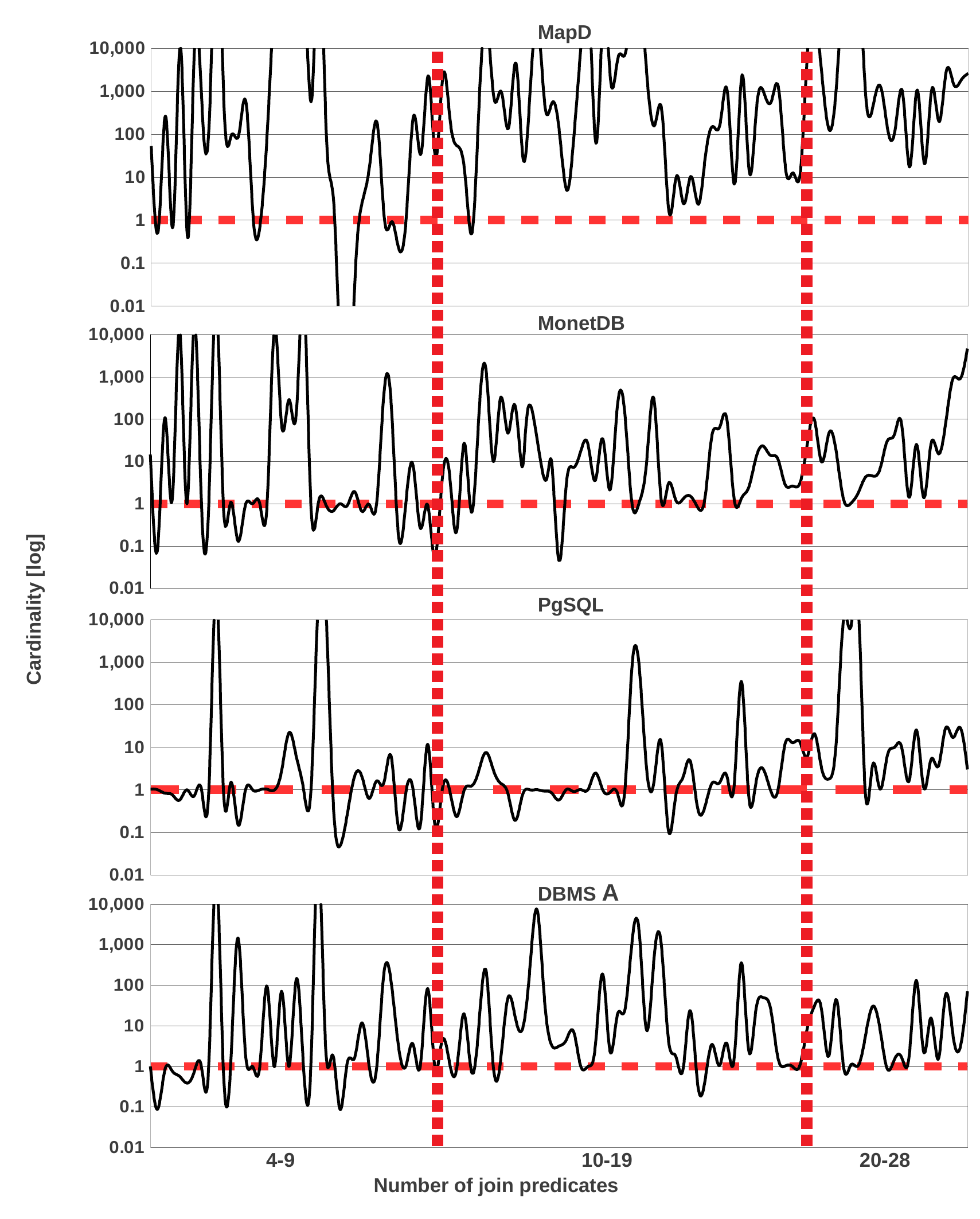}
  \caption{Cardinality (in PostgreSQL) as a normalized ratio to COMPASS.}
  \label{fig:system_inter_comparison}
\end{figure*}
%%%%%%%%%%%%%%%%%%%%%%%%%%%%%%%%%%%%%%%%%%%%%%%%%%%%%%%%%%%%%%%%%

%%%%%%%%%%%%%%%%%%%%%%%%%%%%%%%%%%%%
\paragraph*{Dataset and query workload.}
We perform the experiments on the IMDB dataset~\cite{Boncz:imdb-data} which has been used extensively to evaluate query optimizers~\cite{Leis:QOREALLY:pvldb-2015} and has become a de-facto standard. The JOB benchmark~\cite{JOB-github} defines 113 queries -- grouped into 33 families -- over the IMDB dataset. These queries vary significantly in their complexity, with the simplest one having 4 joins and the most complex one having 28 joins. This variability manifests itself in execution times that are highly-different. To compensate for this, we split the queries into three groups and examine each group separately. These groups are based on the number of joins in the query: group1 contains queries with 4-9 joins; group2, 10-19 joins; and group3, 20-28 joins. We organize the results according to these groups.

%%%%%%%%%%%%%%%%%%%%%%%%%%%%%%%%%%%%
\paragraph*{Methodology.}
To quantitatively assess the quality of a join order plan, we use two metrics---intermediate result cardinality and query execution time. The total cardinality of the intermediate results quantifies how many tuples are produced by all the joins in the plan. The lower this number is, the better the plan. This is the primary metric used in logical query optimization to estimate the cost of a plan. However, the actual execution time depends on specific query processing optimizations. Thus, the execution time is not entirely correlated with the cardinality.

In order to fairly evaluate the join orders produced by every database, we use both PostgreSQL and DBMS A as common ground. First, we run the queries in each database and collect their join plans. Then, we inject these plans into PostgreSQL and DBMS A, respectively, and measure their runtime. Moreover, we execute all the subqueries in the plans to compute the intermediate cardinality. Notice that every system generates its plan independently based on its own algorithm and statistics. PostgreSQL and DBMS A serve as common execution engines for all the plans. This procedure allows for a holistic comparison of the query optimizers---independent of the execution engine.

%%%%%%%%%%%%%%%%%%%%%%%%%%%%%%%%%%%%%%%%%%%%%%%%%%%%%%%%%%%%%%%%%
\begin{figure*}[htbp]
  \centering
  \includegraphics[width=\textwidth]{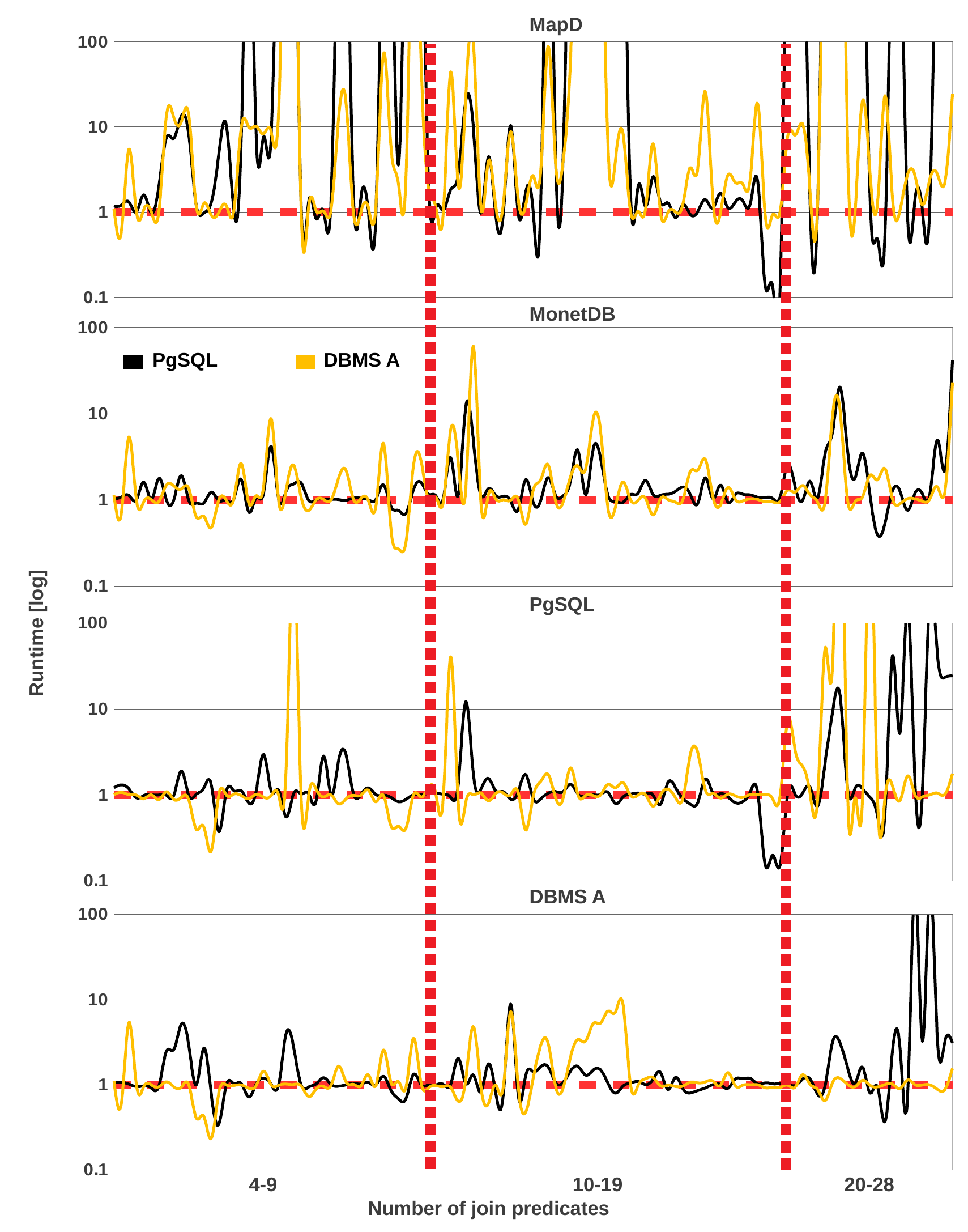}
  \caption{Runtime (in PosgreSQL and DBMS A) as a normalized ratio to COMPASS.}
  \label{fig:system_runtime_comparison}
\end{figure*}
%%%%%%%%%%%%%%%%%%%%%%%%%%%%%%%%%%%%%%%%%%%%%%%%%%%%%%%%%%%%%%%%%

%%%%%%%%%%%%%%%%%%%%%%%%%%%%%%%%%%%%
\subsection{Results}\label{experimental-study-section}

We present the results of our extensive experimental evaluation, organized based on the investigation questions defined at the beginning of this section. The answers to the questions are summarized after the presentation of the results.

%%%%%%%%%%%%%%%%%%%%%%%%%%%%%%%%%%%%
\subsubsection{Query-level Analysis}

In this experiment, we compare COMPASS against every other system for each query in the JOB benchmark. We measure both the intermediate result cardinality, as well as the execution time---taken as the median value over 9 runs. The execution plans are obtained by performing the query in each system. These are subsequently injected in PostgreSQL and DBMS A, and executed on the same execution engine. The cardinalities are generated by executing all the subqueries in the plan in the corresponding order---which is done in PostgreSQL. This information is extracted from the individual plans. Figure~\ref{fig:system_inter_comparison} and~\ref{fig:system_runtime_comparison} depict the results normalized to COMPASS. All the values are divided by the COMPASS results---represented as a horizontal dotted line at position 1 on the y-axis. A point below this line means that the other system has a better result, otherwise, COMPASS performs better. The results are grouped by the number of joins in the JOB queries (x-axis) and separated by two dotted vertical lines.

\paragraph*{MapD.}
MapD consistently produces execution plans that have cardinality two orders of magnitude or larger than COMPASS. With a few exceptions, all MapD plans are worse. There is one such query -- the discontinuity going to zero in the figure -- that indeed has cardinality zero and MapD correctly detects it. However, this is only a matter of chance because the first join in the plan -- between the largest tables in the query -- does not produce any results. The reason for this poor plan quality is the lack of statistics in the MapD query optimizer. Decisions are taken solely based on the full table cardinality---the number of tuples before any selection predicate. Therefore, the resulting plans are highly sub-optimal. While runtime follows cardinality -- with many results 100X slower than COMPASS -- the correlation between the two is not complete. There are several queries for which the MapD cardinality is considerably worse, while the execution time is similar or better than COMPASS. This is the case for some of the complex queries with 20 or more joins executed in PostgreSQL. In this situation, MapD chooses a large well-connected table early in the plan. This allows it to check many join predicates at the beginning and prune a large number of tuples. On the other hand, COMPASS -- and the other systems -- start from small tables on the periphery of the join graph and make their way to the highly-connected tables in the center. This strategy produces many staged intermediate results that increase the runtime. While the runtime trend across PgSQL and DBMS A is similar, we observe that queries with 20 or more joins are handled better by DBMS A, while queries with less than 20 joins are faster in PgSQL. This is an indication that DBMS A is better optimized for complex queries.

\paragraph*{MonetDB.}
The trend of the cardinality results in MonetDB follows the one in MapD. While the majority of the results are worse than COMPASS, the ratio is smaller than for MapD. This improvement is due to the more advanced rule-based MonetDB query optimizer with limited statistics support. However, compared to the full sketch-based COMPASS, the MonetDB cardinalities are considerably worse---many times an order of magnitude or more. Interestingly enough, though, the corresponding query runtimes fare much better than predicted by the cardinality. With few exceptions, they are always within a factor of 10 -- more often less -- off of COMPASS. Moreover, they are independent of query complexity and do not exhibit spikes. Overall, the MonetDB runtimes are the most consistent with COMPASS across both PgSQL and DBMS A. This is because the MonetDB query optimizer finds plans that are executed similarly to COMPASS---albeit they have higher cardinality.

\paragraph*{PostgreSQL.}
The cardinality results for PostgreSQL -- PgSQL in the figure -- are the closest to COMPASS among all the systems. This is entirely due to the advanced statistics the PostgreSQL optimizer employs. While mildly better than COMPASS for several queries, PostgreSQL still exhibits spikes that go beyond a factor of 1000X. The reason is the failure to detect correlations between join attributes. Since the plans are optimized for the PostgreSQL execution engine in this case, we expect the runtimes to be optimal. This is indeed the case for queries with less than 20 joins. However, for 20 or more joins, the PostgreSQL runtime is considerably worse compared to COMPASS. This is where the PostgreSQL optimizer drops dynamic programming in plan search. With a few exceptions where there are dramatic spikes that go beyond 100X, the PgSQL plans executed in DBMS A perform as well as or better than in PostgreSQL itself. This is especially true for the complex queries having 20 or more joins. Overall, COMPASS generates more stable plans than PostgreSQL. Although not specifically optimized for it, PostgreSQL executes them as fast -- or faster -- than its own plans.

\paragraph*{DBMS A.}
The commercial DBMS A produces plans that have consistently higher cardinality than COMPASS across all the JOB queries. This clearly shows that the employed statistics do a poor job at estimating the join cardinality. However, when executed in PostgreSQL, these plans have unexpectedly good runtimes---except for queries with more than 20 joins. This is likely due to the more complex cost function that considers other parameters beyond cardinality in determining the optimal plan. Interestingly enough, when executing its own plans, DBMS A does not fare better than PostgreSQL, except for the complex queries with more than 20 joins. In fact, DBMS A has worse runtime for queries with 10 to 20 joins. The runtimes of DBMS A and COMPASS are close to each other and always within a factor of 10X. This confirms that the COMPASS plans are also optimal for DBMS A.

%%%%%%%%%%%%%%%%%%%%%%%%%%%%%%%%%%%%%%%%%%%%%%%%%%%%%%%%%%%%%%%%%
\begin{figure*}[htbp]
\begin{subfigure}{0.48\textwidth}
\centering
\includegraphics[width=0.85\linewidth]{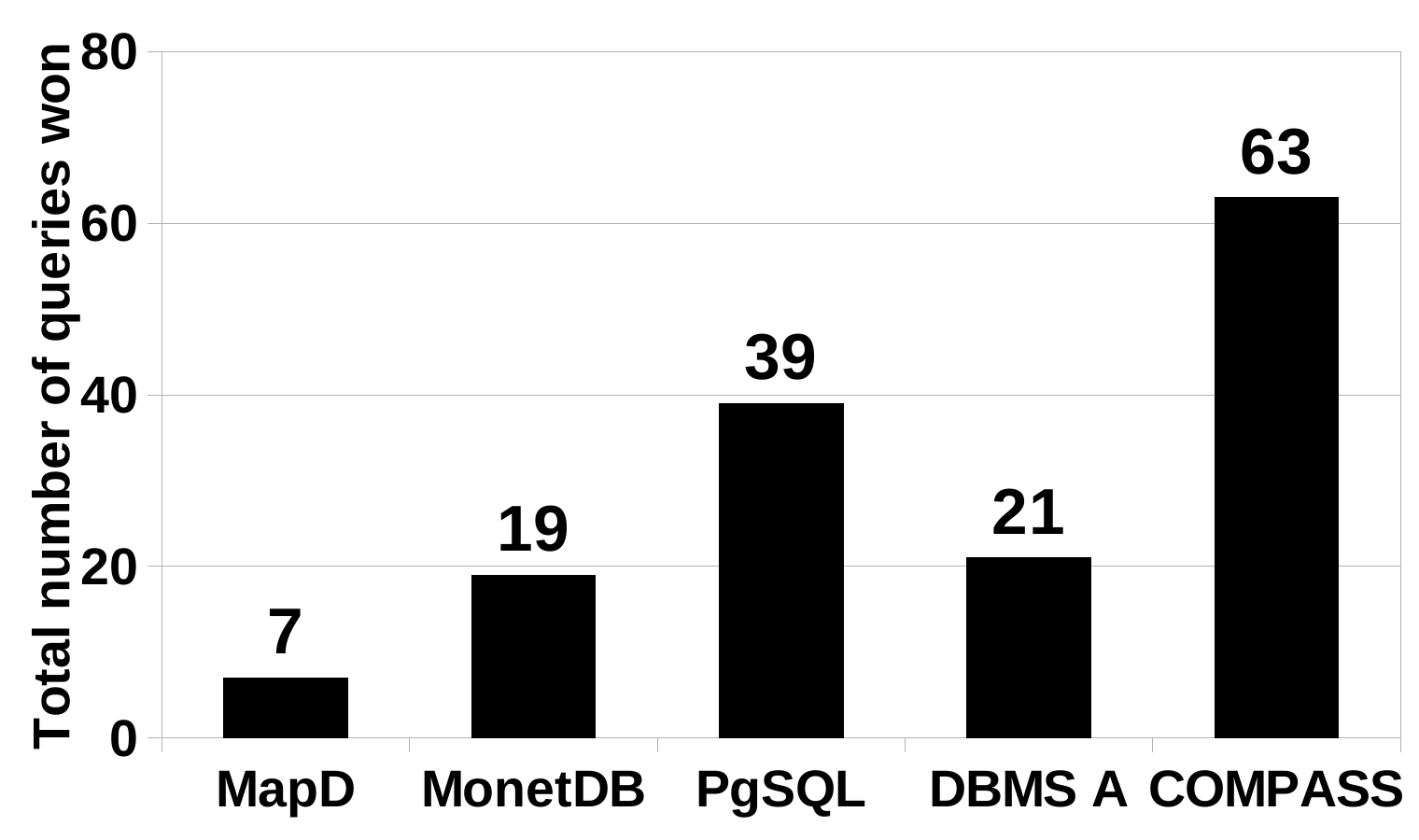}
\caption{Cardinality comparison}
\label{fig:cardinality_piechart}
\end{subfigure}
%\hfill
\begin{subfigure}{0.48\textwidth}
\centering
\includegraphics[width=0.85\linewidth]{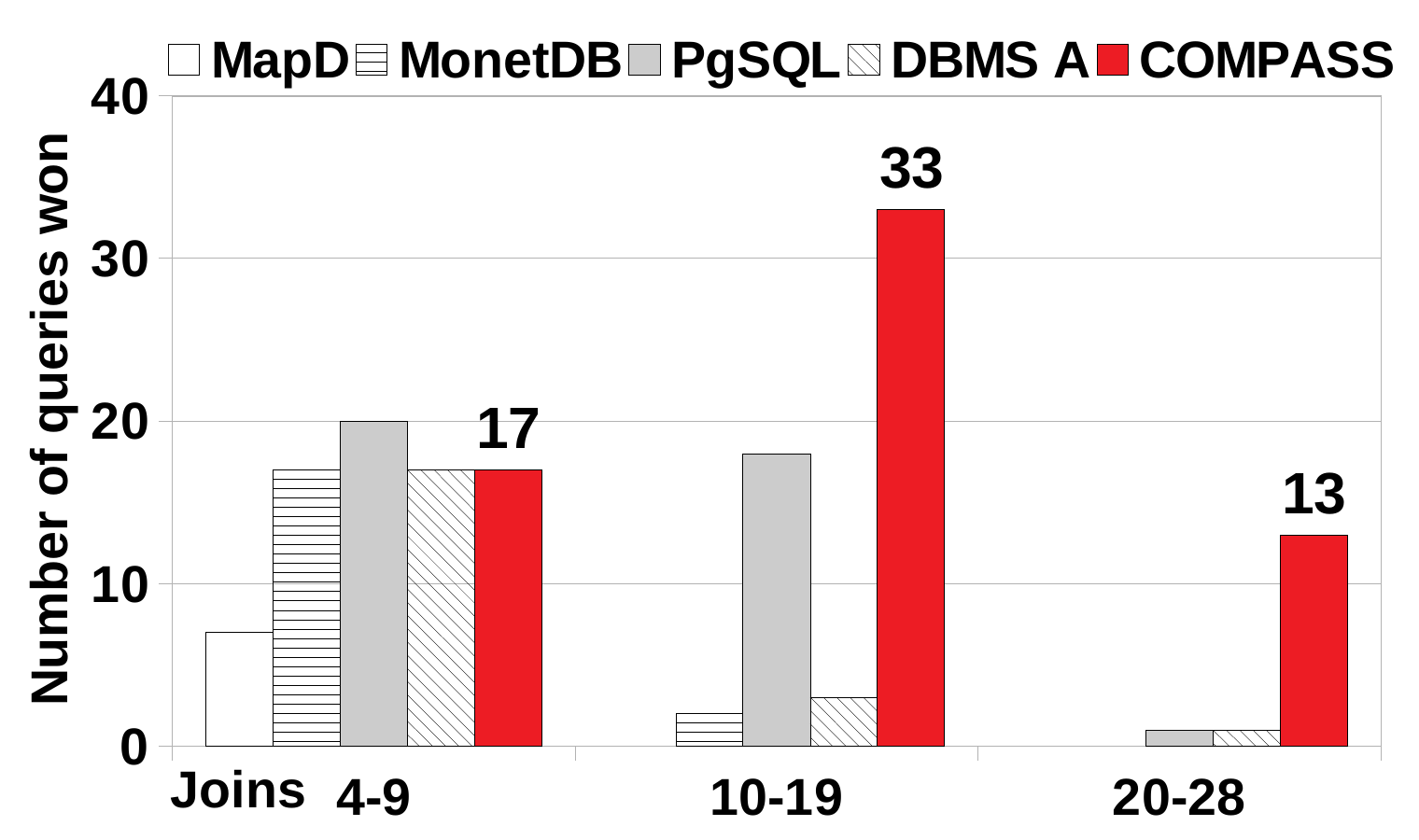}
\caption{Cardinality vs number of joins}
\label{fig:cardinality_histogram}
\end{subfigure}
\caption{Distribution of winning queries across the databases in terms of intermediate cardinality. The databases with the lowest cardinality are the winners. The total (149) is larger than 113 -- the number of queries in JOB -- because there are multiple queries for which more than one database is the winner.}
\label{fig:db_comparison_cardinality}
\end{figure*}
%%%%%%%%%%%%%%%%%%%%%%%%%%%%%%%%%%%%%%%%%%%%%%%%%%%%%%%%%%%%%%%%%

%%%%%%%%%%%%%%%%%%%%%%%%%%%%%%%%%%%%%%%%%%%%%%%%%%%%%%%%%%%%%%%%%
\subsubsection{Aggregated Workload Statistics}

\begin{figure*}[htbp]
\begin{subfigure}{0.48\textwidth}
\centering
\includegraphics[width=0.85\linewidth]{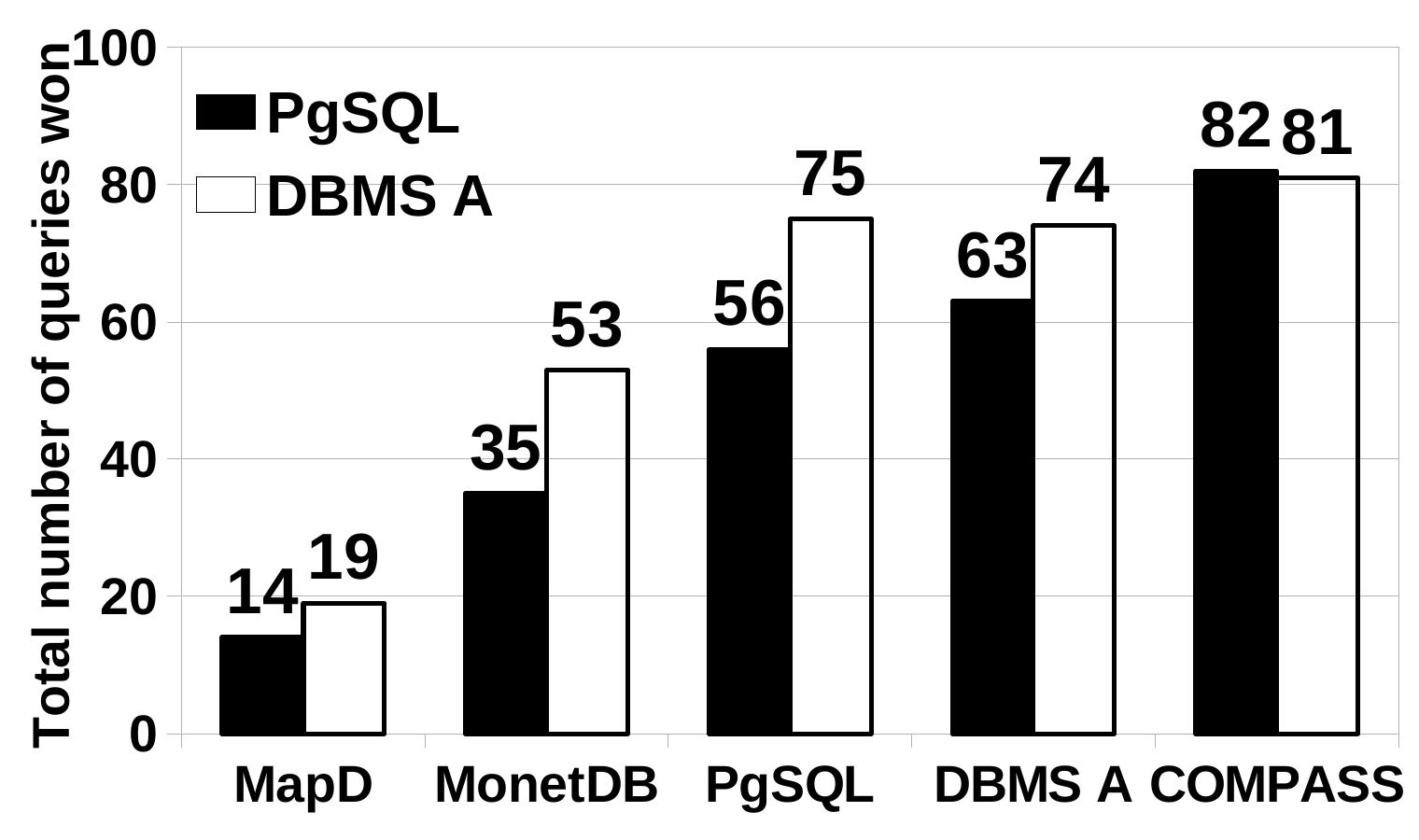}
\caption{Execution time comparison}
\label{fig:runtime_piechart}
\end{subfigure}
\hfill
\begin{subfigure}{0.48\textwidth}
\centering
\includegraphics[width=0.85\linewidth]{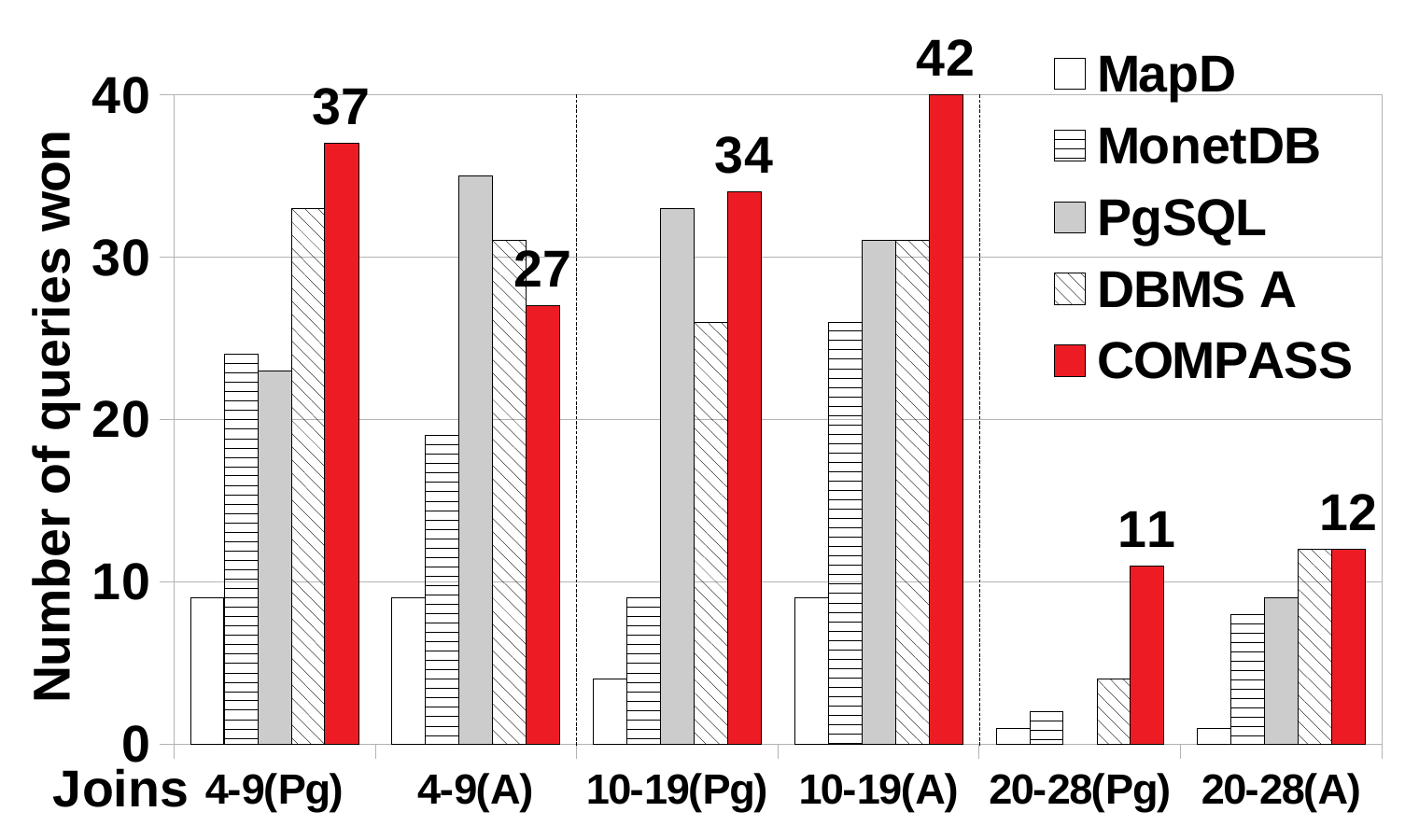}
\caption{Execution time vs number of joins}
\label{fig:runtime_histogram}
\end{subfigure}
\caption{Distribution of winning queries across the databases in terms of execution time when the optimal plans are plugged-in and executed in PostgreSQL and DBMS A. We obtain the optimal plan for every database from its query optimizer and execute it in PostgreSQL and DBMS A. All the databases within 10\% of the fastest execution time are considered as winners. Thus, the total (250) is larger than 113---the number of queries in JOB.}
\label{fig:db_comparison_runtime}
\end{figure*}

We aggregate the query-level results (Figure~\ref{fig:system_inter_comparison} and \ref{fig:system_runtime_comparison}) in order to obtain an overall view of the relative performance of the compared systems. These aggregated results are depicted in Figure~\ref{fig:db_comparison_cardinality} and \ref{fig:db_comparison_runtime}, respectively. They give the total number of queries for which a database performs the best, as well as the distribution as a function of the number of joins in the query. In the case of cardinality, a database is counted if it achieves the minimum cardinality among all the databases. For runtime, a database is counted if it comes within 10\% of the fastest runtime---computed as the median of 9 runs. This bound compensates for variations in the environment.

%%%%%%%%%%%%%%%%%%%%%%%%%%%%%%%%%%%%%%%%%%%%%%%%%%%%%%%%%%%%%%%%%
%%%%%%%%%%%%%%%%%%%%%%%%%%%%%%%%%%%%%%%%%%%%%%%%%%%%%%%%%%%%%%%%%

Based on Figure~\ref{fig:cardinality_piechart}, COMPASS achieves the plan with the minimum cardinality for 63 out of the 113 JOB queries. This represents approximately 56\% of the workload. PostgreSQL (PgSQL) comes in second place with 39 queries. The other three databases obtain the best cardinality in less than 20\% of the queries each, with MapD winning only 7 queries. The careful reader notices that the sum of the winning queries is larger than 113. This is because there are queries for which two or more systems achieve the same best cardinality---case in which we count each of them. The distribution of the winning queries in terms of the number of joins is depicted in Figure~\ref{fig:cardinality_histogram}. While for the simpler queries with less than 10 joins all the systems perform similarly, COMPASS clearly dominates the others when the complexity increases. PostgreSQL is the only other database that performs sufficiently well, however, only for queries with a moderate number of joins. These results prove the benefit of using statistics in query optimization, especially for complicated queries. While the PostgreSQL statistics perform well for simple to moderate queries, COMPASS sketches are less sensitive to the number of joins in the query---they provide more consistent estimates. Moreover, COMPASS is not heavily impacted by the greedy join enumeration algorithm. When PostgreSQL switches from dynamic programming -- more than 18 joins -- it fails to find any best plan.

The aggregated runtime results in PgSQL and DBMS A are depicted in Figure~\ref{fig:runtime_piechart} and~\ref{fig:runtime_histogram}. They follow closely the corresponding cardinality results---with one exception. The runtime for the commercial DBMS A is much better than its cardinalities anticipate---DBMS A has the best runtime for 63 and 74 queries, while its cardinality is best only for 21 queries. The reasons are outlined when the individual query results are discussed. Additionally, DBMS A benefits from the bound on runtime since it often comes within the fastest system. Overall, COMPASS achieves the fastest runtime for 82 (PgSQL) and 81 (DBMS A) out of the 113 JOB queries -- 72\% of the workload -- which is more than any other database. This proves the superiority of the identified plans and confirms the correlation between cardinality and runtime. The correlation manifests more clearly for queries with a larger number of joins because of the higher runtime, which makes ties more unlikely. Moreover, the correlation is stronger for PgSQL than for DBMS A since the number of winning queries is higher in DBMS A for all systems except COMPASS. A careful reader observes that the runtime results are higher than the cardinality results for all the systems---and larger than 113 when summed up. This is because it is more common to have close-enough runtimes than it is to have the same cardinality---multiple counting is more frequent. Based on these results, we conclude that COMPASS is the optimizer with the most consistent and resilient plans on the JOB benchmark.

%%%%%%%%%%%%%%%%%%%%%%%%%%%%%%%%%%%%%%%%%%%%%%%%%%%%%%%%%%%%%%%%%
\subsubsection{Total Workload Runtime}

The runtimes for the complete JOB workload execution in PostgreSQL and DBMS A using the plans generated by each database are included in Table~\ref{table:total_runtime_compare-pgsql}. Given the high variance among queries, these numbers have to be taken with a grain of salt since they may be dominated by a few complex queries with a large number of joins. Nonetheless, we follow prior art~\cite{Cai:PCETUB:sigmod-2019,Trummer:SKINNERDB:sigmod-2019} and include them together with the aggregated workload statistics. As expected, COMPASS has the overall fastest runtime. Somewhat unexpectedly, MonetDB comes in second for the PgSQL execution with a runtime that is almost twice as large as that of COMPASS. The reason is because MonetDB does not fail dramatically for any of the JOB queries. While it performs consistently slower, it never derails on heavily sub-optimal plans. The runtime for PgSQL and DBMS A in PgSQL is dominated by the long-running queries with 20 or more joins, which pull the total time to more than 8X and 5X that of COMPASS. These outliers are sufficient to skew the overall runtime. In the case of MapD, there are 30 queries that do not finish execution even after a timeout of 20 minutes per query. Thus, the very large runtime. When the workload is executed in DBMS A, all systems except DBMS A incur an increase in runtime. The increase is most significant for COMPASS as it stands at 50\% more than in PgSQL. On the other hand, DBMS A has a reduction of more than 50\% of its PgSQL runtime. Nonetheless, COMPASS still has the overall fastest runtime, which is 35\% faster than DBMS A.

%%%%%%%%%%%%%%%%%%%%%%%%%%%%%%%%%%%%
\begin{table}[htbp]
  \centering
  \begin{tabular}{|l||rr|rr|} \hline
    \multirow{2}{*}{\textbf{Database}} & \multicolumn{2}{c|}{\textbf{Runtime (minutes)}} & \multicolumn{2}{c|}{\textbf{Ratio to COMPASS}} \\
    & PgSQL & DBMS A & PgSQL & DBMS A \\
    \hline
   MapD & \textgreater 300 & \textgreater 300 & \textbf{\textgreater 23} & \textbf{\textgreater 13} \\
   MonetDB & 27.52 & 35.71 & \textbf{2.19} & \textbf{1.65} \\
   PgSQL & 103.00 & 244.31 & \textbf{8.20} & \textbf{11.28} \\
   DBMS A & 70.72 & 29.22 & \textbf{5.63} & \textbf{1.35} \\
   \textbf{COMPASS} & \textbf{12.56} & \textbf{21.66} & 1.00 & 1.00 \\ \hline
  \end{tabular}
  \caption{JOB benchmark runtime in PgSQL and DBMS A.}
  \label{table:total_runtime_compare-pgsql}
\end{table}
%%%%%%%%%%%%%%%%%%%%%%%%%%%%%%%%%%%%

%%%%%%%%%%%%%%%%%%%%%%%%%%%%%%%%%%%%%%%%%%%%%%%%%%%%%%%%%%%%%%%%%
\subsubsection{Comparison with Pessimistic Plans}

We compare the plans produced by COMPASS against the pessimistic plans generated in~\cite{Hertzschuch:simplicity:cidr-2021}. The pessimistic plans are determined by minimizing the worst case cardinality estimates. Thus, they always produce over-estimates of the true cardinality. This is in contrast to COMPASS, which generates both over- and under-estimates. The pessimistic plans for all the JOB queries in PostgreSQL are available at~\cite{Hertzschuch:simplicity-github-repo:2020}. They are generated by rewriting the SQL statements such that selection predicates and one-to-many -- key/foreign-key -- joins are evaluated before the many-to-many joins. Moreover, ordering is performed separately for one-to-many and many-to-many joins. This partitioning of the search space results in a massive reduction of the number of considered join orders. A similar idea is employed in~\cite{Cai:PCETUB:sigmod-2019}, where only at most 2-D partitioned sketches are built.

%%%%%%%%%%%%%%%%%%%%%%%%%%%%%%%%%%%%%%%%%%%%%%%%%%%%%%%%%%%%%%%%%
\begin{figure*}[htbp]
\centering
\includegraphics[width=\textwidth]{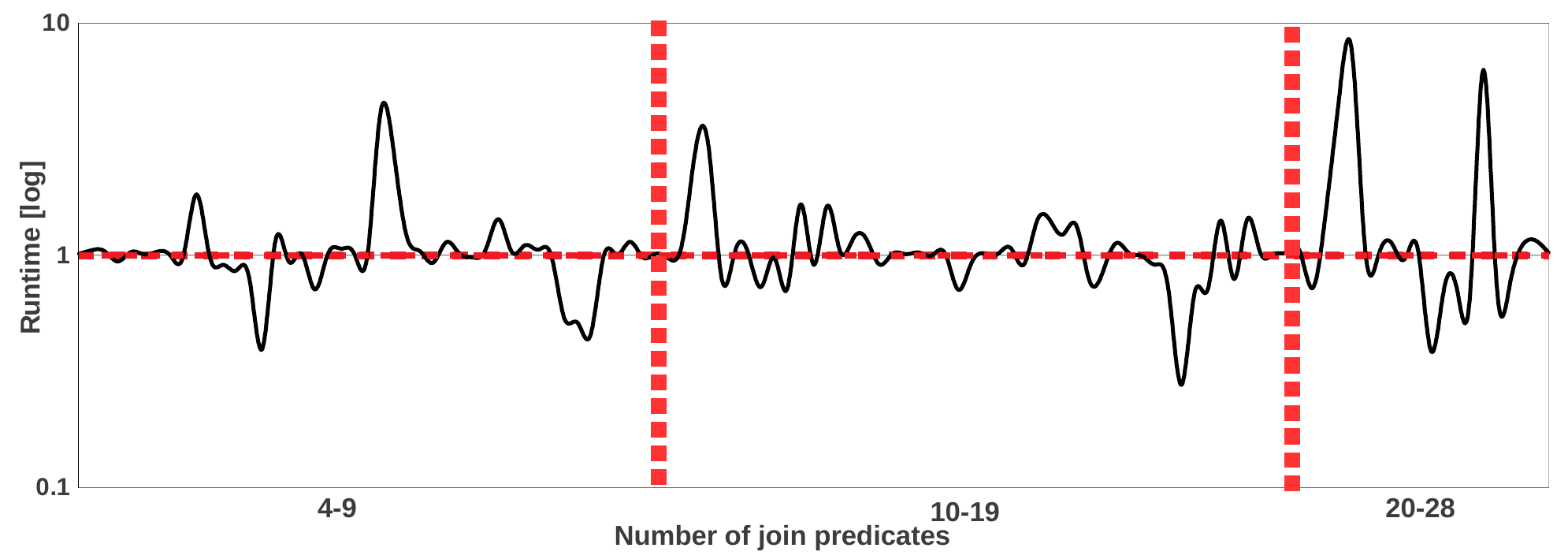}
\caption{Runtime of pessimistic plans (in PosgreSQL) as a normalized ratio to COMPASS.}
\label{fig:pessimistic_vs_compass}
\end{figure*}
%%%%%%%%%%%%%%%%%%%%%%%%%%%%%%%%%%%%%%%%%%%%%%%%%%%%%%%%%%%%%%%%%

The comparison between the runtime of the COMPASS plans and that of the pessimistic plans executed in PostgreSQL is depicted in Figure~\ref{fig:pessimistic_vs_compass}. The results are normalized to the runtime of the COMPASS plans. We observe that the difference between these plans is smaller than for the other systems---an indication that the plans have closer runtime. There are queries for which COMPASS generates faster plans and queries for which the pessimistic plans are better. Overall, there is a slight advantage for COMPASS since the curve is above the horizontal 1-axis more often. Moreover, the gap is higher for COMPASS, reaching a factor of almost 10X for certain queries. In terms of number of joins, the best plans are almost evenly distributed among the two methods. Table~\ref{table:total_runtime_compare-pessimistic} summarizes the individual query results from Figure~\ref{fig:pessimistic_vs_compass}. The timing results are higher than previously published~\cite{Hertzschuch:simplicity:cidr-2021} because no indexes are defined over the key attributes. COMPASS achieves a slightly better performance both in the number of queries won -- 88 vs. 83 -- as well as in the cumulative runtime---COMPASS is faster by approximately 25 seconds. The main reason for the better performance of pessimistic plans -- compared to other systems -- is the separate optimization of the joins. The partition of the join graph based on the many-to-many joins and their independent ordering reduces the multi-way join estimation error significantly. The separate evaluation of the one-to-many joins in every partition reduces the error further. Moreover, the estimation for these simpler joins is more accurate. While partitioning the join order space reduces estimation complexity, it also ignores orderings that can result in better plans. For example, the cardinality of a many-to-many join can be smaller than that of a one-to-many join. Pessimistic plans ignore these interleaved orders. Given the holistic approach that considers the complete join graph, the COMPASS results are quite impressive given the size of the multi-way joins. This is possible because of the good accuracy -- and consistency -- of the merged Fast-AGMS sketches. We conjecture that COMPASS can be improved by adopting a similar tiered approach to join ordering. We plan to explore this idea in future work.

%%%%%%%%%%%%%%%%%%%%%%%%%%%%%%%%%%%%
\begin{table}[htbp]
  \centering
  \begin{tabular}{|l||r|r|r|}
  	\hline
    \textbf{Query plans} & \textbf{Queries won} & \textbf{Runtime (seconds)} & \textbf{Ratio to COMPASS} \\
    \hline
   Pessimistic & 83 & 777.10 & \textbf{1.03} \\
   \textbf{COMPASS} & \textbf{88} & \textbf{753.86} & 1.00 \\ \hline
  \end{tabular}
  \caption{JOB benchmark execution for pessimistic~\cite{Cai:PCETUB:sigmod-2019,Hertzschuch:simplicity:cidr-2021} and COMPASS plans in PostgreSQL.}
  \label{table:total_runtime_compare-pessimistic}
\end{table}
%%%%%%%%%%%%%%%%%%%%%%%%%%%%%%%%%%%%

%%%%%%%%%%%%%%%%%%%%%%%%%%%%%%%%%%%%%%%%%%%%%%%%%%%%%%%%%%%%%%%%%
\begin{figure*}[htbp]
\centering
\includegraphics[width=\textwidth]{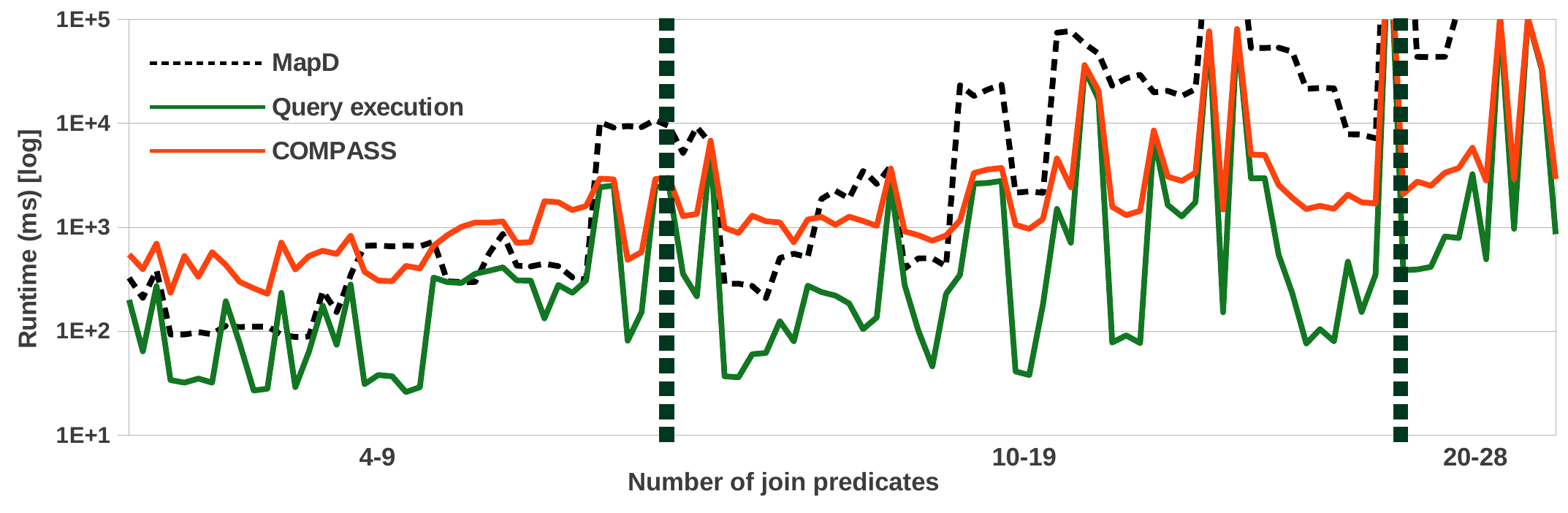}
\caption{Query runtime in MapD with the COMPASS query optimizer.}
\label{fig:mapd_vs_compass}
\end{figure*}
%%%%%%%%%%%%%%%%%%%%%%%%%%%%%%%%%%%%%%%%%%%%%%%%%%%%%%%%%%%%%%%%%

%%%%%%%%%%%%%%%%%%%%%%%%%%%%%%%%%%%%%%%%%%%%%%%%%%%%%%%%%%%%%%%%%
\begin{figure*}[htbp]
\centering
\begin{subfigure}{\textwidth}
\includegraphics[width=\textwidth]{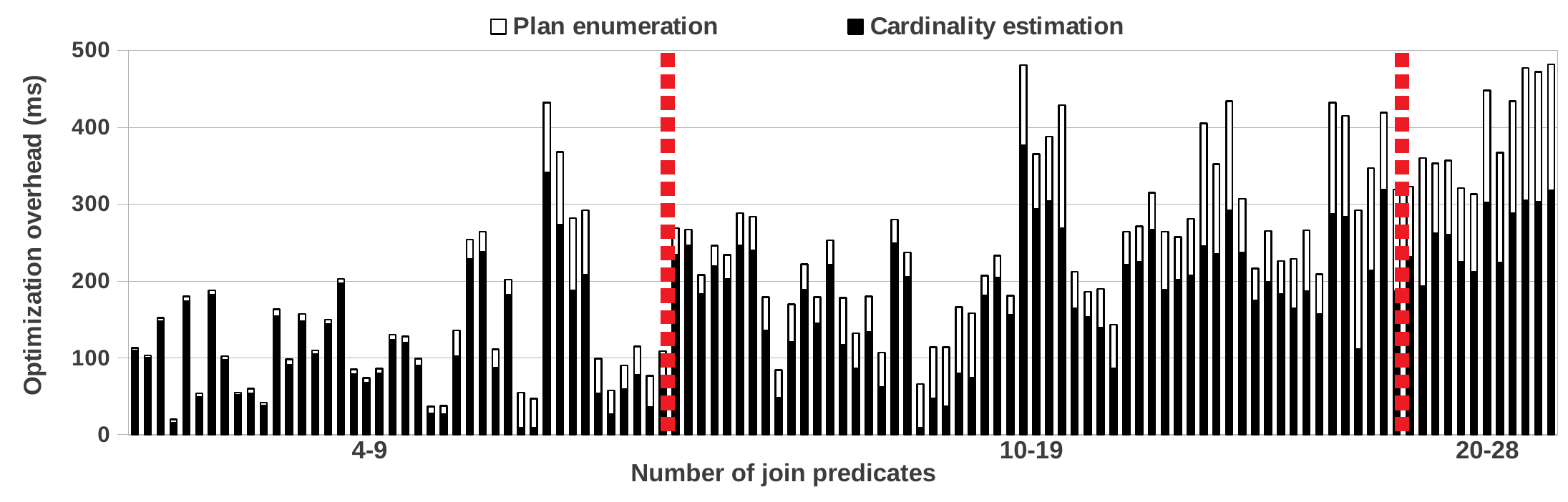}
\caption{Optimization overhead on GPU.}
\label{fig:overhead_absolute}
\end{subfigure}
\begin{subfigure}{\textwidth}
\includegraphics[width=\textwidth]{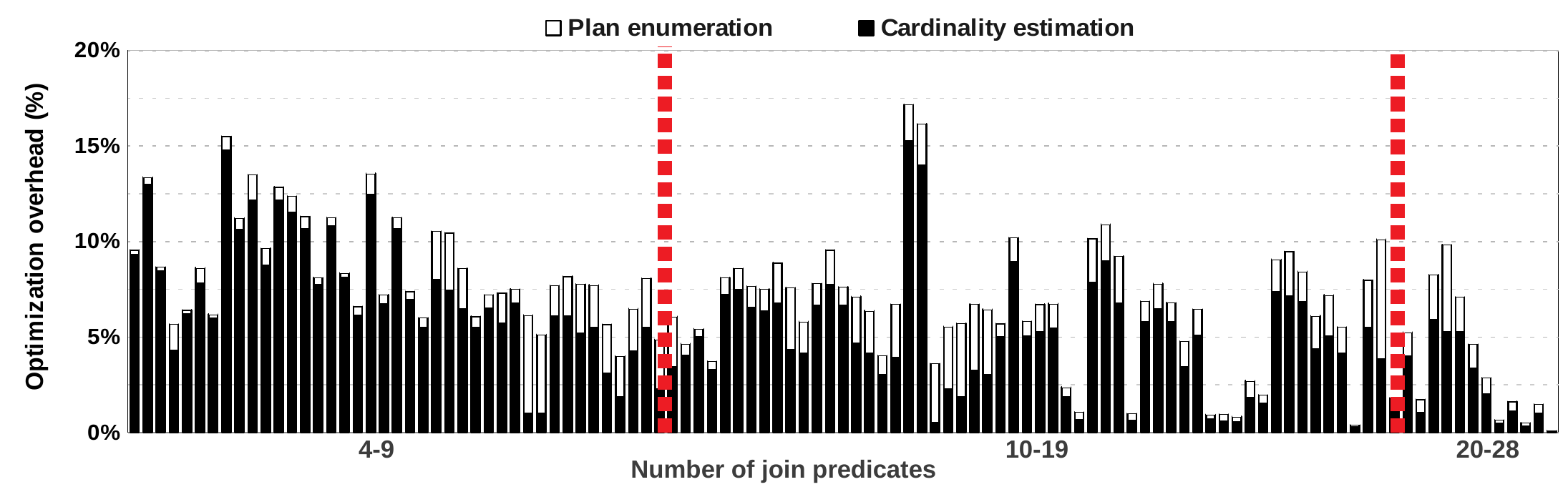}
\caption{Optimization overhead on GPU as a percentage from the total runtime.}
\label{fig:overhead_percent}
\end{subfigure}
\begin{subfigure}{\textwidth}
\includegraphics[width=\textwidth]{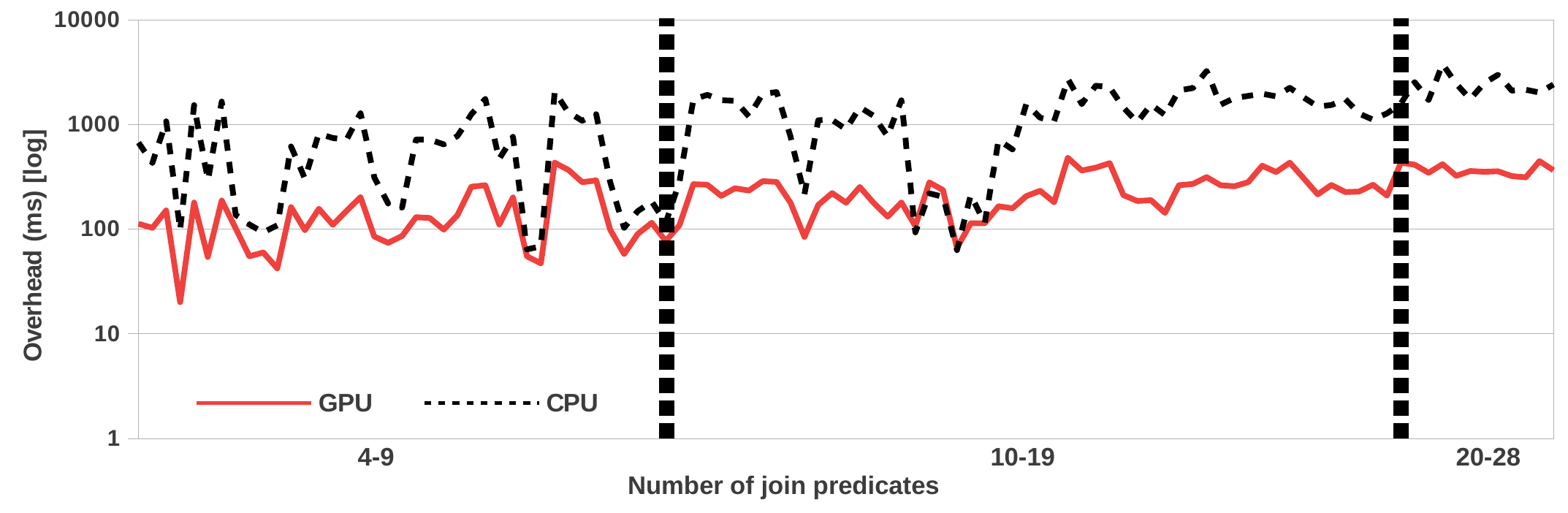}
\caption{Optimization overhead on GPU and CPU.}
\label{fig:overhead_cpu_gpu}
\end{subfigure}
\caption{The overhead of the COMPASS optimizer implemented in MapD.}
\label{fig:overhead_all}
\end{figure*}
%%%%%%%%%%%%%%%%%%%%%%%%%%%%%%%%%%%%%%%%%%%%%%%%%%%%%%%%%%%%%%%%%

\subsubsection{Runtime in MapD}

In this experiment, we evaluate the impact COMPASS has on the MapD database. For this, we replace the default MapD query optimizer with COMPASS and execute the JOB benchmark in both scenarios. We measure the end-to-end query runtime, as well as only the query execution time without optimization---these are the same in MapD. We report the median over 9 runs. Figure~\ref{fig:mapd_vs_compass} depicts the results for every query. We observe that MapD outperforms COMPASS for simple and some moderate queries. However, the differences are not significant, as opposed to the difference for more complex queries. This may be surprising given the primitive MapD query optimizer. However, its execution engine is quite different from PostgreSQL. It is highly-optimized for parallel in-memory processing. This alleviates the need for careful optimization on simple queries. For more complicated queries, though, sketch-based optimization pays off as COMPASS finds considerably better plans. In fact, MapD fails on 8 queries and times out after 30 minutes on 8 other queries. COMPASS finishes all the queries and is faster than MapD for 74 of them, which represents 65\% of the workload. The total runtime for the 97 queries MapD successfully runs is included in Table~\ref{table:total_runtime_compare-mapd}. COMPASS has a runtime of 6.21 minutes to MapD's 47.64---which is a net speedup of 7.67X. This proves both that sketches can be effectively computed at runtime, as well as their benefit to generate better query plans, which result in faster execution. The last point is clear when we compare only the execution time, without optimization overhead---less than ten COMPASS plans have execution time larger than MapD.

%%%%%%%%%%%%%%%%%%%%%%%%%%%%%%%%%%%%
\begin{table}[htbp]
  \centering
  \begin{tabular}{|l||r|r|r|}
  	\hline
    \textbf{Database} & \textbf{Queries won} & \textbf{Runtime (minutes)} & \textbf{Ratio to COMPASS} \\
    \hline
   MapD & 42 & 47.64 & \textbf{7.67} \\
   \textbf{COMPASS} & \textbf{74} & \textbf{6.21} & 1.00 \\ \hline
  \end{tabular}
  \caption{JOB benchmark execution in MapD.}
  \label{table:total_runtime_compare-mapd}
\end{table}
%%%%%%%%%%%%%%%%%%%%%%%%%%%%%%%%%%%%

%%%%%%%%%%%%%%%%%%%%%%%%%%%%%%%%%%%%
\subsubsection{COMPASS Overhead}

We measure the optimization overhead of building Fast-AGMS sketches, as well as that of sketch-based plan enumeration, for the COMPASS MapD implementation. Sketch building can be performed either on GPU or CPU, while merging and plan enumeration are performed on CPU. The results are depicted in Figure~\ref{fig:overhead_all}. Figure~\ref{fig:overhead_absolute} and~\ref{fig:overhead_percent} show the absolute and relative overhead, respectively, for the GPU execution.  The overhead in queries with up to 9 joins is at most 15\%---or at most 420 ms. At a first glance, the overhead may seem significant. However, this time is not spent in vain since the plans selected by COMPASS are quite fast even with the overhead included. For these simpler queries, there is not significant difference between plans. Thus, a primitive optimizer as in MapD is sufficient. The overhead for the rest of the workload is at most 17\%---or 500 ms. This overhead becomes negligible in the overall execution time. As a result, COMPASS largely outperforms the other four databases. Figure~\ref{fig:overhead_cpu_gpu} shows that -- as expected -- sketch building is more efficient on GPU than on CPU due to the higher degree of parallelism. In both cases, the optimization overhead increases with the number of joins in the query. For GPU, the overhead is in the order of hundreds of milliseconds (ms), with a maximum of around 500 ms for certain complex queries. For CPU, the overhead is always below 5 seconds, which is relatively small for queries that take minutes to run. Given that this is only a prototype, we believe that the sketch overhead can be further reduced with more optimized code.

\subsubsection{Comparison with State-of-the-art Synopses for Join Cardinality Estimation}

\begin{figure*}[htbp]
\centering
\hspace{-0.1em}
\vspace{-0.5em}
\begin{subfigure}{0.32\textwidth}
\vspace{-0.5em}
\caption*{IMDB Q1 Uniform}
\vspace{-0.3em}
\includegraphics[width=\linewidth]{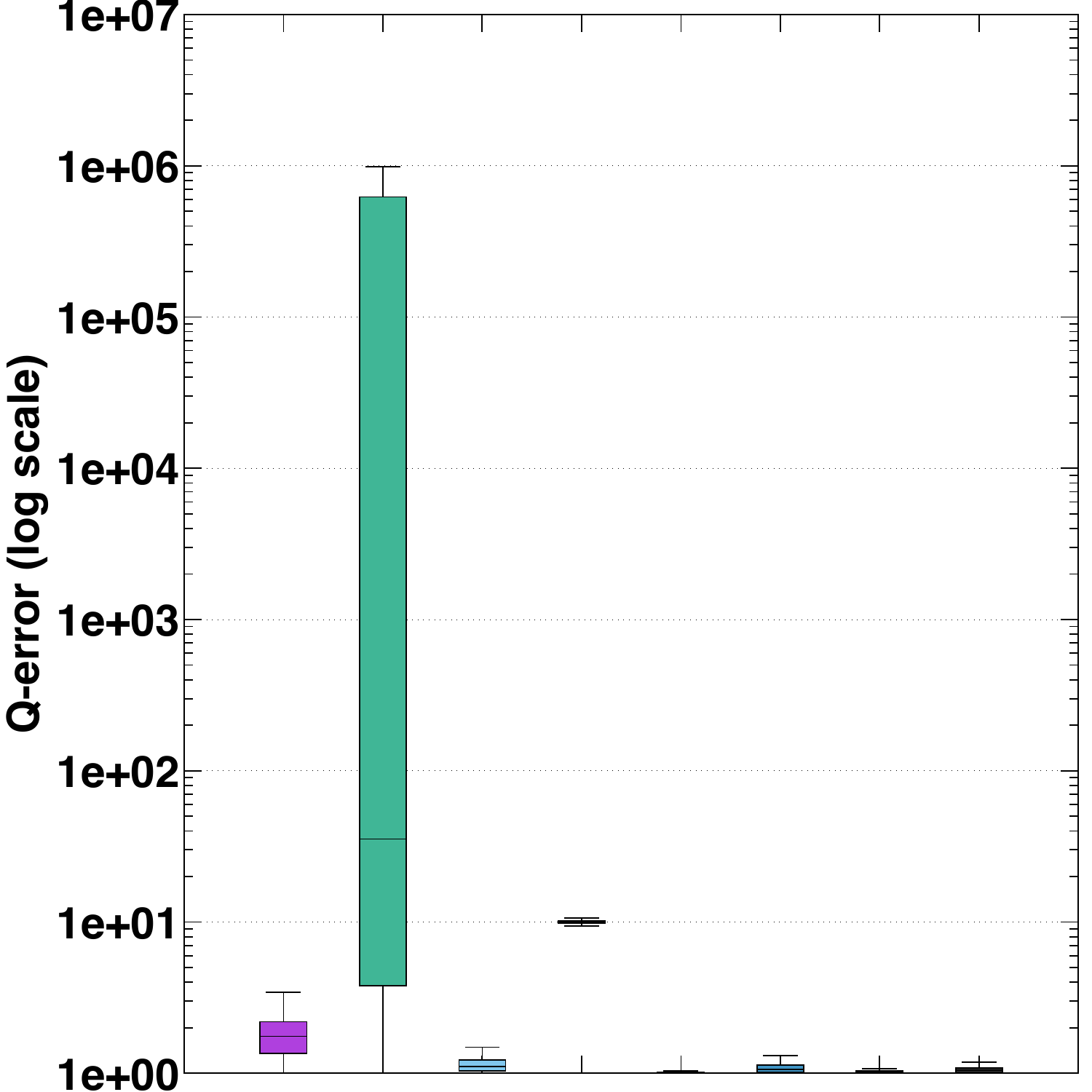}
\label{fig:est_accuracy_1a}
\end{subfigure}
\vspace{-0.1em}
\begin{subfigure}{0.32\textwidth}
\vspace{-0.5em}
\caption*{IMDB Q2 Uniform}
\vspace{-0.3em}
\includegraphics[width=\linewidth]{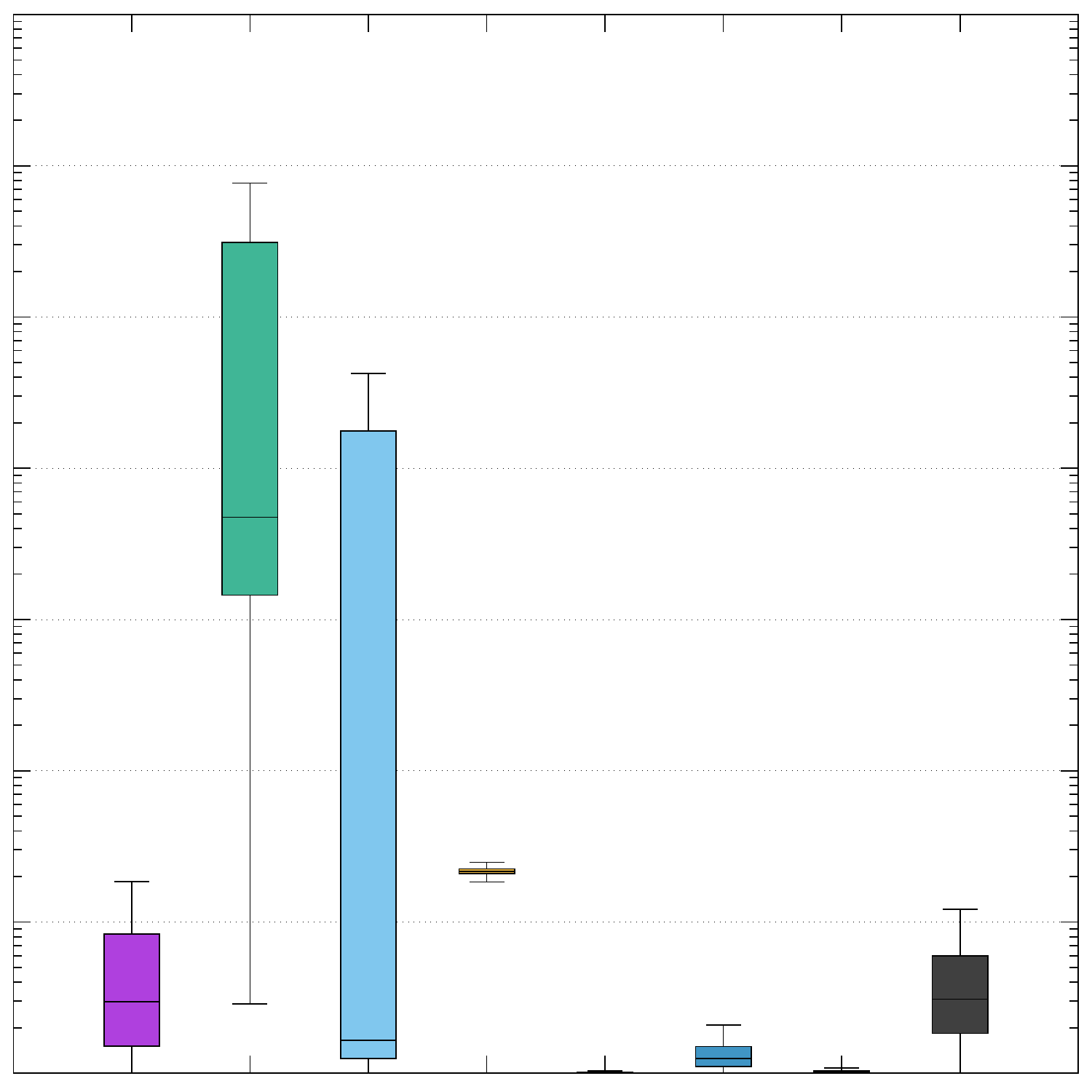}
\label{fig:est_accuracy_2a}
\end{subfigure}
\vspace{-0.5em}
\begin{subfigure}{0.32\textwidth}
\vspace{-0.5em}
\caption*{IMDB Q3 Uniform}
\vspace{-0.4em}
\includegraphics[width=\linewidth]{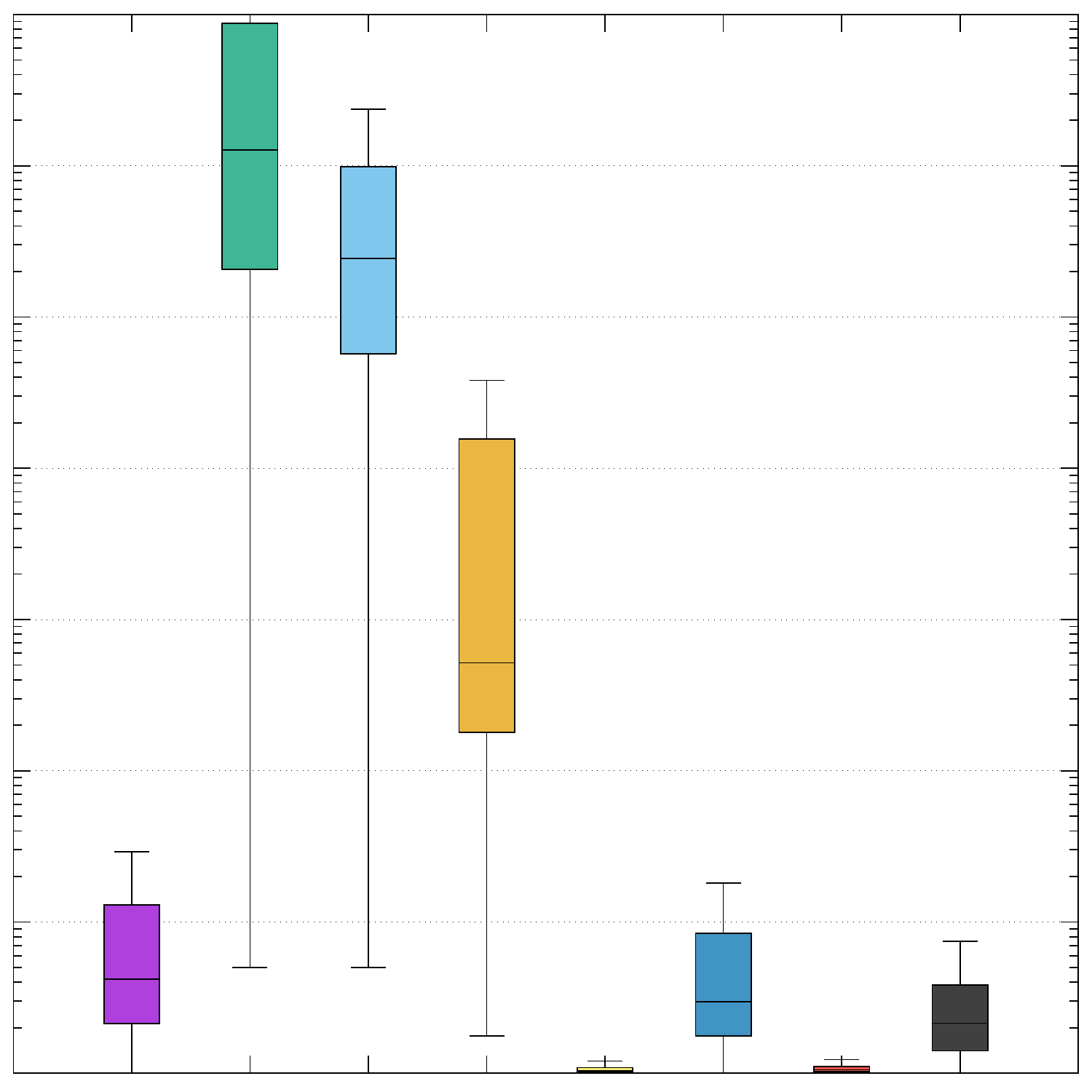}
\label{fig:est_accuracy_3a}
\end{subfigure}
\hfill
\hspace{-0.2em}
\vspace{-0.5em}
\begin{subfigure}{0.32\textwidth}
\vspace{-0.5em}
\caption*{IMDB Q1 Distinct}
\vspace{-0.3em}
\includegraphics[width=\linewidth]{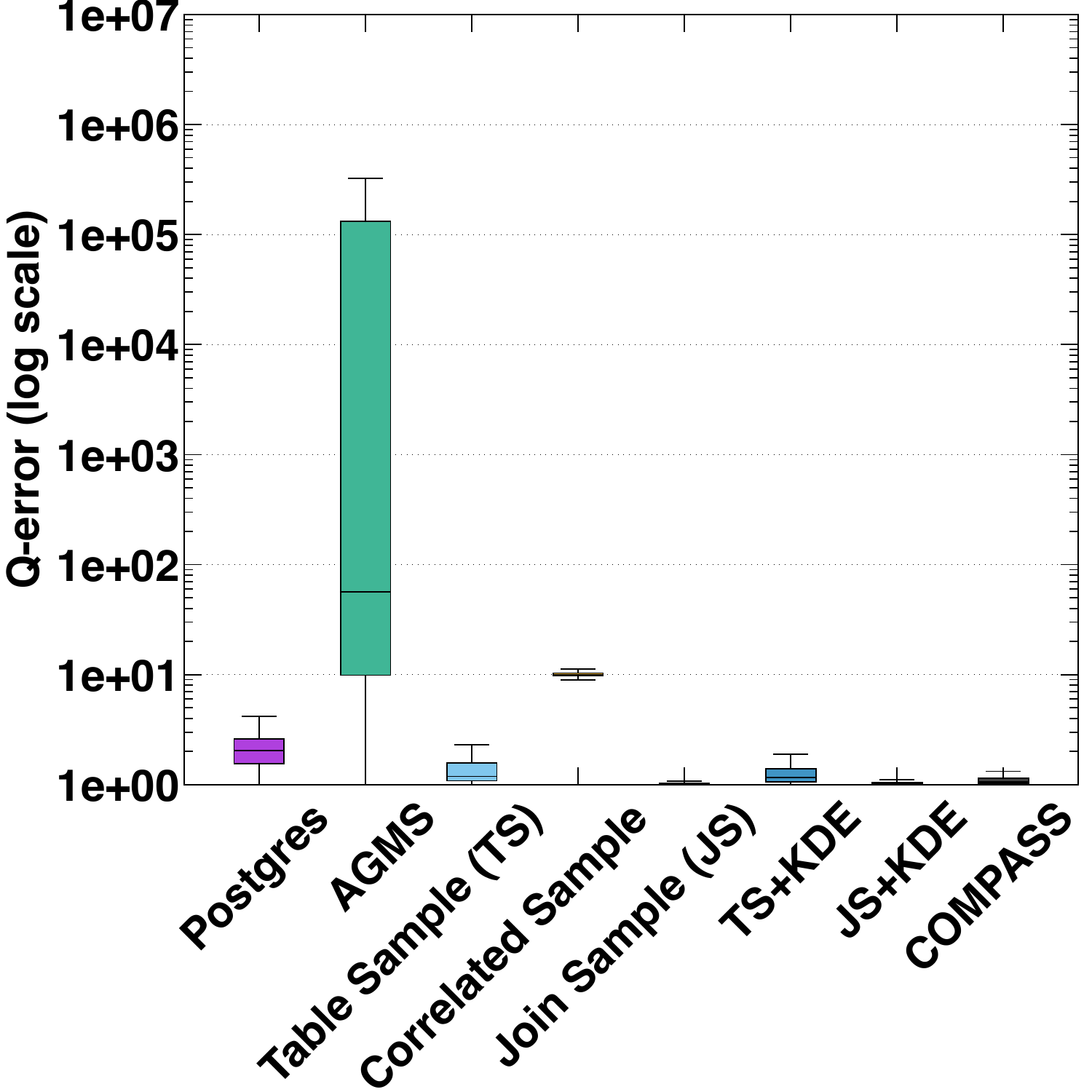}
\label{fig:est_accuracy_1b}
\end{subfigure}
\vspace{-0.5em}
\begin{subfigure}{0.32\textwidth}
\vspace{-0.5em}
\caption*{IMDB Q2 Distinct}
\vspace{-0.4em}
\includegraphics[width=\linewidth]{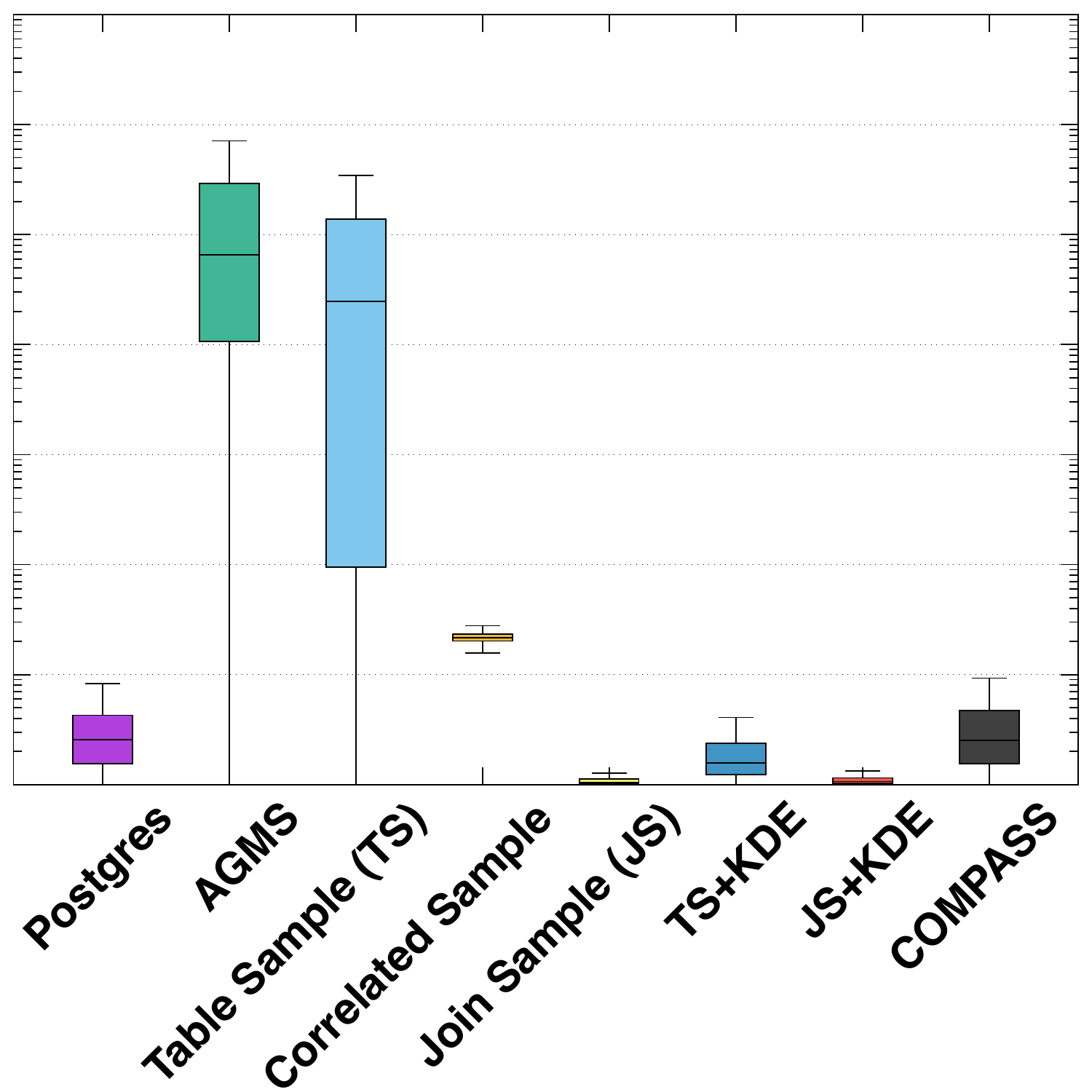}
\label{fig:est_accuracy_2b}
\end{subfigure}
\vspace{-0.5em}
\begin{subfigure}{0.32\textwidth}
\vspace{-0.5em}
\caption*{IMDB Q3 Distinct}
\vspace{-0.4em}
\includegraphics[width=\linewidth]{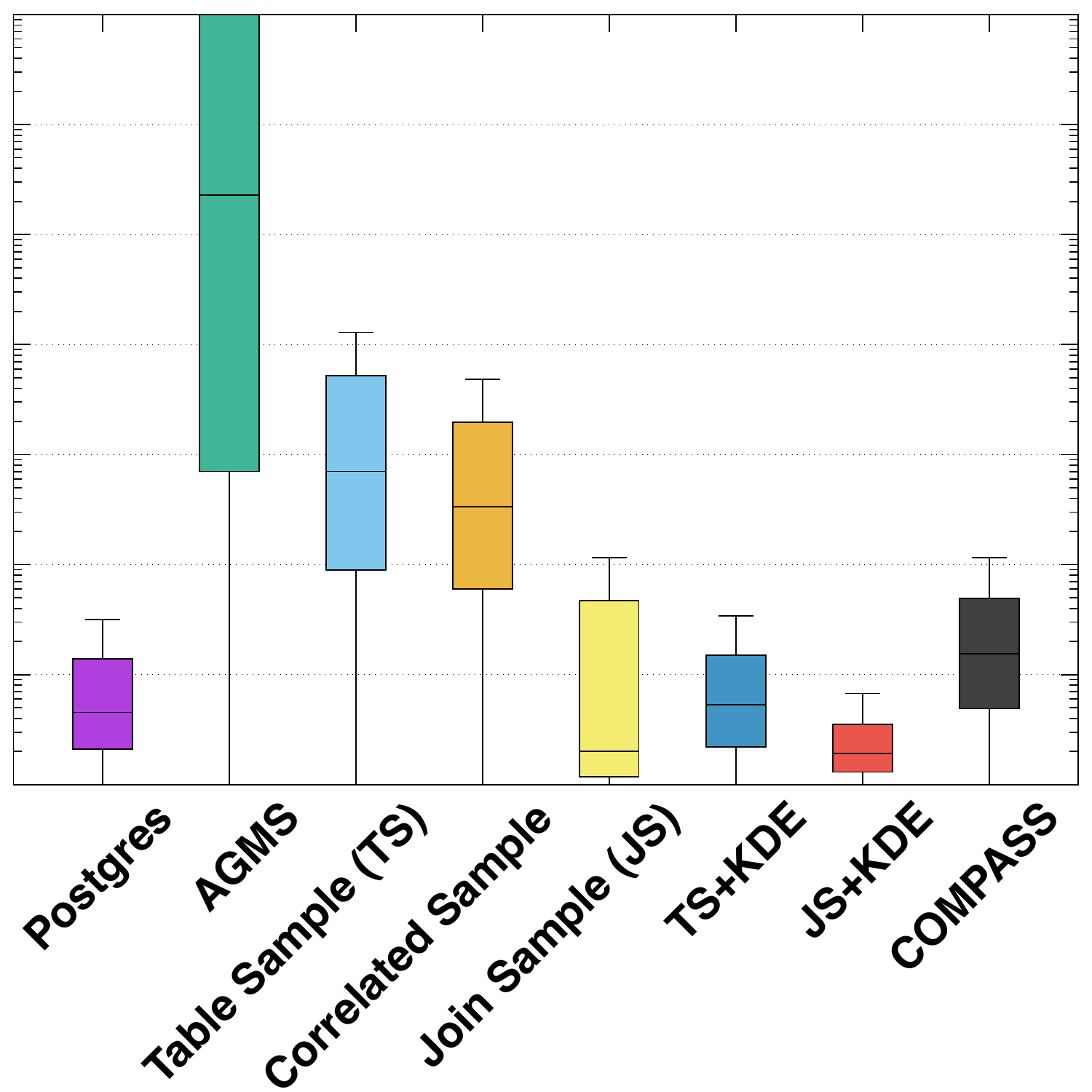}
\label{fig:est_accuracy_3b}
\end{subfigure}
\vspace{-0.5em}
\caption{Comparison with state-of-the-art techniques for cardinality estimation.}
\vspace{-0.5em}
\label{fig:est_accuracy}
\end{figure*}

We compare COMPASS against seven methods for join cardinality estimation following the study presented in~\cite{Kiefer:KDE:pvldb-2017}. These methods are PostgreSQL, AGMS sketches~\cite{Alon:AMS:stoc-1996,Alon:AGMS:pods-1999}, random table samples (TS), correlated samples~\cite{Vengerov:JSEFC:pvldb-2015}, join samples (JS), KDE with table samples (TS+KDE), and KDE with join samples (JS+KDE)~\cite{Kiefer:KDE:pvldb-2017}. We perform all the experiments for join cardinality estimation on three types of JOB queries over at most five tables -- the simplest in the benchmark -- as presented in~\cite{Kiefer:KDE:pvldb-2017}. We use the publicly available code, workload, and data from ~\cite{kiefer:kde-github-repo:2017}. The results are depicted in Figure~\ref{fig:est_accuracy}. We observe that the COMPASS accuracy matches that of the best estimators closely for all the queries. The only two estimators that always outperform COMPASS are based on join samples (Join Sample and JS+KDE). This type of estimators require indexes on all possible join attribute combinations, thus, they have a high set-up and maintenance cost. Moreover, the KDE models are trained on query samples with the same set of selection and join predicates, i.e., same type of training queries with different constant values. In addition, JS+KDE requires training for every join size. Thus, it is not clear what is the behavior of the KDE estimators on different types of queries---the sub-queries enumerated by the optimizer, in particular. Notice that the results also include sketches (AGMS). However, these are the AGMS sketches, not the Fast-AGMS sketches on which COMPASS is built upon. Given the detailed sketch comparison in~\cite{Rusu:SAS:sigmod-2007,Rusu:SJSE:tods-2008}, the difference between the two is expected. Since we perform a thorough query plan evaluation with PostgreSQL (Postgres), the comparison in terms of accuracy is interesting. While COMPASS and Postgres have very similar accuracy -- with an advantage for COMPASS -- our holistic results prove that COMPASS generates better query plans both in terms of quality and execution time. This proves that other factors beyond single query accuracy -- such as sub-plan enumeration and estimator composition -- have to be considered by the optimizer. Overall, the comparison in terms of accuracy is limited to a series of relatively simple hand-picked queries with at most four joins---nothing close to the full JOB benchmark. While useful, it fails to confirm the practicality of the considered approaches in a complete query optimizer, which COMPASS does.

%%%%%%%%%%%%%%%%%%%%%%%%%%%%%%%%%%%%%%%%%%%%%%%%%%%%%%%%%%%%%%%%%
%%%%%%%%%%%%%%%%%%%%%%%%%%%%%%%%%%%%%%%%%%%%%%%%%%%%%%%%%%%%%%%%%

%%%%%%%%%%%%%%%%%%%%%%%%%%%%%%%%%%%%
\subsubsection{Plan Enumeration Analysis}

We assess the performance of the join order enumeration algorithm by varying the search space. We set four different dimensions in the left-deep search space--\textit{greedy}, \textit{full-greedy}, \textit{limit-10}, and \textit{exhaustive}. The \textit{greedy} solution traverses the join graph from a single source and greedily adds nodes relying on the sketch estimations. The source node is chosen as the smallest table from the two-way join that has the smallest cardinality estimation. \textit{full-greedy} executes the \textit{greedy} traversal from every node in the join graph. The join order with the smallest overall estimation is selected. \textit{limit-10} enhances \textit{full-greedy} with backtracking. Rather than stopping after the first greedy plan is generated, \textit{limit-10} traverses additional paths in the join graph until the first 10 join orders are generated, i.e., $max\_plans$ is set to $10$ in Algorithm~\ref{alg:dfs}. Lastly, \textit{exhaustive} traverses the entire search space to find the optimal plan, i.e., $max\_plans$ is set to $\infty$ in Algorithm~\ref{alg:dfs}. Sketch merging and estimation are arithmetic operations amenable to parallelization. Thus, the traversal algorithm quickly explores even the large spaces enumerated by \textit{exhaustive}. To further speed up the enumeration phase, two practical heuristics are applied in Algorithm~\ref{alg:dfs} (see Section~\ref{sec:enumeration}).

\begin{figure*}[htbp]
\centering
\includegraphics[width=.9\textwidth]{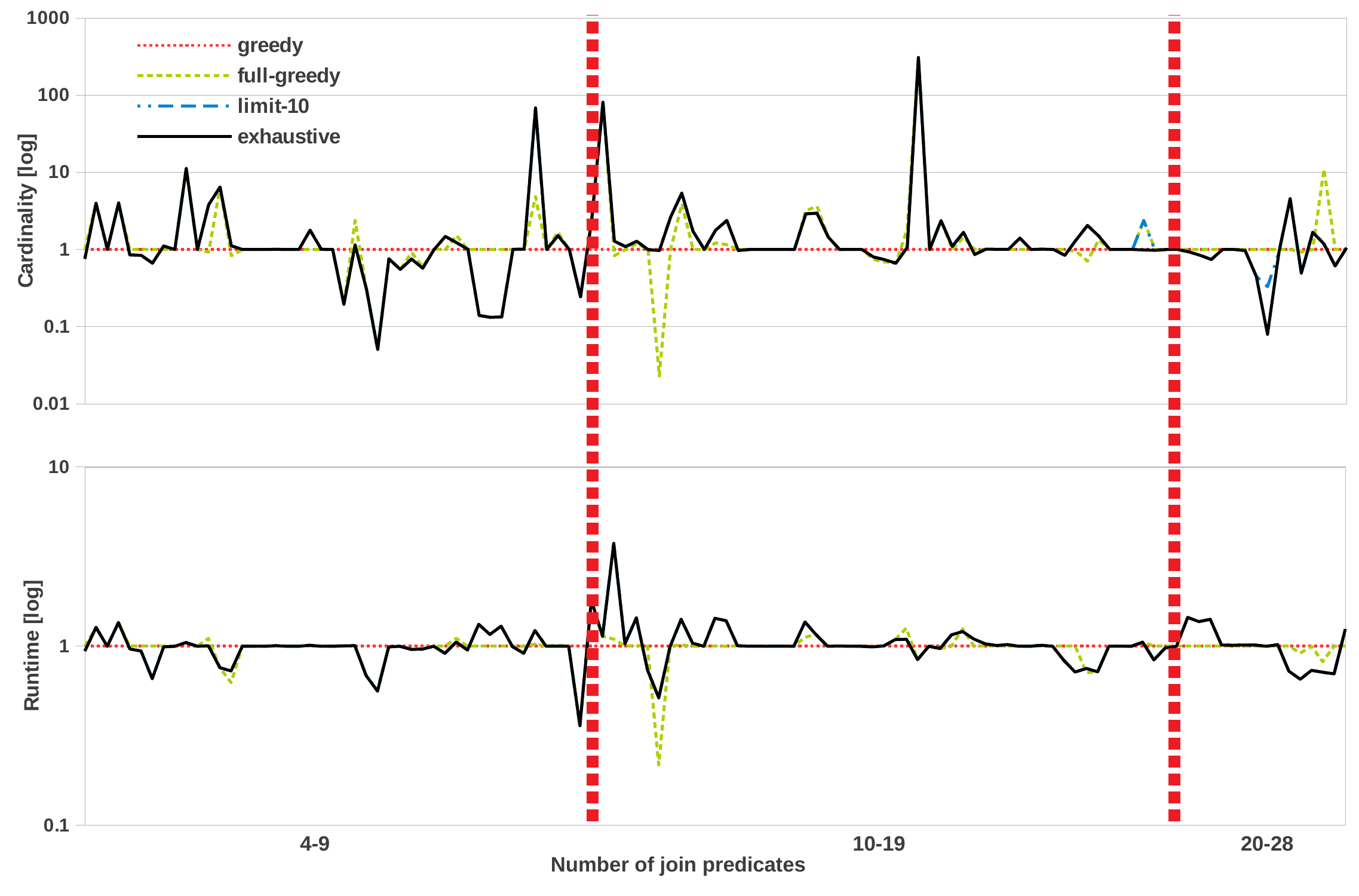}
\caption{The effect of plan enumeration algorithms on the cardinality and runtime of the selected plan. Values are normalized with respect to the cardinality and runtime of the plan determined by the greedy algorithm. Values above the ``1'' horizontal line are worse while values below are better.}
\label{fig:enum_compare}
\end{figure*}

We collect the join order corresponding to every approach for all the JOB queries and compare their intermediate cardinality and runtime. The results are depicted in Figure~\ref{fig:enum_compare}. Intuitively, we expect that the larger the search space is, the higher the odds to find a better join order plan. However, as discussed in~\cite{Leis:JOB:vldb-2018,Leis:QOREALLY:pvldb-2015}, the plan enumeration algorithm has a relatively small impact on plan quality. Our results confirm this hypothesis. We observe that no particular approach is significantly better for the proposed sketch-driven traversal algorithm. In larger search spaces, sub-optimal plans are pruned by the early stopping criteria since the overall cardinality estimates exceed the current minimum cost. The plan search space is not only limited by the join graph. The search space may shrink because of the early stopping criteria that compare the overall cardinality estimates for the sub-queries. For example, it is possible the execution plans for \textit{greedy} and \textit{full-greedy} are the same, although \textit{full-greedy} searches from each node separately. The reason is that all the other execution plans selected by \textit{full-greedy} may have larger overall cost. In fact, even the sub-queries may have already larger cost than the plan selected by \textit{greedy} solution---thus the enumeration algorithm backtracks. As a result, there is no particular preference for the search space size since the cardinality differences are not significant in Figure~\ref{fig:enum_compare} (upper part of the figure). Although there are some outliers -- spikes in the figure -- they are mostly selected by \textit{full-greedy}, \textit{limit-10}, and \textit{exhaustive}. This behavior is caused by the over- and under-estimates. Also, there is no particular trend when comparing the runtime corresponding to the different plans---except for a few outliers in \textit{exhaustive}. Moreover, there is no correlation between cardinality and runtime results for the outliers. We believe the plan differences chosen by these four different solutions are not significantly different in order to notice changes in terms of the runtime. The results confirm that the plan chosen by \textit{greedy} is not necessarily slow and, thus, we conclude that COMPASSS is not heavily impacted by the join enumeration algorithm. In fact, COMPASS outperforms the other systems, especially for complex queries with a large number of joins.

%%%%%%%%%%%%%%%%%%%%%%%%%%%%%%%%%%%%%%%%%%%%%%%%%%%%%%%%%%%%%%%%%
%%%%%%%%%%%%%%%%%%%%%%%%%%%%%%%%%%%%%%%%%%%%%%%%%%%%%%%%%%%%%%%%%

%%%%%%%%%%%%%%%%%%%%%%%%%%%%%%%%%%%%
\subsection{Summary}

Based on the presented results, we can answer the questions raised at the beginning of the experimental section:

\begin{itemize}[leftmargin=*,noitemsep,nolistsep]
\item COMPASS generates query plans with the lowest cardinality among all the considered systems for 56\% of the queries in the workload. This percentage increases to 65\% for complicated queries with 10 or more joins.

\item The better plans identified by COMPASS translate into faster query runtimes in PostgreSQL, DBMS A, and MapD. Out of the 113 JOB queries, COMPASS achieves the fastest runtime for more than 80 in PostgreSQL and DBMS A, and 74 in MapD. This confirms the correlation between cardinality and runtime. DBMS A is the only database that does not satisfy this correlation, which can be problematic for a user.

\item COMPASS and MonetDB are the only databases that perform all the JOB queries without serious hiccups both in PostgreSQL and DBMS A. The other systems have several queries for which the runtime ``explodes''. This results in significantly higher workload runtime. On the PostgreSQL engine, COMPASS outperforms MonetDB by a factor of 2.19X, while on DBMS A by 1.65X. DBMS A optimizes queries specifically for its engine, resulting in a significant reduction in runtime compared to PostgreSQL. However, COMPASS is faster by a factor of 1.35X. Moreover, COMPASS is at least 7.67X faster than MapD.

\item Even though COMPASS treats the complete join graph in ordering -- which results in a much larger search space -- it manages to outperform the pessimistic optimizers---which perform join ordering in a tiered approach. While only 3\% reduction in the workload runtime, this improvement is significant because COMPASS achieves a faster runtime and higher speedup for more queries.

\item The overhead incurred by the COMPASS optimizer in MapD is less than 500 milliseconds on GPU and less than 5 seconds on CPU. While this may be too large for simple queries, it results in faster execution for more than 91\% of the queries. We plan to optimize our implementation in the future.

\item The accuracy achieved by COMPASS matches -- and often surpasses -- that of most of the state-of-the-art methods for join cardinality estimation. The only methods that have better accuracy are based on join samples, which have never been fully-integrated in a query optimizer because of their complexity. If we consider only methods implemented in existent query optimizers, COMPASS has better overall accuracy.

\item The COMPASS plan enumeration algorithm can be customized to perform the search for the optimal join order from a limited greedy to an exhaustive left-deep tree. However, the difference between these alternatives is not significant enough to compensate for their gap in overhead. Thus, the fast greedy join order enumeration is sufficient to achieve the optimal COMPASS plans.
\end{itemize}

%%%%%%%%%%%%%%%%%%%%%%%%%%%%%%%%%%%%%%%%%%%%%%%%%%%%%%%%%%%%%%%%%
%\input{related_work}
\section{RELATED WORK}\label{sec:rel-work}

%%%%%%%%%%%%%%%%%%%%%%%%%%%%%%%%%%%%%%%%%%%%%%%%%%%%%%%%%%%%%%%%%
%\textbf{Cardinality estimation.}
\paragraph*{Cardinality estimation.}
While exhaustive surveys on query optimization~\cite{Chaudhuri:OQO:pods-1998,Lohman:QOSP:2014} argue that each component is important in finding the optimal plan, Leis et al.~\cite{Leis:QOREALLY:pvldb-2015,Leis:JOB:vldb-2018} show experimentally that cardinality estimation is the most dominant component in query optimization. However, consistency in estimations is more important than high accuracy only for a limited number of instances. There are four mainstream cardinality estimation approaches in the literature---histograms, sampling techniques, sketches, and, more recently, machine learning models. While histograms can provide accurate selectivity estimation for a single attribute in a relation~\cite{Ioannidis:PESJR:sigmodrec-1991}, it is difficult for them to capture correlations between cross-join attributes~\cite{Poosala:SEW:vldb-1997}, thus reducing their applicability to joins. Unlike histograms, sampling techniques~\cite{Leis:index-join-sample:cidr-2017,Muller:ISECKSS:pvldb-2018} can detect arbitrary correlations for common values. However, samples are sensitive to skewed and sparse data when few tuples are selected by a query~\cite{Wu:SCEA:2012}. As the query optimizer estimates a large number of joins, the cardinality drops quickly, causing wrong estimates for intermediate results. Estimating the cardinality of multi-way joins with AGMS sketches is introduced in~\cite{Dobra:PCAQDS:sigmod-2002,Dobra:SBMQPDS:edbt-2004}, while a statistical analysis of two-way join sketch-based techniques is performed by Rusu and Dobra~\cite{Rusu:SAS:sigmod-2007,Rusu:SJSE:tods-2008}. Their results show that Fast-AGMS sketches are clearly superior to other sketches. In this work, we extend Fast-AGMS sketches to capture all the join attributes involved in a given query within a single sketch and efficiently estimate multi-way joins. Vengerov et al.~\cite{Vengerov:JSEFC:pvldb-2015} present an extension to AGMS sketches that captures selection predicates, while Cai et al.~\cite{Cai:PCETUB:sigmod-2019} introduce bound sketches that provide theoretical upper bounds for cardinality estimation. The problem with these approaches is that the online sketch building process is not scalable. Hertzschuch et al.~\cite{Hertzschuch:simplicity:cidr-2021} maintain the pessimistic property for cardinality estimation, while replacing sketches with a simple formula based on statistics already available to the PostgreSQL query optimizer. This eliminates the sketch overhead, while preserving the quality of the pessimistic plans---as long as the optimizer statistics estimate predicate selectivity accurately. Kernel density models for cardinality estimation (KDE) are introduced in~\cite{Heimel:KDE:sigmod-2015,Kiefer:KDE:pvldb-2017}. They are built on samples extracted either from the base tables or the join. While their accuracy is shown to be superior to any other method on JOB queries over at most five tables -- the simplest in the benchmark -- it is not clear how to generalize and fully integrate KDE models in plan enumeration. Specifically, the KDE implementation~\cite{kiefer:kde-github-repo:2017} builds a separate estimator for every query. No details are provided on how to apply the estimator to query sub-plans derived from the main query, which is the centerpiece of plan enumeration.

%%%%%%%%%%%%%%%%%%%%%%%%%%%%%%%%%%%%%%%%%%%%%%%%%%%%%%%%%%%%%%%%%
%\textbf{Query reoptimization.}
\paragraph*{Query reoptimization.}
In order to overcome the inherent mis-estimations in the query optimizer, Adaptive Query Processing~\cite{Deshpande:AQP:fnt-2007} allows the query processor to modify the optimal query plan computed by the optimizer in case of large deviations from the true cardinality values detected at runtime. The Mid-Query Re-Optimizer~\cite{Kabra:MQO:sigmod-1998}, ROX~\cite{Kader:ROX:sigmod-2009}, and SkinnerDB~\cite{Trummer:SKINNERDB:sigmod-2019} re-run the query optimizer at runtime in the case of large differences between estimations and the true cardinalities. Wu et al.~\cite{Wu:SampleReOptS:sigmod-2016} apply online sampling to correct the errors in the plans generated by the query optimizer. These approaches use the output of the query executor and sampling techniques to re-estimate the cardinalities based on already computed intermediate join outputs and change the query plan whenever the estimated values deviate significantly. In the self-adaptable LEO optimizer~\cite{Markl:LEO:ibmsys-2003}, the query engine monitors and uses the feedback from the execution engine in order to adjust the histogram-based synopses for better performance in subsequent queries. Eddies~\cite{Avnur:Eddies:sigmod-2000} process batches of tuples by following dynamic routing policies during query execution. Unlike these systems, COMPASS performs query optimization as a single stage, while query execution is partitioned into two phases---before and after the optimization. As in query reoptimization, COMPASS uses the intermediates -- sketches -- produced by the first phase of execution. However, this process is performed only once, thus its overhead is smaller compared to continuous reoptimization.

%%%%%%%%%%%%%%%%%%%%%%%%%%%%%%%%%%%%%%%%%%%%%%%%%%%%%%%%%%%%%%%%%
%\textbf{Machine learning for query optimization.}
\paragraph*{Machine learning for query optimization.}
Using machine learning techniques and deep neural networks is a recent trend in query optimization. Join order enumeration~\cite{Marcus:DRLJOE:aiDM-2018,Krishnan:LOJQDRL:arxiv-2018,Marcus:NEO:arxiv-2019}, cardinality estimation~\cite{Malik:BBAQCE:cidr-2007,Liu:CEUNN:cascon-2015,Kipf:LCECJDL:cidr-2019,Kipf:ECDL:arxiv-2019,Woltmann:CEL:aiDM-2019,Ortiz:EADLCE:arxiv-2019}, selectivity estimation~\cite{Yang:SEDLM:arxiv-2019,Hasan:MASEDL:arxiv-2019,Dutt:SER:pvldb-2019}, and index structures~\cite{Kraska:CLI:sigmod-2018} have been active research directions. Regarding the cardinality estimation problem, Malik et al.~\cite{Malik:BBAQCE:cidr-2007} propose to train neural network models based on cardinality distributions for a separate class of similar queries and estimate overall query cardinalities. Yang et al.~ \cite{Yang:SFSKETCH:arxiv-2017} utilize neural networks to learn a function to estimate cardinalities of queries with range selection predicates. Kipf et al.~\cite{Kipf:LCECJDL:cidr-2019} use multi-set convolutional neural networks in order to model join and selection predicates, and capture join correlations. Woltmann et al.~\cite{Woltmann:CEL:aiDM-2019} propose to train neural network models to estimate cardinalities in equi-joins. Marcus et al.~\cite{Marcus:DRLJOE:aiDM-2018} use reinforcement learning in order to efficiently explore the search space and find optimal join order plans. Different from these approaches, COMPASS uses traditional randomized algorithms to estimate cardinality. Moreover, COMPASS is fully-integrated into a database engine, which is often not the case for these machine learning solutions.

%%%%%%%%%%%%%%%%%%%%%%%%%%%%%%%%%%%%%%%%%%%%%%%%%%%%%%%%%%%%%%%%%
\paragraph*{Plan enumeration.}
In plan enumeration, multiple semantically equivalent plans are explored in order to identify the optimal execution plan. Different exhaustive~\cite{Vance:RBJ:sigmodrec-1996,Moerkotte:ATE:vldb-2006} and heuristic-based~\cite{Steinbrunn:HRO:jvldb-1997} algorithms have been proposed. They consider different tree shapes -- such as left-deep and bushy trees -- in the search space. Leis et al.~\cite{Leis:QOREALLY:pvldb-2015,Leis:JOB:vldb-2018} evaluate the influence of several plan enumeration algorithms and the impact of considering bushy trees. Several recent approaches have been proposed to optimize the plan enumeration phase by using GPUs~\cite{Meister-GPUJOO:phdvldb-2015,Meister:GPUDP:gvd-2016-1} and deep reinforcement learning models~\cite{Marcus:DRLJOE:aiDM-2018,Marcus:THFQODL:arxiv-2018}. In this work, we propose an adaptive graph traversal algorithm that efficiently explores the search space. This algorithm can be configured to cover plan enumeration from greedy to exhaustive search.

%%%%%%%%%%%%%%%%%%%%%%%%%%%%%%%%%%%%%%%%%%%%%%%%%%%%%%%%%%%%%%%%%
%\input{conclusion}
\section{CONCLUSIONS}\label{sec:conclusions}

We introduce the online sketch-based COMPASS query optimizer, which uses exclusively Fast-AGMS sketches for cardinality estimation and plan enumeration. Fast-AGMS sketches are computed online by leveraging the optimized parallel execution engine in modern databases. Selection predicates and sketch updates are pushed-down and evaluated online during query optimization. Plan enumeration is performed over the query join graph by incrementally composing the corresponding sketches. We prototype COMPASS in MapD and perform extensive experiments over the complete JOB benchmark. The results prove the reduced overhead COMPASS incurs, while generating better execution plans than four other database systems---COMPASS outperforms four other databases on all the considered metrics over the JOB benchmark. In future work, we plan to investigate alternative merging strategies for Fast-AGMS sketches in order to support multi-way joins. SIMD-optimized sketch algorithms -- for CPU and GPU -- with lower overhead and alternative plan enumeration strategies are other directions we plan to pursue.

%%%%%%%%%%%%%%%%%%%%%%%%%%%%%%%%%%%%%%%%%%%%%%%%%%%%%%%
\paragraph*{Acknowledgments.}
This work is supported by NSF award number 2008815 and by a U.S. Department of Energy Early Career Award (DOE Career). The authors want to thank Alex Suhan -- one of the MapD architects -- for explaining the internals of the MapD execution engine, as well as Hung Ngo and Mahmoud Abo Khamis from relationalAI for the discussions on query optimization. Lastly, the authors acknowledge the insightful comments made by the SIGMOD 2021 anonymous reviewers that helped improve the quality of the paper.

%%%%%%%%%%%%%%%%%%%%%%%%%%%%%%%%%%%%%%%%%%%%%%%%%%%%%%%%%%%%%%%%%
\bibliographystyle{abbrv}
%\bibliography{sketch_based_join_estimation}

\begin{thebibliography}{10}

\bibitem{Alon:AGMS:pods-1999}
N.~Alon, P.~B. Gibbons, Y.~Matias, and M.~Szegedy.
\newblock {Tracking Join and Self-Join Sizes in Limited Storage}.
\newblock In {\em PODS 1999}, pages 10--20.

\bibitem{Alon:AMS:stoc-1996}
N.~Alon, Y.~Matias, and M.~Szegedy.
\newblock {The Space Complexity of Approximating the Frequency Moments}.
\newblock In {\em STOC 1996}, pages 20--29.

\bibitem{Avnur:Eddies:sigmod-2000}
R.~Avnur and J.~M. Hellerstein.
\newblock {Eddies: Continuously Adaptive Query Processing}.
\newblock In {\em SIGMOD 2000}, pages 261--272.

\bibitem{Bress:GPUADS:tlkds-2014}
S.~Bre{\ss}, M.~Heimel, N.~Siegmund, L.~Bellatreche, and G.~Saake.
\newblock {GPU-accelerated Database Systems: Survey and Open Challenges}.
\newblock {\em TLKDS}, pages 1--35, 2014.

\bibitem{Cai:PCETUB:sigmod-2019}
W.~Cai, M.~Balazinska, and D.~Suciu.
\newblock {Pessimistic Cardinality Estimation: Tighter Upper Bounds for
  Intermediate Join Cardinalities}.
\newblock In {\em SIGMOD 2019}, pages 18--35.

\bibitem{Chaudhuri:OQO:pods-1998}
S.~Chaudhuri.
\newblock {An Overview of Query Optimization in Relational Systems}.
\newblock In {\em PODS 1998}, pages 34--43.

\bibitem{Cormode:FAGMS:vldb-2005}
G.~Cormode and M.~Garofalakis.
\newblock {Sketching Streams Through the Net: Distributed Approximate Query
  Tracking}.
\newblock In {\em VLDB 2005}, pages 13--24.

\bibitem{Cormode:SMD:fnt-2012}
G.~Cormode, M.~Garofalakis, P.~J. Haas, and C.~Jermaine.
\newblock {Synopses for Massive Data: Samples, Histograms, Wavelets, Sketches}.
\newblock {\em Foundation and Trends in Databases}, 4:1--294, 2012.

\bibitem{Deshpande:AQP:fnt-2007}
A.~Deshpande, Z.~Ives, and V.~Raman.
\newblock {Adaptive Query Processing}.
\newblock {\em Foundations and Trends in Databases}, 1(1):1--140, 2007.

\bibitem{Dobra:PCAQDS:sigmod-2002}
A.~Dobra, M.~Garofalakis, J.~Gehrke, and R.~Rastogi.
\newblock {Processing Complex Aggregate Queries over Data Streams}.
\newblock In {\em SIGMOD 2002}, pages 61--72.

\bibitem{Dobra:SBMQPDS:edbt-2004}
A.~Dobra, M.~Garofalakis, J.~Gehrke, and R.~Rastogi.
\newblock {Sketch-Based Multi-query Processing over Data Streams}.
\newblock In {\em EDBT 2004}, pages 551--568.

\bibitem{Dutt:SER:pvldb-2019}
A.~Dutt, C.~Wang, A.~Nazi, S.~Kandula, V.~Narasayya, and S.~Chaudhuri.
\newblock {Selectivity Estimation for Range Predicates Using Lightweight
  Models}.
\newblock {\em PVLDB}, 12(9):1044--1057, 2019.

\bibitem{Funke:PQP:sigmod-2018}
H.~Funke, S.~Bre\ss, S.~Noll, V.~Markl, and J.~Teubner.
\newblock {Pipelined Query Processing in Coprocessor Environments}.
\newblock In {\em SIGMOD 2018}, pages 1603--1618.

\bibitem{Ullman:db-book}
H.~Garcia-Molina, J.~D. Ullman, and J.~Widom.
\newblock {\em {Database Systems: The Complete Book}}.
\newblock Prentice Hall, 2008.

\bibitem{Hasan:MASEDL:arxiv-2019}
S.~{Hasan}, S.~{Thirumuruganathan}, J.~{Augustine}, N.~{Koudas}, and G.~{Das}.
\newblock {Multi-Attribute Selectivity Estimation Using Deep Learning}.
\newblock {\em CoRR}, arXiv:1903.09999v2, 2019.

\bibitem{He:RQC:tods-2009}
B.~He, M.~Lu, K.~Yang, R.~Fang, N.~K. Govindaraju, Q.~Luo, and P.~V. Sander.
\newblock {Relational Query Coprocessing on Graphics Processors}.
\newblock {\em TODS}, 34(4):1--39, 2009.

\bibitem{Heimel:KDE:sigmod-2015}
M.~Heimel, M.~Kiefer, and V.~Markl.
\newblock {Self-Tuning, GPU-Accelerated Kernel Density Models for
  Multidimensional Selectivity Estimation}.
\newblock In {\em SIGMOD 2015}, pages 1477--1492.

\bibitem{Hertzschuch:simplicity:cidr-2021}
A.~Hertzschuch, C.~Hartmann, D.~Habich, and W.~Lehner.
\newblock {Simplicity Done Right for Join Ordering}.
\newblock In {\em CIDR 2021}.

\bibitem{Stratos:MONETDB:deb-2012}
S.~Idreos, F.~Groffen, N.~Nes, S.~Manegold, S.~Mullender, and M.~Kersten.
\newblock {MonetDB: Two Decades of Research in Column-oriented Database
  Architectures}.
\newblock {\em IEEE Data Engineering Bulletin}, 35(1):40--45, 2012.

\bibitem{Ioannidis:PESJR:sigmodrec-1991}
Y.~E. Ioannidis and S.~Christodoulakis.
\newblock {On the Propagation of Errors in the Size of Join Results}.
\newblock {\em SIGMOD Record}, 20(2):268--277, 1991.

\bibitem{Kabra:MQO:sigmod-1998}
N.~Kabra and D.~J. DeWitt.
\newblock {Efficient Mid-query Re-optimization of Sub-optimal Query Execution
  Plans}.
\newblock In {\em SIGMOD 1998}, pages 106--117.

\bibitem{Kader:ROX:sigmod-2009}
A.~R. Kader, P.~Boncz, S.~Manegold, and M.~van Keulen.
\newblock {ROX: Run-time Optimization of XQueries}.
\newblock In {\em SIGMOD 2009}, pages 615--626.

\bibitem{Kiefer:KDE:pvldb-2017}
M.~Kiefer, M.~Heimel, S.~Bre{\ss}, and V.~Markl.
\newblock {Estimating Join Selectivities using Bandwidth-Optimized Kernel
  Density Models}.
\newblock {\em PVLDB}, 10(13):2085--2096, 2017.

\bibitem{Kipf:LCECJDL:cidr-2019}
A.~Kipf, T.~Kipf, B.~Radke, V.~Leis, P.~Boncz, and A.~Kemper.
\newblock {Learned Cardinalities: Estimating Correlated Joins with Deep
  Learning}.
\newblock In {\em CIDR 2019}.

\bibitem{Kipf:ECDL:arxiv-2019}
A.~Kipf, D.~Vorona, J.~Muller, T.~Kipf, B.~Radke, V.~Leis, P.~Boncz,
  T.~Neumann, and A.~Kemper.
\newblock {Estimating Cardinalities with Deep Sketches}.
\newblock {\em CoRR}, arXiv:1904.08223v1, 2019.

\bibitem{Kraska:CLI:sigmod-2018}
T.~Kraska, A.~Beutel, E.~H. Chi, J.~Dean, and N.~Polyzotis.
\newblock {The Case for Learned Index Structures}.
\newblock In {\em SIGMOD 2018}, pages 489--504.

\bibitem{Krishnan:LOJQDRL:arxiv-2018}
S.~Krishnan, Z.~Yang, K.~Goldberg, J.~Hellerstein, and I.~Stoica.
\newblock {Learning to Optimize Join Queries With Deep Reinforcement Learning}.
\newblock {\em CoRR}, arXiv:1808.03196v2, 2018.

\bibitem{Leis:QOREALLY:pvldb-2015}
V.~Leis, A.~Gubichev, A.~Mirchev, P.~Boncz, A.~Kemper, and T.~Neumann.
\newblock {How Good Are Query Optimizers, Really?}
\newblock {\em PVLDB}, 9(3):204--215, 2015.

\bibitem{Leis:index-join-sample:cidr-2017}
V.~Leis, B.~Radke, A.~Gubichev, A.~Kemper, and T.~Neumann.
\newblock {Cardinality Estimation Done Right: Index-Based Join Sampling}.
\newblock In {\em CIDR 2017}.

\bibitem{Leis:JOB:vldb-2018}
V.~Leis, B.~Radke, A.~Gubichev, A.~Mirchev, P.~Boncz, A.~Kemper, and
  T.~Neumann.
\newblock {Query Optimization Through the Looking Glass, and What We Found
  Running the Join Order Benchmark}.
\newblock {\em VLDB Journal}, 27:643--668, 2018.

\bibitem{Liu:CEUNN:cascon-2015}
H.~Liu, M.~Xu, Z.~Yu, V.~Corvinelli, and C.~Zuzarte.
\newblock {Cardinality Estimation Using Neural Networks}.
\newblock In {\em CASCON 2015}, pages 53--59.

\bibitem{Malik:BBAQCE:cidr-2007}
T.~Malik, R.~C. Burns, and N.~V. Chawla.
\newblock {A Black-Box Approach to Query Cardinality Estimation}.
\newblock In {\em CIDR 2007}.

\bibitem{Marcus:NEO:arxiv-2019}
R.~Marcus, P.~Negi, H.~Mao, C.~Zhang, M.~Alizadeh, T.~Kraska, O.~Papaemmanouil,
  and N.~Tatbul.
\newblock {Neo: A Learned Query Optimizer}.
\newblock {\em VLDB Journal}, 12(11), 2019.

\bibitem{Marcus:DRLJOE:aiDM-2018}
R.~Marcus and O.~Papaemmanouil.
\newblock {Deep Reinforcement Learning for Join Order Enumeration}.
\newblock In {\em aiDM 2018}.

\bibitem{Marcus:THFQODL:arxiv-2018}
R.~Marcus and O.~Papaemmanouil.
\newblock {Towards a Hands-Free Query Optimizer through Deep Learning}.
\newblock {\em CoRR}, arXiv:1809.10212v2, 2018.

\bibitem{Markl:LEO:ibmsys-2003}
V.~Markl, G.~M. Lohman, and V.~Raman.
\newblock {LEO: An Autonomic Query Optimizer for DB2}.
\newblock {\em IBM Systems Journal}, 42(1):98--106, 2003.

\bibitem{Meister-GPUJOO:phdvldb-2015}
A.~Meister.
\newblock {GPU-Accelerated Join-Order Optimization}.
\newblock In {\em PhD Workshop @ VLDB 2015}.

\bibitem{Meister:GPUDP:gvd-2016-1}
A.~Meister and G.~Saake.
\newblock {Challenges for a GPU-Accelerated Dynamic Programming Approach for
  Join-Order Optimization}.
\newblock In {\em GvD 2016}.

\bibitem{Moerkotte:ATE:vldb-2006}
G.~Moerkotte and T.~Neumann.
\newblock {Analysis of Two Existing and One New Dynamic Programming Algorithm
  for the Generation of Optimal Bushy Join Trees Without Cross Products}.
\newblock In {\em VLDB 2006}, pages 930--941.

\bibitem{Muller:ISECKSS:pvldb-2018}
M.~Muller, G.~Moerkotte, and O.~Kolb.
\newblock {Improved Selectivity Estimation by Combining Knowledge from Sampling
  and Synopses}.
\newblock {\em PVLDB}, 9(11):1016--1028, 2018.

\bibitem{Ngo:WORST:pods-2012}
H.~Q. Ngo, E.~Porat, C.~R\'{e}, and A.~Rudra.
\newblock {Worst-Case Optimal Join Algorithms}.
\newblock In {\em PODS 2012}, pages 37--48.

\bibitem{Ortiz:EADLCE:arxiv-2019}
J.~{Ortiz}, M.~{Balazinska}, J.~{Gehrke}, and S.~{Sathiya Keerthi}.
\newblock {An Empirical Analysis of Deep Learning for Cardinality Estimation}.
\newblock {\em CoRR}, arXiv:1905.06425v2, 2019.

\bibitem{Poosala:SEW:vldb-1997}
V.~Poosala and Y.~E. Ioannidis.
\newblock {Selectivity Estimation Without the Attribute Value Independence
  Assumption}.
\newblock In {\em VLDB 1997}, pages 486--495.

\bibitem{Alonso:AugSketch:sigmod-2016}
P.~Roy, A.~Khan, and G.~Alonso.
\newblock {Augmented Sketch: Faster and More Accurate Stream Processing}.
\newblock In {\em SIGMOD 2016}, pages 1449--1463.

\bibitem{Rusu:FRSRV:sigmod-2006}
F.~Rusu and A.~Dobra.
\newblock {Fast Range-Summable Random Variables for Efficient Aggregate
  Estimation}.
\newblock In {\em SIGMOD 2006}, pages 193--204.

\bibitem{Rusu:SSD:icde-2009}
F.~Rusu and A.~Dobra.
\newblock {Sketching Sampled Data Streams}.
\newblock In {\em ICDE 2009}, pages 381--392.

\bibitem{Rusu:SAS:sigmod-2007}
F.~Rusu and A.~Dobra.
\newblock {Statistical Analysis of Sketch Estimators}.
\newblock In {\em SIGMOD 2007}, pages 187--198.

\bibitem{Rusu:PRNG:tods-2007}
F.~Rusu and A.~Dobra.
\newblock {Pseudo-Random Number Generation for Sketch-Based Estimations}.
\newblock {\em TODS}, 32(2), 2007.

\bibitem{Rusu:SJSE:tods-2008}
F.~Rusu and A.~Dobra.
\newblock {Sketches for Size of Join Estimation}.
\newblock {\em TODS}, 33(15), 2008.

\bibitem{Selinger:APSRDMS:sigmod-1979}
P.~G. Selinger, M.~M. Astrahan, D.~D. Chamberlain, R.~A. Lorie, and T.~G.
  Price.
\newblock {Access Path Selection in a Relational Database Management System}.
\newblock In {\em SIGMOD 1979}, pages 23--34.

\bibitem{Shin:ESC:arxiv-2019}
J.~H. Shin, F.~Rusu, and A.~Suhan.
\newblock {Exact Selectivity Computation for Modern In-Memory Database Query
  Optimization}.
\newblock {\em CoRR}, arXiv:1901.01488v1, 2019.

\bibitem{Steinbrunn:HRO:jvldb-1997}
M.~Steinbrunn, G.~Moerkotte, and A.~Kemper.
\newblock {Heuristic and Randomized Optimization for the Join Ordering
  Problem}.
\newblock {\em VLDB Journal}, 6(3):191--208, 1997.

\bibitem{Charalampos:DelSketch:eurosys-2020}
C.~Stylianopoulos, I.~Walulya, M.~Almgren, O.~Landsiedel, and
  M.~Papatriantafilou.
\newblock {Delegation Sketch: A Parallel Design with Support for Fast and
  Accurate Concurrent Operations}.
\newblock In {\em EuroSys 2020}.

\bibitem{Trummer:SKINNERDB:sigmod-2019}
I.~Trummer, J.~Wang, D.~Maram, S.~Moseley, S.~Jo, and J.~Antonakakis.
\newblock {SkinnerDB: Regret-Bounded Query Evaluation via Reinforcement
  Learning}.
\newblock In {\em SIGMOD 2019}, pages 1153--1170.

\bibitem{Vance:RBJ:sigmodrec-1996}
B.~Vance and D.~Maier.
\newblock {Rapid Bushy Join-order Optimization with Cartesian Products}.
\newblock {\em SIGMOD Record}, 25(2):35--46, 1996.

\bibitem{Vengerov:JSEFC:pvldb-2015}
D.~Vengerov, A.~C. Menck, M.~Zait, and S.~P. Chakkappen.
\newblock {Join Size Estimation Subject to Filter Condition}.
\newblock {\em PVLDB}, 8(12):1530--1541, 2015.

\bibitem{Woltmann:CEL:aiDM-2019}
L.~Woltmann, C.~Hartmann, M.~Thiele, D.~Habich, and W.~Lehner.
\newblock {Cardinality Estimation with Local Deep Learning Models}.
\newblock In {\em aiDM 2019}, pages 1--8.

\bibitem{Wu:SCEA:2012}
W.~Wu.
\newblock {Sampling-Based Cardinality Estimation Algorithms: A Survey and An
  Empirical Evaluation}, 2012.

\bibitem{Wu:SampleReOptS:sigmod-2016}
W.~Wu, J.~F. Naughton, and H.~Singh.
\newblock {Sampling-Based Query Re-Optimization}.
\newblock In {\em SIGMOD 2016}, pages 1721--1736.

\bibitem{Yang:SFSKETCH:arxiv-2017}
T.~Yang, L.~Liu, Y.~Yan, M.~Shahzad, Y.~Shen, X.~Li, B.~Cui, and G.~Xie.
\newblock {SF-sketch: A Two-stage Sketch for Data Streams}.
\newblock {\em CoRR}, arXiv:1701.04148v3, 2017.

\bibitem{Yang:SEDLM:arxiv-2019}
Z.~{Yang}, E.~{Liang}, A.~{Kamsetty}, C.~{Wu}, Y.~{Duan}, X.~{Chen},
  P.~{Abbeel}, J.~M. {Hellerstein}, S.~{Krishnan}, and I.~{Stoica}.
\newblock {Selectivity Estimation with Deep Likelihood Models}.
\newblock {\em CoRR}, arXiv:1905.04278v2, 2019.

\bibitem{Yu:CS2:sigmod-2013}
F.~Yu, W.~Hou, C.~Luo, D.~Che, and M.~Zhu.
\newblock {CS2: A New Database Synopsis for Query Estimation}.
\newblock In {\em SIGMOD 2013}.

\bibitem{Boncz:imdb-data}
P.~Boncz.
\newblock {The IMDB Dataset}.
\newblock \url{http://homepages.cwi.nl/~boncz/job/imdb.tgz}.

\bibitem{Hertzschuch:simplicity-github-repo:2020}
A.~Hertzschuch.
\newblock {SimplicityDoneRight}.
\newblock \url{https://github.com/axhertz/SimplicityDoneRight}.

\bibitem{Izenov:compass-github:2020}
Y.~Izenov.
\newblock {The COMPASS Query Optimizer}.
\newblock \url{https://github.com/yizenov/compass_query_optimizer}.

\bibitem{kiefer:kde-github-repo:2017}
M.~Kiefer.
\newblock {join-kde}.
\newblock \url{https://github.com/martinkiefer/join-kde}.

\bibitem{Lohman:QOSP:2014}
G.~Lohman.
\newblock {Is Query Optimization a Solved Problem?}
\newblock \url{https://wp.sigmod.org/?p=1075}, 2014.

\bibitem{JOB-github}
G.~Rahn.
\newblock {Join Order Benchmark (JOB)}.
\newblock \url{https://github.com/gregrahn/join-order-benchmark}.

\bibitem{Rusu:sketch-library}
F.~Rusu.
\newblock {Sketches for Size of Join Estimation}.
\newblock \url{https://faculty.ucmerced.edu/frusu/Projects/Sketches}.

\bibitem{ocelot}
M.~Saecker.
\newblock {MonetDB Ocelot}.
\newblock
  \url{https://bitbucket.org/msaecker/monetdb-opencl/src/simple_mem_manager/}.

\bibitem{stack:perm-l1:2017}
StackExchange.
\newblock {Distance Between Two Permutations?}
\newblock
  \url{https://math.stackexchange.com/questions/2492954/distance-between-two-permutations}.

\bibitem{brytlyt}
{Brytlyt}.
\newblock \url{https://www.brytlyt.com/gpu-accelerated-database/#}.

\bibitem{calcite}
{Apache Calcite}.
\newblock \url{https://calcite.apache.org}.

\bibitem{cogadb}
{CoGaDB}.
\newblock \url{http://cogadb.cs.tu-dortmund.de/wordpress/download/}.

\bibitem{kinetica}
{Kinetica}.
\newblock \url{https://www.kinetica.com/products/gpu-accelerated-database/}.

\bibitem{mapd}
{MapD}.
\newblock \url{www.omnisci.com}.

\bibitem{monetdb-web}
{MonetDB}.
\newblock \url{www.monetdb.org}.

\bibitem{postgres}
{PostgreSQL}.
\newblock \url{www.postgresql.org}.

\end{thebibliography}

\end{document}